\documentclass{article}
\usepackage{hyperref}
\usepackage{amsmath}
\usepackage{graphicx}
\usepackage{amsfonts}
\usepackage{amssymb}
\renewcommand{\theequation}{\arabic{section}.\arabic{equation}}
\begin{document}

\title{Analytical renormalization of large-size expansion for polygonal Wilson loops
in effective string theory}
\author{P.V. Pobylitsa\\\emph{Petersburg Nuclear Physics Institute}\\\emph{Gatchina, 188300, St. Petersburg, Russia}}
\date{}
\maketitle

\textbf{Abstract}

Schwarz-Christoffel (SC) mapping plays a crucial role in the calculation of
the large-size expansion for polygonal Wilson loops in confining gauge
theories using effective string theory (EST). Recently a new analytical
regularization based on SC\ mapping was suggested and successfully applied to
the calculation of the two-loop contribution of EST in the case of triangular
Wilson loops. We prove that this analytical renormalization produces finite
results for arbitrary polygonal Wilson loops and show that the result of the
analytical renormalization for a given polygonal contour is independent of the
choice of SC\ mapping for this polygon.

\newpage
\tableofcontents
\newpage

\renewcommand{\arraystretch}{1.2} 

\section{Introduction}

\setcounter{equation}{0} 

\subsection{Large-size expansion for Wilson loops}

The problem of quark confinement in Quantum Chromodynamics (QCD) still remains
a challenge. This theoretical problem simplifies if one turns from the
full-fledged QCD with light quarks to the gauge theory containing only heavy
quarks. The interaction between a heavy quark-antiquark pair allows for a
non-relativistic description in terms of quark-antiquark potential that can be
expressed via Wilson loops \cite{Wilson-74}

\begin{equation}
W\left(  C\right)  =\left\langle \mathrm{Tr\,}\text{P}\,\exp\left[  i\oint
_{C}dx^{\mu}A_{\mu}\left(  x\right)  \right]  \right\rangle
\end{equation}
computed in the pure gauge theory with Euclidean time using the path-ordered
exponent taken along a closed contour $C$.

In a wide class of gauge theories Wilson loops obey area law \cite{Wilson-74}
\begin{equation}
\ln W(C)\overset{\left|  C\right|  \rightarrow\infty}{\sim}-\sigma S\left(
C\right)  \label{area-law-1}
\end{equation}
where $S(C)$ is the area of the minimal surface spanned on (generally
non-flat) contour $C$ and notation $\left|  C\right|  \rightarrow\infty$
stands for the large-size limit of contour $C$.

It is a tradition to refer to gauge theories obeying area law
(\ref{area-law-1}) as confining gauge theories for brevity (implying the
confinement of a heavy quark-antiquark pair embedded in the pure gauge
theory), although the problem of quark confinement in real QCD is much more complicated.

Area law (\ref{area-law-1}) sends us a message that some sort of string theory
may stand behind Wilson loops in confining gauge theories. Starting from
qualitative and heuristic arguments \cite{Polyakov-1980}, one tried to justify
the stringy approach to Wilson loops using various limits and expansions:
large size, large number of colors \cite{Hooft-1974}-\cite{Migdal-1984}, large
number of space-time dimensions \cite{Alvarez}-\cite{Makeenko-2012-a}, Regge
limit \cite{Makeenko-2012-a}-\cite{Makeenko-2012-b}.

If one is interested only in large-distance properties in gauge theories
satisfying area law, it is natural to expect that the stringy picture needs
for its justification only the large-size limit and nothing else. This point
of view is implemented in \emph{effective string theory} (EST). EST assumes
that the asymptotic behavior of Wilson loops in the limit of large size of
contour $C$ can be described by a functional integral over surfaces $\Sigma$
bounded by contour $C$:
\begin{equation}
W\left(  C\right)  \rightarrow\mathrm{const}\int_{\partial\Sigma=C}D\Sigma
\exp\left(  -\mathcal{S}_{\mathrm{EST}}\left[  \Sigma\right]  \right)  .
\label{W-C-EST}
\end{equation}
The idea that one can go beyond naive string models and can interpret
functional integral (\ref{W-C-EST}) as a theoretical tool for the construction
of a systematic large-size expansion of Wilson loops has a long history. An
important step in this direction was made by M.~L\"{u}scher, G.~M\"{u}nster,
K.~Symanzik and P.~Weisz \cite{Luscher:1980fr}-\cite{Luscher:1980iy}. In the
computation of the first terms of the large-size expansion, one can
approximate $\mathcal{S}_{\mathrm{EST}}\left[  \Sigma\right]  $ by Nambu
action
\begin{equation}
\mathcal{S}_{\text{Nambu}}\left(  \Sigma\right)  =\sigma_{0}S\left(
\Sigma\right)  \label{S-Nambu}
\end{equation}
where $S\left(  \Sigma\right)  $ is the area of surface $\Sigma$ and
$\sigma_{0}$ is the bare string tension (different from the renormalized
physical string tension $\sigma$ appearing in area law (\ref{area-law-1} )).
If one wants to use EST for the construction of higher terms of the large-size
expansion then $S_{\mathrm{EST}}\left[  \Sigma\right]  $ must be understood as
an infinite series containing all possible terms compatible with the
symmetries of the problem.

EST has developed in various directions including

\begin{itemize}
\item derivation of general constraints on terms appearing in EST action
$S_{\mathrm{EST}}\left[  \Sigma\right]  $ \cite{Luscher:2004ib}-\cite{GM-12},

\item calculation of EST loop corrections for rectangular Wilson loops
\cite{BCGMP-12}, \cite{Filk-preprint}-\cite{Billo:2011fd} and for other
quantities like correlation functions of Polyakov lines and spectra of closed
and open strings \cite{Luscher:2004ib}, \cite{Aharony:2009gg}-\cite{AK-2013},

\item analysis of string finite-width effects \cite{Luscher:1980iy},
\cite{Gliozzi:2010zt}, \cite{Meyer-2010}.
\end{itemize}

EST was successfully tested by lattice Monte Carlo simulations (see
\cite{BCGMP-12}, \cite{BM-16}-\cite{Pobylitsa-2016} and references therein).

EST leads to the following expansion of Wilson loops in the large-size limit:

\begin{align}
&  \ln W\left(  \lambda C\right)  \overset{\lambda\rightarrow\infty}{=}
f_{-2}\left(  C\right)  \lambda^{2}+f_{-1}\left(  C\right)  \lambda\nonumber\\
&  +f_{\ln}\left(  C\right)  \ln\lambda\nonumber\\
&  +f_{0}\left(  C\right)  +f_{2}\left(  C\right)  \lambda^{-2}+f_{3}\left(
C\right)  \lambda^{-3}+\ldots\label{W-lambda-C-general-expansion-0}
\end{align}
Here $\lambda C$ stands for contour $C$ uniformly rescaled by factor $\lambda
$. In this paper we assume that $C\;$is a flat polygon. A detailed discussion
of this expansion can be found in ref. \cite{Pobylitsa-2019}. A short summary is:

\begin{itemize}
\item Coefficient $f_{-2}\left(  C\right)  $ corresponds to area law
(\ref{area-law-1}).

\item Coefficient $f_{-1}\left(  C\right)  $ is proportional to the length of
contour $C$.

\item Coefficients $f_{\ln}\left(  C\right)  $, $f_{0}\left(  C\right)  $ can
be expressed via the functional determinant of the two-dimensional Laplace
operator in the region bounded by contour $C$ with Dirichlet boundary
conditions. This Laplace determinant was computed in ref. \cite{AS-93-journal}
for arbitrary polygons using Schwarz-Christoffel (SC)\ mapping.

\item Coefficient $f_{2}\left(  C\right)  $ was computed for rectangular
contours $C$ in refs. \cite{Filk-preprint}, \cite{Dietz-83} but later
Bill\'{o} et al. \cite{BCVG-2010}, \cite{Billo:2011fd} detected an arithmetic
error in the result of refs. \cite{Filk-preprint}, \cite{Dietz-83} and
corrected it.

\item In ref. \cite{Pobylitsa-2019} coefficient $f_{2}(C)$ was computed for
triangular diagrams and a certain progress was made towards the calculation of
$f_{2}(C)$ for arbitrary polygonal contours $C$.
\end{itemize}

EST leads to the following general expression for $f_{2}\left(  C\right)  $ \cite{Pobylitsa-2019}

\begin{equation}
f_{2}\left(  C\right)  =\frac{1}{\sigma}\frac{D-2}{\left(  8\pi\right)  ^{2}
}\left[  I_{1}^{\text{ren}}\left(  C\right)  +\frac{2}{9}\left(  D-2\right)
I_{2}^{\text{ren}}\left(  C\right)  \right]  \,. \label{f2-via-I-I2-ren}
\end{equation}

Here

\begin{itemize}
\item $\sigma$ is the string tension appearing in Wilson law (\ref{area-law-1}),

\item $D$ is the space-time dimension of the confining gauge theory,

\item quantities $I_{m}^{\text{ren}}\left(  C\right)  $ depend only contour
$C$ and on nothing else.
\end{itemize}

The structure of expression (\ref{f2-via-I-I2-ren}) is determined by a certain
(8-shaped) Feynman diagram of EST \cite{Filk-preprint}, \cite{Dietz-83}. The
dynamics of the underlying confining gauge theory enters only via string
tension $\sigma$. The dependence on dimension $D$ comes from the tensor
algebra of the Feynman diagram of EST for $f_{2}\left(  C\right)  $. Common
factor $D-2$ in (\ref{f2-via-I-I2-ren}) reflects the triviality of $D=2$ pure
gauge theories and appears in EST via the number of transverse degrees of freedom.

Once $\sigma$ and $D$ dependence is factored out (\ref{f2-via-I-I2-ren}), the
nontrivial part of the calculation of $f_{2}\left(  C\right)  $ is localized
in $I_{m}^{\text{ren}}\left(  C\right)  $.

The Feynman diagram of EST for $f_{2}(C)$ has ultraviolet divergences that
must be renormalized (in the sense of effective field theories). When choosing
an ultraviolet regularization, one has to find a reasonable compromise between
a solid theoretical status of the calculation and computational efficiency.

If one is interested in the calculation of $f_{2}\left(  C\right)  $ for
arbitrary polygons $C$ then as shown in ref. \cite{Pobylitsa-2019} a certain
progress may be achieved by using

\begin{itemize}
\item Schwarz-Christoffel (SC) mapping for polygon $C$,

\item analytical regularization formulated in terms of parameters of SC mapping.
\end{itemize}

\subsection{From triangles to general polygons}

In ref. \cite{Pobylitsa-2019} the problem of the computation of $f_{2}\left(
C\right)  $ was completely solved for triangular contours $C$ but with only a
partial progress for general polygonal contours $C$. In case of arbitrary
polygons the results of ref. \cite{Pobylitsa-2019} were limited to

\begin{itemize}
\item derivation of naive \emph{ultraviolet divergent} integral
representations for nonrenormalized analogs $I_{m}\left(  C\right)  $ of
renormalized functions $I_{m}^{\text{ren}}\left(  C\right)  $ appearing in
(\ref{f2-via-I-I2-ren}),

\item suggesting a procedure of regularization for these divergent integrals
$I_{m}\left(  C\right)  $ using analytical continuation \emph{without proving}
that this analytical continuation exists and that it gives a finite result for
$I_{m}^{\text{ren}}\left(  C\right)  $,

\item demonstration that in the special case of \emph{triangular} contours $C$
this analytical continuation really exists and produces a finite result (which
was explicitly computed).
\end{itemize}

In the current work we prove that the analytical continuation suggested in
ref. \cite{Pobylitsa-2019} exists and leads to finite values of $I_{m}
^{\text{ren}}\left(  C\right)  $ for \emph{arbitrary} polygons $C$.

\subsection{SC\ mapping}

\label{SC-section}

SC\ transformation $\zeta(z)$ is a conformal mapping of the upper complex $z$
semiplane
\begin{equation}
\mathbb{C}_{+}=\left\{  z\in\mathbb{C}:\operatorname{Im}z>0\right\}
\label{C-plus}
\end{equation}
to a polygon in complex $\zeta$ plane. The real axis of the $z$ plane is
mapped onto the boundary of the polygon whereas $n_{\text{v}}$ real points
$z_{1},\ldots,z_{n_{\text{v}}}$ are mapped to vertices $\zeta_{1},\ldots
,\zeta_{n_{\text{v}}}$ of the polygon.

For a given polygon there exists an infinite set of conformal mappings from
$\mathbb{C}_{+}$ to this polygon because linear fractional transformations
\begin{equation}
z^{\prime}=\frac{az+b}{cz+d}\quad(a,b,c,d\,\in\,\mathbb{R})
\label{z-prime-z-linear-fractional}
\end{equation}
with real coefficients $a,b,c,d$ map upper semiplane $\mathbb{C}_{+}$ to itself.

Using this freedom of linear fractional transformations, one can always choose
SC\ mapping so that
\begin{equation}
z_{n_{\text{v}}}=\infty\label{z-n-v-infinity}
\end{equation}
i.e. infinity of the $z$ plane is mapped to vertex $\zeta_{n_{\text{v}}}$ of
the polygon.

This special type of SC\ mappings simplifies calculations and is always
assumed in our analytical renormalization procedure. We will refer to SC
mappings with $z_{n_{\text{v}}}=\infty$ as SC$_{\infty}$ mappings.
SC$_{\infty}$\ mapping is described by differential equation

\begin{equation}
\frac{d\zeta}{dz}=\tilde{A}\prod_{k=1}^{n_{\text{v}}-1}\left(  z-z_{k}\right)
^{-\beta_{k}} \label{SC-diff-eq}
\end{equation}
where
\begin{equation}
\tilde{A},\left\{  \beta_{k}\right\}  _{k=1}^{n_{\text{v}}-1},\left\{
z_{k}\right\}  _{k=1}^{n_{\text{v}}-1} \label{SC-parameters}
\end{equation}
are SC$_{\infty}$\ parameters of the polygon.

Although condition (\ref{z-n-v-infinity}) restricts the freedom of linear
fractional transformations (\ref{z-prime-z-linear-fractional}), for any given
polygon we still have an infinite amount of SC$_{\infty}$ mappings from
$\mathbb{C}_{+}$ to this polygon (see section \ref{linear-fractional-section}).

Interior angles $\theta_{k}$ of the polygon associated with SC\ vertices
$z_{k}$ are given by
\begin{equation}
\theta_{k}=\pi\left(  1-\beta_{k}\right)  \quad\left(  1\leq k\leq
n_{\text{\textbf{v}}}\right)  . \label{theta-k-beta-k}
\end{equation}
Parameters $\beta_{k}$ appear in SC\ equation (\ref{SC-diff-eq}) only for
$1\leq k\leq n_{\text{\textbf{v}}\mathbf{-1}}$ but we define also
$\beta_{n_{\text{v}}}$ by extrapolating relation (\ref{theta-k-beta-k}) to
$k=n_{\text{\textbf{v}}}$. Then the geometric property
\begin{equation}
\sum_{k=1}^{n_{\text{v}}}\left(  \pi-\theta_{k}\right)  =2\pi
\end{equation}
leads to

\begin{equation}
\beta_{n_{\text{v}}}=2-\sum_{k=1}^{n_{\text{v}}-1}\beta_{k}. \label{beta-n-v}
\end{equation}
Interior angles of the polygon belong to the range
\[
0<\theta_{k}<2\pi\,,\quad\theta_{k}\neq\pi.
\]
Together with eq. (\ref{theta-k-beta-k}) this leads to
\begin{equation}
0<\beta_{k}<2\,,\,\beta_{k}\neq1\quad(1\leq k\leq n_{\text{v}})\,.
\label{beta-k-range}
\end{equation}

\subsection{Functions $\Pi_{P}^{(n)}$}

In ref. \cite{Pobylitsa-2019} quantities $I_{m}^{\text{ren}}$ appearing in eq.
(\ref{f2-via-I-I2-ren}) were expressed via certain functions $\Pi_{P}^{(n)}$.
The explicit expression for $I_{m}^{\text{ren}}$ via functions $\Pi_{P}^{(n)}$
will be given below in eqs. (\ref{I-1-ren-via-Pi}), (\ref{I-2-ren-via-Pi}).
But first it makes sense to discuss the status of arguments of these functions
$\Pi_{P}^{(n)}\left(  \alpha,\left\{  \gamma_{k}\right\}  _{k=1}^{n_{\text{v}
}-1},\left\{  z_{k}\right\}  _{k=1}^{n_{\text{v}}-1}\right)  $:

\begin{itemize}
\item $\alpha,\left\{  \gamma_{k}\right\}  _{k=1}^{n_{\text{v}}-1}$ are
generally complex variables,

\item $\left\{  z_{k}\right\}  _{k=1}^{n_{\text{v}}-1}$ are real variables,

\item $P\left(  z,z^{\ast}\right)  $ is a polynomial of two variables.
\end{itemize}

We use a widespread but slightly misleading notation: from \emph{algebraic}
point of view $P\left(  z,z^{\ast}\right)  $ is a polynomial of two
\emph{independent} variables $z,z^{\ast}$, e.g. this interpretation can used
in eq. (\ref{P-symmetric-1}), however, quite often, e.g. in eq.
(\ref{Pi-P-n-def}), $z$ and $z^{\ast}$ are treated as complex conjugate variables.

We assume that polynomial $P\left(  z,z^{\ast}\right)  $ has property

\begin{equation}
P\left(  z,z^{\ast}\right)  =P\left(  z^{\ast},z\right)  \,.
\label{P-symmetric-1}
\end{equation}

Functions $\Pi_{P}^{(n)}$ are formally defined by the integral
\begin{equation}
\Pi_{P}^{(n)}\left(  \alpha,\left\{  \gamma_{k}\right\}  _{k=1}^{n},\left\{
z_{k}\right\}  _{k=1}^{n}\right)  =\int_{\mathbb{C}_{+}}d^{2}z\,\left|
\operatorname{Im}\,z\right|  ^{\alpha-1}\prod_{k=1}^{n}\left|  z-z_{k}\right|
^{\gamma_{k}-\alpha-1}P\left(  z,z^{\ast}\right)  \, \label{Pi-P-n-def}
\end{equation}
running over upper complex semiplane $\mathbb{C}_{+}$ (\ref{C-plus}). Our
normalization of the integration measure is
\begin{equation}
d^{2}z=(d\operatorname{Re}z)(d\operatorname{Im}z)\,. \label{d-2-z-measure}
\end{equation}

The integral on the RHS of (\ref{Pi-P-n-def}) may be divergent. In ref.
\cite{Pobylitsa-2019} it was suggested to understand function $\Pi_{P}^{(n)}$
as follows:

1) concentrate on the dependence of $\Pi_{P}^{(n)}$ on complex variables
$\alpha,\left\{  \gamma_{k}\right\}  _{k=1}^{n}$ but fix real parameters
$\left\{  z_{k}\right\}  _{k=1}^{n}$ and polynomial $P$,

2) first define function $\Pi_{P}^{(n)}$ in the subset of complex space
$\mathbb{C}^{n+1}$ of variables $\alpha,\left\{  \gamma_{k}\right\}
_{k=1}^{n}$ where integral (\ref{Pi-P-n-def}) is convergent,

3) then analytically continue $\Pi_{P}^{(n)}$ in $\alpha,\left\{  \gamma
_{k}\right\}  _{k=1}^{n}$ as far as possible (expecting to obtain a
meromorphic function in $\mathbb{C}^{n+1}$).

These steps are rather nontrivial:

\begin{itemize}
\item One must prove that integral (\ref{Pi-P-n-def}) is convergent in some
non-empty region of variables $\alpha,\left\{  \gamma_{k}\right\}  _{k=1}^{n}$.

\item One must prove that analytical continuation of integral
(\ref{Pi-P-n-def}) leads to function $\Pi_{P}^{(n)}\left(  \alpha,\left\{
\gamma_{k}\right\}  _{k=1}^{n},\left\{  z_{k}\right\}  _{k=1}^{n}\right)  $
that is meromorphic in $\alpha,\left\{  \gamma_{k}\right\}  _{k=1}^{n}$ (in
the sense of the theory of functions of several complex variables).

\item One must prove that function $\Pi_{P}^{(n)}\left(  \alpha,\left\{
\gamma_{k}\right\}  _{k=1}^{n},\left\{  z_{k}\right\}  _{k=1}^{n}\right)  $ is
regular at those values of $\alpha,\left\{  \gamma_{k}\right\}  _{k=1}^{n}$
arguments which are used for the computation of $I_{m}^{\text{ren}}$ in eqs.
(\ref{I-1-ren-via-Pi}), (\ref{I-2-ren-via-Pi}).
\end{itemize}

These properties of functions $\Pi_{P}^{(n)}$ were announced in ref.
\cite{Pobylitsa-2019} for arbitrary polygons but checked only for triangles.
The proof of these statements for arbitrary polygons is one of the main
subjects of this paper.

\subsection{Expressions for $I_{m}^{\text{ren}}$}

In ref. \cite{Pobylitsa-2019} it was suggested

\begin{itemize}
\item to compute Feynman diagram of EST contributing to $f_{2}\left(
C\right)  $ (\ref{f2-via-I-I2-ren}) using SC$_{\infty}$\ mapping
(\ref{SC-diff-eq}) for the polygon bounded by contour $C$,

\item to perform renormalization of ultraviolet\ divergences of this Feynman
diagram using analytical renormalization formulated in terms of SC$_{\infty}
$\ parameters (\ref{SC-parameters}).
\end{itemize}

In this approach quantities $I_{m}^{\text{ren}}\left(  C\right)  $ arise as
functions of SC$_{\infty}$\ parameters (\ref{SC-parameters}). In ref.
\cite{Pobylitsa-2019} $I_{m}^{\text{ren}}$ were expressed via functions
$\Pi_{P}^{(n)}$ (\ref{Pi-P-n-def})
\begin{equation}
I_{1}^{\text{ren}}\left(  \tilde{A},\left\{  \beta_{k}\right\}  _{k=1}
^{n_{\text{v}}-1},\left\{  z_{k}\right\}  _{k=1}^{n_{\text{v}}-1}\right)
=\left|  \tilde{A}\right|  ^{-2}\Pi_{1}^{(n_{\text{v}}-1)}\left(  -3,\left\{
2\beta_{k}-2\right\}  _{k=1}^{n_{\text{v}}-1},\left\{  z_{k}\right\}
_{k=1}^{n_{\text{v}}-1}\right)  \,,\, \label{I-1-ren-via-Pi}
\end{equation}
\begin{equation}
I_{2}^{\text{ren}}\left(  \tilde{A},\left\{  \beta_{k}\right\}  _{k=1}
^{n_{\text{v}}-1},\left\{  z_{k}\right\}  _{k=1}^{n_{\text{v}}-1}\right)
=\left|  \tilde{A}\right|  ^{-2}\Pi_{P_{2}}^{(n_{\text{v}}-1)}\left(
1,\left\{  2\beta_{k}-2\right\}  _{k=1}^{n_{\text{v}}-1},\left\{
z_{k}\right\}  _{k=1}^{n_{\text{v}}-1}\right)  \,. \label{I-2-ren-via-Pi}
\end{equation}
Here $\Pi_{1}^{(n_{\text{v}}-1)}$ stands for \ $\Pi_{P}^{(n_{\text{v}}-1)}$
with trivial polynomial $P\left(  z,z^{\ast}\right)  =1$ and in $\Pi_{P_{2}
}^{(n_{\text{v}}-1)}$ polynomial $P_{2}$ is given by
\begin{equation}
P_{2}\left(  z,z^{\ast}\right)  =T_{2}\left(  z\right)  T_{2}\left(  z^{\ast
}\right)  \label{P2-via-T2}
\end{equation}
where polynomial $T_{2}$ is defined by
\begin{align}
T_{2}\left(  z\right)   &  =T_{2}\left(  z;\left\{  \beta_{k}\right\}
_{k=1}^{n_{\text{v}}-1},\left\{  z_{k}\right\}  _{k=1}^{n_{\text{v}}-1}\right)
\nonumber\\
&  =\left\{  \left[  \sum_{k=1}^{n_{\text{v}}-1}\frac{\beta_{k}}{\left(
z-z_{k}\right)  ^{2}}\right]  -\frac{1}{2}\left[  \sum_{k=1}^{n_{\text{v}}
-1}\frac{\beta_{k}}{z-z_{k}}\right]  ^{2}\right\}  \prod_{k=1}^{n_{\text{v}
}-1}\left(  z-z_{k}\right)  ^{2}\,. \label{T2-def}
\end{align}
Although the expression on the RHS contains terms with negative powers of
$\left(  z-z_{k}\right)  $, after the expansion of brackets one arrives at a
polynomial in $z$.

\subsection{Polynomials $T_{2}$ and $P_{2}$}

This section can be ignored at the first reading. Here we discuss some
subtleties concerning polynomials $P_{2}$ and $T_{2}$. Index $2$ in notation
$P_{2},T_{2}$ comes from the fact that these polynomials are associated with
quantity $I_{2}^{\text{ren}}$ (\ref{I-2-ren-via-Pi}). Usually we use compact
notation $T_{2}\left(  z\right)  $ for the RHS of (\ref{T2-def}) concentrating
on the $z$ dependence at fixed $\left\{  \beta_{k}\right\}  _{k=1}
^{n_{\text{v}}-1},\left\{  z_{k}\right\}  _{k=1}^{n_{\text{v}}-1}$. But in
some cases expanded notation $T_{2}\left(  z;\left\{  \beta_{k}\right\}
_{k=1}^{n_{\text{v}}-1},\left\{  z_{k}\right\}  _{k=1}^{n_{\text{v}}
-1}\right)  $ containing the full set of arguments is preferable. The expanded
notation may be also useful for $P_{2}$:
\begin{align}
&  P_{2}\left(  z,z^{\ast};\left\{  \beta_{k}\right\}  _{k=1}^{n_{\text{v}}
-1},\left\{  z_{k}\right\}  _{k=1}^{n_{\text{v}}-1}\right) \nonumber\\
&  =T_{2}\left(  z;\left\{  \beta_{k}\right\}  _{k=1}^{n_{\text{v}}
-1},\left\{  z_{k}\right\}  _{k=1}^{n_{\text{v}}-1}\right)  T_{2}\left(
z^{\ast};\left\{  \beta_{k}\right\}  _{k=1}^{n_{\text{v}}-1},\left\{
z_{k}\right\}  _{k=1}^{n_{\text{v}}-1}\right)  \,. \label{P2-T2-expanded}
\end{align}

Obviously $T_{2}\left(  z\right)  $ is a holomorphic polynomial of $z$ at any
fixed $\left\{  \beta_{k}\right\}  _{k=1}^{n_{\text{v}}-1},\left\{
z_{k}\right\}  _{k=1}^{n_{\text{v}}-1}$. If parameters $\left\{  \beta
_{k}\right\}  _{k=1}^{n_{\text{v}}-1},\left\{  z_{k}\right\}  _{k=1}
^{n_{\text{v}}-1}$ are real then
\begin{equation}
T_{2}\left(  z^{\ast}\right)  =\left[  T_{2}\left(  z\right)  \right]  ^{\ast
}\,. \label{T-2-conjugation}
\end{equation}
We (almost) always keep parameters $\left\{  \beta_{k}\right\}  _{k=1}
^{n_{\text{v}}-1},\left\{  z_{k}\right\}  _{k=1}^{n_{\text{v}}-1}$ real so
that relation (\ref{T-2-conjugation}) holds (almost) always. Whenever we
mention analytical continuation of functions $\Pi_{P}^{(n)}\left(
\alpha,\left\{  \gamma_{k}\right\}  _{k=1}^{n},\left\{  z_{k}\right\}
_{k=1}^{n}\right)  $, we imply analytical continuation in complex variables
$\alpha,\left\{  \gamma_{k}\right\}  _{k=1}^{n}$ at fixed real $\left\{
z_{k}\right\}  _{k=1}^{n}$ and at fixed polynomial $P$. In particular,
expressions (\ref{I-1-ren-via-Pi}), (\ref{I-2-ren-via-Pi}) must be understood
as follows:

1) construct function $\Pi_{P}^{(n)}\left(  \alpha,\left\{  \gamma
_{k}\right\}  _{k=1}^{n},\left\{  z_{k}\right\}  _{k=1}^{n}\right)  $ starting
from its integral representation (\ref{Pi-P-n-def}) for those complex
$\alpha,\left\{  \gamma_{k}\right\}  _{k=1}^{n}$ and real $\left\{
z_{k}\right\}  _{k=1}^{n}$ where the integral is convergent,

2) next continue function $\Pi_{P}^{(n)}\left(  \alpha,\left\{  \gamma
_{k}\right\}  _{k=1}^{n},\left\{  z_{k}\right\}  _{k=1}^{n}\right)  $ in
complex $\alpha,\left\{  \gamma_{k}\right\}  _{k=1}^{n}$ at fixed real
$\left\{  z_{k}\right\}  _{k=1}^{n}$ and at fixed polynomial $P$,

3) only after that use the resulting (analytically continued) functions
$\Pi_{P}^{(n)}\left(  \alpha,\left\{  \gamma_{k}\right\}  _{k=1}^{n},\left\{
z_{k}\right\}  _{k=1}^{n}\right)  $ in relations (\ref{I-1-ren-via-Pi}),
(\ref{I-2-ren-via-Pi}), substituting \emph{real} values
\begin{equation}
\left\{  2\beta_{k}-2\right\}  _{k=1}^{n_{\text{v}}-1}\,,\,\left\{
z_{k}\right\}  _{k=1}^{n_{\text{v}}-1}
\end{equation}
for arguments of $\Pi_{P}^{(n)}$.

We never use analytical continuation in $\beta_{k}$ and we (almost) always
keep $\beta_{k},z_{k}$ real.

For real $\left\{  \beta_{k}\right\}  _{k=1}^{n_{\text{v}}-1},\left\{
z_{k}\right\}  _{k=1}^{n_{\text{v}}-1}$ (which is almost always our case) we
have
\begin{equation}
P_{2}\left(  z,z^{\ast}\right)  =P_{2}\left(  z^{\ast},z\right)  \,,
\end{equation}
i.e. $P_{2}$ obeys general condition (\ref{P-symmetric-1}).

\subsection{Outline of the work}

\subsubsection{Analytical continuation}

In ref. \cite{Pobylitsa-2019} some preliminary comments were made about the
construction of the rigorous definition of functions $\Pi_{P}^{(n)}$ and some
basic properties of functions $\Pi_{P}^{(n)}$ were announced without a proof
for arbitrary polygons, including

\begin{itemize}
\item meromorphy of $\Pi_{P}^{(n)}$ in $\alpha,\left\{  \gamma_{k}\right\}
_{k=1}^{n}$,

\item regularity of $\Pi_{P}^{(n)}$ at points appearing in the above
expressions for $I_{m}^{\text{ren}}$ (\ref{I-1-ren-via-Pi}),
(\ref{I-2-ren-via-Pi}).
\end{itemize}

The aim of the current paper is to provide proofs of these properties.

The main steps of our work are as follows:

\textbf{1.} In section \ref{convergence-of-integrals-section} we prove that
integral on the RHS of eq. (\ref{Pi-P-n-def}) defining function $\Pi_{P}
^{(n)}$ is convergent in a certain non-empty region of complex parameters
$\alpha,\left\{  \gamma_{k}\right\}  _{k=1}^{n}$ for any fixed real parameters
$\left\{  z_{k}\right\}  _{k=1}^{n_{\text{v}}-1}$ and for a fixed polynomial
$P$.

\textbf{2.} In section \ref{Analytical-continuation-section} we prove that
function $\Pi_{P}^{(n)}$ (originally defined for $\alpha,\left\{  \gamma
_{k}\right\}  _{k=1}^{n}$ in the convergence region) allows for an analytical
continuation to a \emph{meromorphic} function in the full $\mathbb{C}^{n+1}$
space of complex variables $\alpha,\left\{  \gamma_{k}\right\}  _{k=1}^{n}$.
In other (slightly oversimplified) words, this analytical continuation of
$\Pi_{P}^{(n)}$ is regular `almost for all complex $\alpha,\left\{  \gamma
_{k}\right\}  _{k=1}^{n}$' except for simple pole singularities.

\textbf{3.} In section \ref{Analytical-continuation-section} we derive a
representation which exhibits the pole structure of $\Pi_{P}^{(n)}$:

\begin{align}
&  \Pi_{P}^{\left(  n\right)  }\left(  \alpha,\left\{  \gamma_{k}\right\}
_{k=1}^{n},\left\{  z_{k}\right\}  _{k=1}^{n}\right) \nonumber\\
&  =\Gamma\left(  \frac{\alpha}{2}\right)  \Gamma\left(  \frac{1}{2}\left(
-2M_{P}+\left(  n-1\right)  \left(  \alpha+1\right)  -\sum_{k=1}^{n}\gamma
_{k}\right)  \right)  \prod_{j=1}^{n}\Gamma\left(  \frac{\gamma_{j}}{2}\right)
\nonumber\\
&  \times H_{P}^{\left(  n\right)  }\left(  \alpha,\left\{  \gamma
_{k}\right\}  _{k=1}^{n},\left\{  z_{k}\right\}  _{k=1}^{n}\right)  \,.
\label{K-rough-pole-structure}
\end{align}
Here

\begin{itemize}
\item $H_{P}^{\left(  n\right)  }\left(  \alpha,\left\{  \gamma_{k}\right\}
_{k=1}^{n},\left\{  z_{k}\right\}  _{k=1}^{n}\right)  $ is an entire function
in $\mathbb{C}^{n+1}$ space of complex variables $\alpha,\left\{  \gamma
_{k}\right\}  _{k=1}^{n}$ (i.e. regular analytical function in the whole space
$\mathbb{C}^{n+1}$),

\item $M_{P}$ is an integer number depending on polynomial $P\left(
z,z^{\ast}\right)  $. $M_{P}$ is defined by eq. (\ref{M-P-min-def}) in section
\ref{parameter-M-P-section}.
\end{itemize}

Representation (\ref{K-rough-pole-structure}) separates $\alpha,\left\{
\gamma_{k}\right\}  _{k=1}^{n}$ singularities of functions $\Pi_{P}^{\left(
n\right)  }\left(  \alpha,\left\{  \gamma_{k}\right\}  _{k=1}^{n},\left\{
z_{k}\right\}  _{k=1}^{n}\right)  $ in terms of Euler $\Gamma$ functions.

In Appendix \ref{pole-structure-Pi-2-1-section} we show that the explicit
expression for the special case of $\Pi_{1}^{(2)}$ computed in ref.
\cite{Pobylitsa-2019} agrees with general pole structure
(\ref{K-rough-pole-structure}).

\textbf{4.} After the derivation of representation
(\ref{K-rough-pole-structure}), the rest of the work is rather simple.
Analytical continuation of function $\Pi_{P}^{(n)}$ from the convergence
region of integral (\ref{Pi-P-n-def}) is unambiguous: representation
(\ref{K-rough-pole-structure}) shows that function $\Pi_{P}^{\left(  n\right)
}$ has no branching singularities so that this analytical continuation does
not depend on the path in the $\mathbb{C}^{n+1}$ space of parameters
$\alpha,\left\{  \gamma_{k}\right\}  _{k=1}^{n}$.

\textbf{5.} The proof of the finiteness of analytical renormalization for
$I_{m}^{\text{ren}}$ reduces to testing that arguments of functions $\Pi
_{P}^{(n_{\text{v}}-1)}$ \ appearing on the RHS of eqs. (\ref{I-1-ren-via-Pi}
), (\ref{I-2-ren-via-Pi}) do not overlap with poles of Euler $\Gamma$
functions on the RHS of eq. (\ref{K-rough-pole-structure}). This work is done
in section \ref{Analytical-renormalization-finite-section}.

\subsubsection{Invariance with respect to SC\ reparametrization}

\label{different-SC-parametrizations-problem}

As discussed above, quantities $I_{m}^{\text{ren}}\left(  C\right)  $
appearing in eq. (\ref{f2-via-I-I2-ren}) must depend only on the geometry of
polygonal contour $C$. But in the computation of the EST Feynman diagram in
SC$_{\infty}$\ representation, quantities $I_{m}^{\text{ren}}$
(\ref{I-1-ren-via-Pi}), (\ref{I-2-ren-via-Pi}) arise as functions of
SC$_{\infty}$\ parameters (\ref{SC-parameters}). As mentioned in section
\ref{SC-section}, any polygon allows for many different SC$_{\infty}
$\ mappings with different SC$_{\infty}$ parameters. A\ priori it is not
obvious that starting from different SC$_{\infty}$ parametrizations
$\tilde{A},\left\{  \beta_{k}\right\}  _{k=1}^{n_{\text{v}}-1},\left\{
z_{k}\right\}  _{k=1}^{n_{\text{v}}-1}$ of the same polygon, one arrives at
the same results for $I_{m}^{\text{ren}}\left(  \tilde{A},\left\{  \beta
_{k}\right\}  _{k=1}^{n_{\text{v}}-1},\left\{  z_{k}\right\}  _{k=1}
^{n_{\text{v}}-1}\right)  $ (\ref{I-1-ren-via-Pi}), (\ref{I-2-ren-via-Pi}).

Fortunately there is no problem: in section
\ref{renormalization-invariance-inversion-section} we prove that the result
for $I_{m}^{\text{ren}}$ depends only on the geometry of the polygon and not
on its SC$_{\infty}$\ parametrization.

\subsubsection{Physics and mathematics of functions $\Pi_{P}^{\left(
n\right)  }$}

The central subject of this paper is properties of analytical regularization
(and renormalization) for the two-loop term $f_{2}(C)$ in EST expansion
(\ref{W-lambda-C-general-expansion-0}) and not mathematics of functions
$\Pi_{P}^{\left(  n\right)  }$. Our approach to the analysis of functions
$\Pi_{P}^{\left(  n\right)  }$ is rather utilitarian and devoid of
mathematical elegance and perfectionism. The arguments use standard and rather
simple mathematical methods. Most of calculations and proofs are described in
detail but experts may easily find their own path to the results of this work
after looking through the basic guidelines.

\subsubsection{Ideas and technical details}

The paper is structured in a way that can help those readers who are
interested more in ideas rather than in technical details: we start from the
discussion of main final results and from basic underlying ideas. Then we
explain how these final statements may be derived from auxiliary technical
results and in the end prove these technical results.

\section{Conventions and assumptions}

\label{conventions-section}

\setcounter{equation}{0} 

\subsection{Functions $\Pi_{P}^{\left(  n\right)  }$}

In the sections devoted to properties of functions $\Pi_{P}^{\left(  n\right)
}\left(  \alpha,\left\{  \gamma_{k}\right\}  _{k=1}^{n},\left\{
z_{k}\right\}  _{k=1}^{n}\right)  $ we make the following assumptions (if the
opposite is not explicitly claimed):

1) $n$ is an integer number obeying condition
\begin{equation}
n\geq2\,. \label{n-ge-2}
\end{equation}

2)\ Parameters $z_{j}$ are real
\begin{equation}
\operatorname{Im}\,z_{j}=0\,. \label{Im-z-k-zero}
\end{equation}

3) All parameters $z_{j}$ are different
\begin{equation}
j\neq k\quad\Longrightarrow\quad z_{j}\neq z_{k}\,. \label{all-z-j-different}
\end{equation}

4) Usually we assume that parameters $z_{j}$ are ordered
\begin{equation}
z_{1}<z_{2}<\ldots<z_{n}\,. \label{z-k-monotonic-order}
\end{equation}
However, in section \ref{renormalization-invariance-inversion-section} we do
not impose this constraint.

5) Polynomial $P\left(  z,z^{\ast}\right)  $ has symmetry property
(\ref{P-symmetric-1}).

7) In most statements about convergence and analytical continuation we assume
that $P\neq0$ and avoid comments about the exceptional but trivial case $P=0$.

\subsection{Convergence of integrals and analytical continuation}

We use notation $\mathbb{C}_{+}$ (\ref{C-plus}) for the upper complex semiplane.

Integration measure $d^{2}z$ in the complex $z$ plane is normalized by
condition (\ref{d-2-z-measure}).

Whenever convergence of integrals is discussed, e.g.
\begin{equation}
\int_{U}d^{2}zf\left(  z\right)  ,
\end{equation}
we imply absolute convergence:
\begin{equation}
\int_{U}d^{2}z\left|  f\left(  z\right)  \right|  <\infty\,.
\end{equation}

When we speak about analytical continuation of functions $\Pi_{P}^{\left(
n\right)  }\left(  \alpha,\left\{  \gamma_{k}\right\}  _{k=1}^{n},\left\{
z_{k}\right\}  _{k=1}^{n}\right)  $ from the convergence region, we always
mean analytical continuation in the $C^{n+1}$ space of complex variables
$\alpha,\left\{  \gamma_{k}\right\}  _{k=1}^{n}$ at fixed $\left\{
z_{k}\right\}  _{k=1}^{n}$ and at fixed $P$ starting from the region of this
$C^{n+1}$ space where integral (\ref{Pi-P-n-def}) defining function $\Pi
_{P}^{\left(  n\right)  }$ is absolutely convergent.

\section{Regions of integration}

\setcounter{equation}{0} 

\subsection{From complex semiplane to complex plane}

Functions $\Pi_{P}^{(n)}$ are formally defined by integral (\ref{Pi-P-n-def})
running over the upper semiplane $\mathbb{C}_{+}$. Changing the integration
variable in (\ref{Pi-P-n-def})
\begin{equation}
z\rightarrow z^{\ast}
\end{equation}
and using property (\ref{P-symmetric-1}), we find a similar representation in
terms of the integration over the lower complex semiplane $\mathbb{C}_{-}$
\begin{equation}
\mathbb{C}_{-}=\left\{  z\in\mathbb{C}:\operatorname{Im}z<0\right\}  \,,
\end{equation}
\begin{equation}
\Pi_{P}^{(n)}\left(  \alpha,\left\{  \gamma_{k}\right\}  _{k=1}^{n},\left\{
z_{k}\right\}  _{k=1}^{n}\right)  =\int_{\mathbb{C}_{-}}d^{2}z\,\left|
\operatorname{Im}\,z\right|  ^{\alpha-1}\prod_{k=1}^{n}\left|  z-z_{k}\right|
^{\gamma_{k}-\alpha-1}P\left(  z,z^{\ast}\right)  \,\,.
\label{Pi-P-n-lower-semiplane}
\end{equation}
Taking the average of (\ref{Pi-P-n-def}) and (\ref{Pi-P-n-lower-semiplane}),
we arrive at
\begin{equation}
\Pi_{P}^{(n)}\left(  \alpha,\left\{  \gamma_{k}\right\}  _{k=1}^{n},\left\{
z_{k}\right\}  _{k=1}^{n}\right)  =\frac{1}{2}\int_{\mathbb{C}}d^{2}z\,\left|
\operatorname{Im}\,z\right|  ^{\alpha-1}\prod_{k=1}^{n}\left|  z-z_{k}\right|
^{\gamma_{k}-\alpha-1}P\left(  z,z^{\ast}\right)  \,. \label{Pi-P-n-plane}
\end{equation}

The problems of convergence and analytical continuation can be studied using
any of integral representations (\ref{Pi-P-n-def}),
(\ref{Pi-P-n-lower-semiplane}) or (\ref{Pi-P-n-plane}). The choice of the
representation is a matter of convenience. Integral (\ref{Pi-P-n-def}) over
upper semiplane $\mathbb{C}_{+}$ is convenient for the determination of
convergence region whereas integral (\ref{Pi-P-n-plane}) over complex plane
$\mathbb{C}$ is preferable at some stages of the study of the analytical
continuation in complex variables $\alpha,\left\{  \gamma_{k}\right\}
_{k=1}^{n}$, e.g. in section \ref{Meromorphic-structure-D-z-j-r-j-section}.

\subsection{Splitting complex plane in regions}

In order to proceed with the analysis of convergence region and with
analytical continuation, we want to split the original semiplane or plane
integration region into subregions. Let us start from the integral over
complex plane $\mathbb{C}$ (\ref{Pi-P-n-plane}) and split $\mathbb{C}$ in
\begin{equation}
\mathbb{C=}\bigcup_{j}D_{j} \label{C-union-D-alpha}
\end{equation}
so that
\begin{equation}
j\neq m\quad\Longrightarrow\quad\text{measure}\left(  D_{j}\cap D_{m}\right)
=0\,. \label{no-overlap-1}
\end{equation}
Then
\begin{align}
&  \Pi_{P}^{(n)}\left(  \alpha,\left\{  \gamma_{k}\right\}  _{k=1}
^{n},\left\{  z_{k}\right\}  _{k=1}^{n}\right) \nonumber\\
&  =\frac{1}{2}\sum_{j}\int_{D_{j}}d^{2}z\,\left|  \operatorname{Im}
\,z\right|  ^{\alpha-1}\prod_{k=1}^{n}\left|  z-z_{k}\right|  ^{\gamma
_{k}-\alpha-1}P\left(  z,z^{\ast}\right)  \,. \label{Pi-D-b}
\end{align}
Our regions $D_{k}$ will be invariant under complex conjugation $z\rightarrow
z^{\ast}$.

\subsection{Case of semiplane}

Let us define
\begin{equation}
D_{j,+}=D_{j}\cap\mathbb{C}_{+}\,, \label{D-plus-def-via-D-b}
\end{equation}
\begin{equation}
\mathbb{C}_{+}\mathbb{=}\bigcup_{j}D_{j,+}\,. \label{C-plus-D-j-plus}
\end{equation}

Starting from (\ref{Pi-P-n-def}) and using (\ref{C-plus-D-j-plus}), we derive
an equivalent representation for $\Pi_{P}^{(n)}$:
\begin{align}
& \Pi_{P}^{(n)}\left(  \alpha,\left\{  \gamma_{k}\right\}  _{k=1}^{n},\left\{
z_{k}\right\}  _{k=1}^{n}\right)  \nonumber\\
& =\sum_{j}\int_{D_{j,+}}d^{2}z\,\left|  \operatorname{Im}\,z\right|
^{\alpha-1}\prod_{k=1}^{n}\left|  z-z_{k}\right|  ^{\gamma_{k}-\alpha
-1}P\left(  z,z^{\ast}\right)  \,.\label{Pi-D-plus-splitting}
\end{align}

\subsection{Choice of $D_{j}$}

\label{choce-D-b-section}

Now we want to specify $n+2$ regions $D_{j}$ labelled by index $j$ running in
the interval
\begin{equation}
0\leq j\leq n+1\,.
\end{equation}

\begin{figure}[ptb]
\begin{center}
\includegraphics[
width=8cm
]
{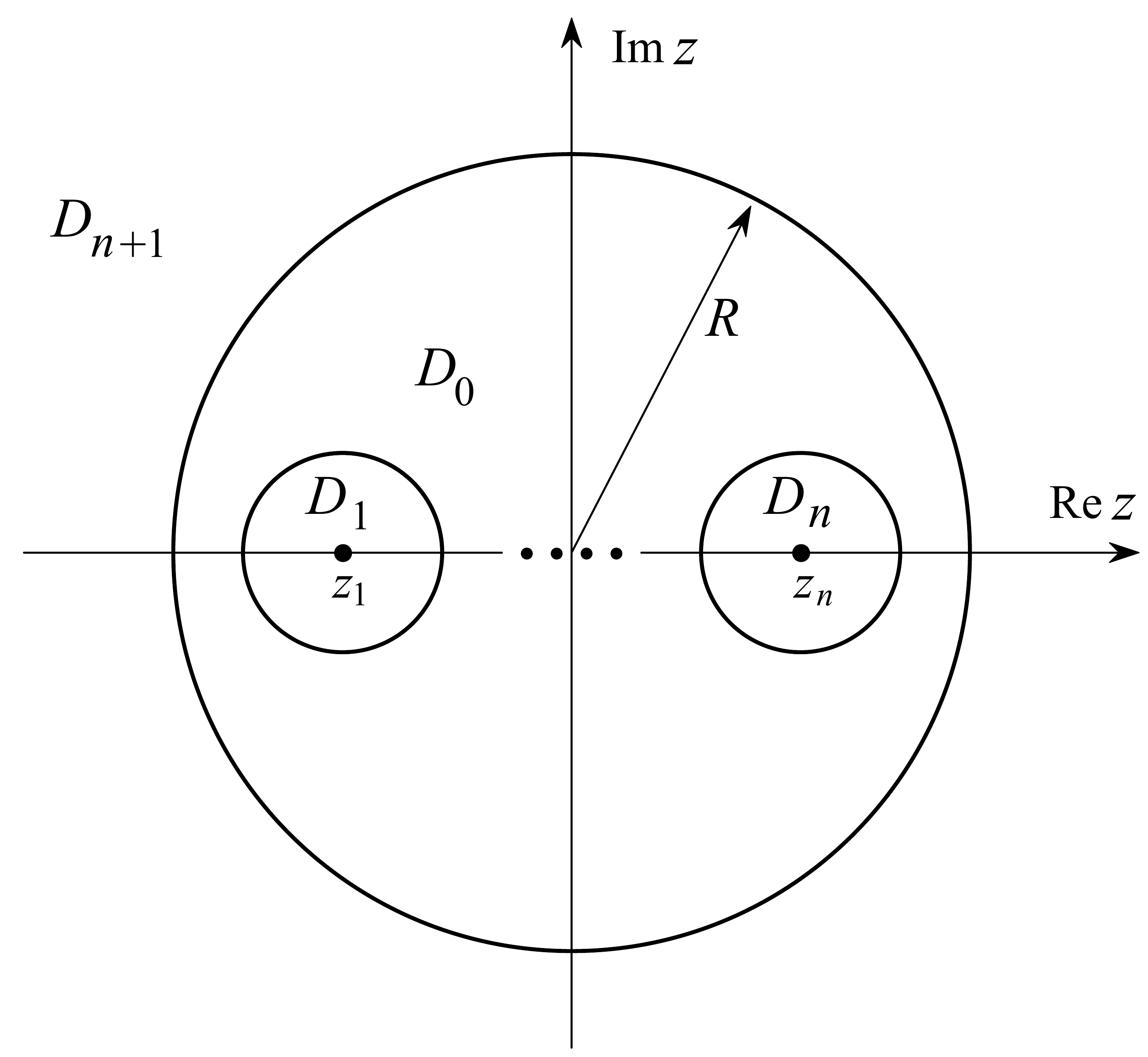}
\end{center}
\caption{Regions $D_{j}$}
\label{fig-1}
\end{figure}

Regions $D_{j}$ will be constructed from circles and their complements. They
are shown in Fig.~\ref{fig-1}. Below a formal description of this region
structure follows.

Let us denote the circle with center at $a$ and with radius $r$ by
\begin{equation}
C(a,r)=\left\{  z:|z-z_{0}|\leq r\right\}  \,.
\end{equation}
We start from the definition of $D_{k}$ with $k$ in the range $1\leq k\leq n$.
For each $z_{k}$ argument of $\Pi_{P}^{(n)}\left(  \alpha,\left\{  \gamma
_{k}\right\}  _{k=1}^{n},\left\{  z_{k}\right\}  _{k=1}^{n}\right)  $ we
define region $D_{k}$ as circle $C(z_{k},r_{k})$
\begin{equation}
D_{k}=C(z_{k},r_{k})\quad(1\leq k\leq n)\,. \label{D-k-def}
\end{equation}
The radii $r_{k}$ are chosen small enough so that each circle $C(z_{k},r_{k})$
contains no $z_{j}$ with $j\neq k$:
\begin{equation}
z_{j}\in D_{k}\Longleftrightarrow j=k\quad(1\leq j,k\leq n)\,
\end{equation}
and so that the circles do not intercept
\begin{equation}
1\leq j,k\leq n;\,\,j\neq k\quad\Longrightarrow\quad D_{j}\cap D_{k}
=\emptyset\,. \label{D-circles-empty-intersection}
\end{equation}

Next, we choose $R$ so that
\begin{equation}
1\leq k\leq n\quad\Longrightarrow\quad\left|  z_{k}\right|  <R.
\end{equation}
We also require that all circles $D_{k}=C(z_{k},r_{k})$ labelled by $1\leq
k\leq n$ are inside circle $C(0,R)$:

\begin{equation}
1\leq k\leq n\quad\Longrightarrow\quad C(z_{k},r_{k})\subset C(0,R)\,.
\label{r-k-inside-R}
\end{equation}
Now \bigskip we define $D_{0}$ as the complement of $\bigcup_{k=1}^{n}D_{k}$
in circle $C(0,R)$:
\begin{align}
D_{0}  &  =C(0,R)\backslash\left(  \bigcup_{k=1}^{n}D_{k}\right) \nonumber\\
&  =\left\{  z:\left(  |z|\leq R\right)  \,\&\,\left(  \forall k|z-z_{k}
|>r_{k}\right)  \right\}  \,. \label{D-0-def}
\end{align}
Finally we define $D_{n+1}$ as the complement to circle $C(0,R)$:
\begin{equation}
D_{n+1}=\mathbb{C}\backslash C(0,R)=\left\{  z:|z|>R\right\}  \,.
\label{D-n-plus-1-def}
\end{equation}

With this choice of regions $D_{j}$ ($0\leq j\leq n+1$) we obey condition
(\ref{C-union-D-alpha})
\begin{equation}
\mathbb{C=}\bigcup_{j=0}^{n+1}D_{j}
\end{equation}
and we also satisfy condition (\ref{no-overlap-1})
\begin{equation}
0\leq j,k\leq n+1,\,j\neq k\quad\Longrightarrow\quad\text{measure}\left(
D_{j}\cap D_{k}\right)  =0\,.
\end{equation}
Now regions $\left\{  D_{k}\right\}  _{k=0}^{n+1}$ can be used in eq.
(\ref{Pi-D-b}).

\subsection{Compact notation for other schemes of splitting in regions}

In the previous section \emph{basic} regions $D_{j}$ were defined with index
$j$ running in the interval
\begin{equation}
0\leq j\leq n+1\,.
\end{equation}
In our work we will sometimes need \emph{other} integration regions which will
denoted as $D_{A}$ with some \emph{multi-index} $A$, e.g. $D_{0,1}$. This
multi-index notation will help us avoid confusion with \emph{basic} regions
$D_{j}$ discussed in the previous section and labelled by a single index $j$.
For both \emph{basic} and \emph{alternative} regions $D_{A}$ (i.e. for simple
indices $A$ and for multi-indices $A$) we will use compact notation
\begin{equation}
S_{A}\left(  \alpha,\left\{  \gamma_{k}\right\}  _{k=1}^{n}\right)
=\int_{D_{A}}d^{2}z\,\left|  \operatorname{Im}\,z\right|  ^{\alpha-1}
\prod_{k=1}^{n}\left|  z-z_{k}\right|  ^{\gamma_{k}-\alpha-1}P\left(
z,z^{\ast}\right)  \,, \label{S-A-def}
\end{equation}
\begin{equation}
U_{A}\left(  \alpha,\left\{  \gamma_{k}\right\}  _{k=1}^{n}\right)
=\int_{D_{A}}d^{2}z\,\left|  \operatorname{Im}\,z\right|  ^{\alpha-1}
\prod_{k=1}^{n}\left|  z-z_{k}\right|  ^{\operatorname{Re}\left(  \gamma
_{k}-\alpha-1\right)  }\left|  P\left(  z,z^{\ast}\right)  \right|  \,.
\label{U-A-def}
\end{equation}
The integrand of (\ref{U-A-def}) is equal to the absolute value of the
integrand of (\ref{S-A-def}). Functions $S_{A}$ arise in the problem of
analytical continuation in $\alpha,\left\{  \gamma_{k}\right\}  _{k=1}^{n}$
whereas functions $U_{A}$ appear in the analysis of convergence of integrals.
In our analysis of convergence conditions we always imply \emph{absolute}
convergence in the sense of integrals (\ref{U-A-def}).

In both case we are interested in dependence of $S_{A}$ and $U_{A}$ on
variables $\alpha,\left\{  \gamma_{k}\right\}  _{k=1}^{n}$ at fixed $D_{A}$,
$P$ and $\left\{  z_{k}\right\}  _{k=1}^{N}$. Therefore for brevity we do not
write fixed objects $D_{A}$, $P$ and $\left\{  z_{k}\right\}  _{k=1}^{n}$
explicitly in the list of arguments of $S_{A}$ and $U_{A}$.

Note that functions $U_{A}\left(  \alpha,\left\{  \gamma_{k}\right\}
_{k=1}^{n}\right)  $ depend on $\alpha,\left\{  \gamma_{k}\right\}  _{k=1}
^{n}$ via real parts $\operatorname{Re}\alpha,\left\{  \operatorname{Re}
\gamma_{k}\right\}  _{k=1}^{n}$. Nevertheless we use a universal notational
scheme for functions $S_{A}$ and $U_{A}$ because this allows for performing
routine calculations in a form common for the problems of convergence and
analytical continuation using `substitution dictionary'
\begin{equation}
\left|  \operatorname{Im}\,z\right|  ^{\alpha-1}\rightarrow\left|
\operatorname{Im}\,z\right|  ^{\operatorname{Re}\alpha-1}\,,
\label{U-to-S-dic-1}
\end{equation}
\begin{equation}
\prod_{k=1}^{n}\left|  z-z_{k}\right|  ^{\gamma_{k}-\alpha-1}\rightarrow
\prod_{k=1}^{n}\left|  z-z_{k}\right|  ^{\operatorname{Re}\left(  \gamma
_{k}-\alpha\right)  -1}\,, \label{U-to-S-dic-2}
\end{equation}
\begin{equation}
P\left(  z,z^{\ast}\right)  \rightarrow\left|  P\left(  z,z^{\ast}\right)
\right|  \,, \label{U-to-S-dic-3}
\end{equation}
\begin{equation}
S_{A}\rightarrow U_{A}\,. \label{U-to-S-dic-4}
\end{equation}

\subsection{Degree of polynomials $P$ and parameter $M_{P}$}

\label{parameter-M-P-section}

The standard definition of the degree of polynomial $P$ is based on assigning
\begin{equation}
\text{degree}\left(  z^{k}z^{\ast m}\right)  =k+m\,,
\end{equation}
\begin{equation}
\text{degree}\left(  \sum_{k,m}c_{km}z^{k}z^{\ast m}\right)  =\max_{c_{km}
\neq0}\text{degree}\left(  z^{k}z^{\ast m}\right)  \,,
\end{equation}
For this standard degree of polynomial $P\left(  z,z^{\ast}\right)  $ we use
notation
\begin{equation}
N_{P}=\text{degree}\left(  P\right)  \,. \label{N-P-def-2}
\end{equation}

In addition to $N_{P}$ we will need a different integer quantity $M_{P}$
characterizing polynomial $P\left(  z,z^{\ast}\right)  $. Let us define
$M_{P}$ as the \emph{minimal} integer number such that that
\begin{equation}
\left(  zz^{\ast}\right)  ^{M}P\left(  z^{-1},\left(  z^{\ast}\right)
^{-1}\right)  =Q_{M}\left(  z,z^{\ast}\right)  \label{Q-polynomial}
\end{equation}
is a polynomial of $z$, $z^{\ast}$, i.e.

\begin{equation}
M_{P}=\min_{M\geq0}M\text{:}\quad\left(  zz^{\ast}\right)  ^{M}P\left(
z^{-1},\left(  z^{\ast}\right)  ^{-1}\right)  =\text{Polynomial}\left(
z,z^{\ast}\right)  \,. \label{M-P-min-def}
\end{equation}

Instructive examples:
\begin{align}
P  &  =z^{\ast}z\quad\Rightarrow\quad N_{P}=2,\,M_{P}
=1\,,\label{N-P-example-1}\\
P  &  =z^{\ast}+z\quad\Rightarrow\quad N_{P}=1,\,M_{P}=1\,.
\label{N-P-example-2}
\end{align}

Note that for any $P$
\begin{equation}
N_{P}\leq2M_{P}\leq2N_{P}\,. \label{M-P-N-P-ineqs}
\end{equation}

In case of factorizable polynomials
\begin{equation}
P\left(  z,z^{\ast}\right)  =T\left(  z\right)  \left[  T\left(  z\right)
\right]  ^{\ast}
\end{equation}
we have
\begin{equation}
N_{P}=2M_{P}\,. \label{M-P-min-N-P-for-special-P}
\end{equation}
The degree of holomorphic polynomial $T_{2}$ can be easily derived from its
definition (\ref{T2-def})
\begin{equation}
N_{T_{2}}=2\left(  n_{\text{v}}-2\right)  \,.
\end{equation}
Factorizable polynomial $P_{2}$ (\ref{P2-via-T2}) defined via holomorphic
polygon $T_{2}$ has degree
\begin{equation}
N_{P_{2}}=2N_{T_{2}}=4\left(  n_{\text{v}}-2\right)  \,.
\end{equation}
Now $M_{P_{2}}$ can be computed using (\ref{M-P-min-N-P-for-special-P})

\begin{equation}
M_{P_{2}}=\frac{1}{2}N_{P_{2}}=2\left(  n_{\text{v}}-2\right)  \,.
\label{M-P2-res}
\end{equation}

\section{Convergence of integrals}

\label{convergence-of-integrals-section}

\setcounter{equation}{0} 

\subsection{Results}

\label{results-convergence-conditions-section}

As discussed above, functions $\Pi_{P}^{\left(  n\right)  }$ are defined first
in the convergence region of integral (\ref{Pi-P-n-def}). More exactly, we
want to start from the region where integral (\ref{Pi-P-n-def}) is
\emph{absolutely} convergent, i.e.
\begin{align}
&  \int_{\mathbb{C}_{+}}d^{2}z\,\left|  \left|  \operatorname{Im}\,z\right|
^{\alpha-1}\prod_{k=1}^{n}\left|  z-z_{k}\right|  ^{\gamma_{k}-\alpha
-1}P\left(  z,z^{\ast}\right)  \right| \nonumber\\
&  =\int_{\mathbb{C}_{+}}d^{2}z\,\left|  \operatorname{Im}\,z\right|
^{\operatorname{Re}\left(  \alpha-1\right)  }\prod_{k=1}^{n}\left|
z-z_{k}\right|  ^{\operatorname{Re}\left(  \gamma_{k}-\alpha-1\right)
}\left|  P\left(  z,z^{\ast}\right)  \right|  \text{ is convergent}
\label{abs-convergence}
\end{align}
Note that the convergence of this integral depends only on real parts of
generally complex parameters $\alpha,\left\{  \gamma_{k}\right\}  $.

\textbf{Statement 1.}

The set of conditions
\begin{equation}
\operatorname{Re}\,\alpha>0\,, \label{convergence-condition-1}
\end{equation}
\begin{equation}
\operatorname{Re}\gamma_{k}>0\,, \label{convergence-condition-2}
\end{equation}
\begin{equation}
\operatorname{Re}\alpha>-1+\frac{1}{n-1}\left(  N_{P}+\sum_{k=1}
^{n}\operatorname{Re}\gamma_{k}\right)  \label{convergence-condition-3}
\end{equation}
is \emph{sufficient} for the absolute convergence (\ref{abs-convergence}).
Here $N_{P}$ is degree (\ref{N-P-def-2}) of polynomial $P$.

For a special case of polynomials $P$ \ one can make a stronger statement:

\textbf{Statement 2.}

If polynomial $P$

1) has the form
\begin{equation}
P\left(  z,z^{\ast}\right)  =T\left(  z\right)  \left[  T\left(  z\right)
\right]  ^{\ast} \label{P-T-factorizable}
\end{equation}
where $T\left(  z\right)  $ is a holomorphic polynomial of $z$,

2) for any $k=1,\ldots,n$
\begin{equation}
T\left(  z_{k}\right)  \neq0 \label{T-z-k-nonzero}
\end{equation}
then the combination of conditions (\ref{convergence-condition-1}),
(\ref{convergence-condition-2}), (\ref{convergence-condition-3}) is not only
\emph{sufficient} but also \emph{necessary} for the absolute convergence
(\ref{abs-convergence}).

\textbf{Remark.} For factorizable polynomials (\ref{P-T-factorizable})
according to (\ref{M-P-min-N-P-for-special-P}) we have $2M_{P}=N_{P}$ where
$N_{P}$ is usual degree (\ref{N-P-def-2}) of polynomial $P$.

\subsection{Naive derivation of convergence conditions}

Conditions (\ref{convergence-condition-1}), (\ref{convergence-condition-2}),
(\ref{convergence-condition-3}) can be easily `explained' in terms of naive
order counting near potentially singular regions of integral
(\ref{abs-convergence}):

\begin{itemize}
\item The singularity at $\left|  \operatorname{Im}\,z\right|  \rightarrow0$
is integrable under condition (\ref{convergence-condition-1}).

\item The convergence of integral at $z\rightarrow z_{j}$ is provided by
condition (\ref{convergence-condition-2}).

\item Convergence at $\left|  z\right|  \rightarrow\infty$ is provided by
condition (\ref{convergence-condition-3}).
\end{itemize}

This order counting is straightforward. In the next section the case
$z\rightarrow z_{j}$ is discussed in detail.

\subsection{Example: convergence at $z\rightarrow z_{j}$}

Decomposing $z$ in real and imaginary parts we have at $z\rightarrow z_{j}$:
\begin{equation}
z=x+iy,
\end{equation}
\begin{equation}
x\rightarrow z_{j}\,,
\end{equation}
\begin{equation}
y\rightarrow0\,.
\end{equation}
We obtain in this limit
\begin{align}
\,\left|  \operatorname{Im}\,z\right|  ^{\operatorname{Re}\left(
\alpha-1\right)  } &  =|y|^{\operatorname{Re}\left(  \alpha-1\right)  }\\
\prod_{k=1}^{n}\left|  z-z_{k}\right|  ^{\operatorname{Re}\left(  \gamma
_{k}-\alpha-1\right)  } &  \sim\left|  z-z_{j}\right|  ^{\operatorname{Re}
\left(  \gamma_{j}-\alpha-1\right)  }\,.
\end{align}
Concentrating on the case
\begin{equation}
\left|  x-z_{j}\right|  \sim y\sim\left|  z-z_{j}\right|  \rightarrow0\,,
\end{equation}
we find
\begin{equation}
\prod_{k=1}^{n}\left|  z-z_{k}\right|  ^{\operatorname{Re}\left(  \gamma
_{k}-\alpha-1\right)  }\sim\left|  z-z_{j}\right|  ^{\operatorname{Re}\left(
\gamma_{k}-\alpha-1\right)  }
\end{equation}
so that
\begin{equation}
\,\left|  \operatorname{Im}\,z\right|  ^{\operatorname{Re}\left(
\alpha-1\right)  }\prod_{k=1}^{n}\left|  z-z_{k}\right|  ^{\operatorname{Re}
\left(  \gamma_{k}-\alpha-1\right)  }\sim|z-z_{j}|^{\operatorname{Re}\left(
\gamma_{j}-2\right)  }\,.
\end{equation}
If
\begin{equation}
P\left(  z_{j},z_{j}^{\ast}\right)  \neq0
\end{equation}
then
\begin{equation}
\,\left|  \operatorname{Im}\,z\right|  ^{\operatorname{Re}\left(
\alpha-1\right)  }\prod_{k=1}^{n}\left|  z-z_{k}\right|  ^{\operatorname{Re}
\left(  \gamma_{k}-\alpha-1\right)  }\left|  P\left(  z,z^{\ast}\right)
\right|  ^{2}\sim|y|^{\operatorname{Re}\left(  \gamma_{k}-2\right)  }
\end{equation}
so that convergence of integral (\ref{abs-convergence}) at $z\rightarrow
z_{k}$ requires condition
\begin{equation}
\operatorname{Re}\gamma_{j}>0
\end{equation}
in agreement with inequality (\ref{convergence-condition-2}) in rigorous
Statement 1,

In the special case when
\begin{equation}
P\left(  z_{j},z_{j}^{\ast}\right)  =0\,
\end{equation}
the situation is different because this zero of polynomial $P$ mitigates the
singularity of the integrand at $z\rightarrow z_{j}$. This explains why in
Statement 1 the set of conditions (\ref{convergence-condition-1}) --
(\ref{convergence-condition-3}) is considered as \emph{sufficient} for the
convergence, whereas in Statement 2, this set of conditions is both
\emph{sufficient} and \emph{necessary}.

\subsection{Careful derivation of convergence conditions}

\begin{figure}[ptb]
\begin{center}
\includegraphics[
width=8cm
]
{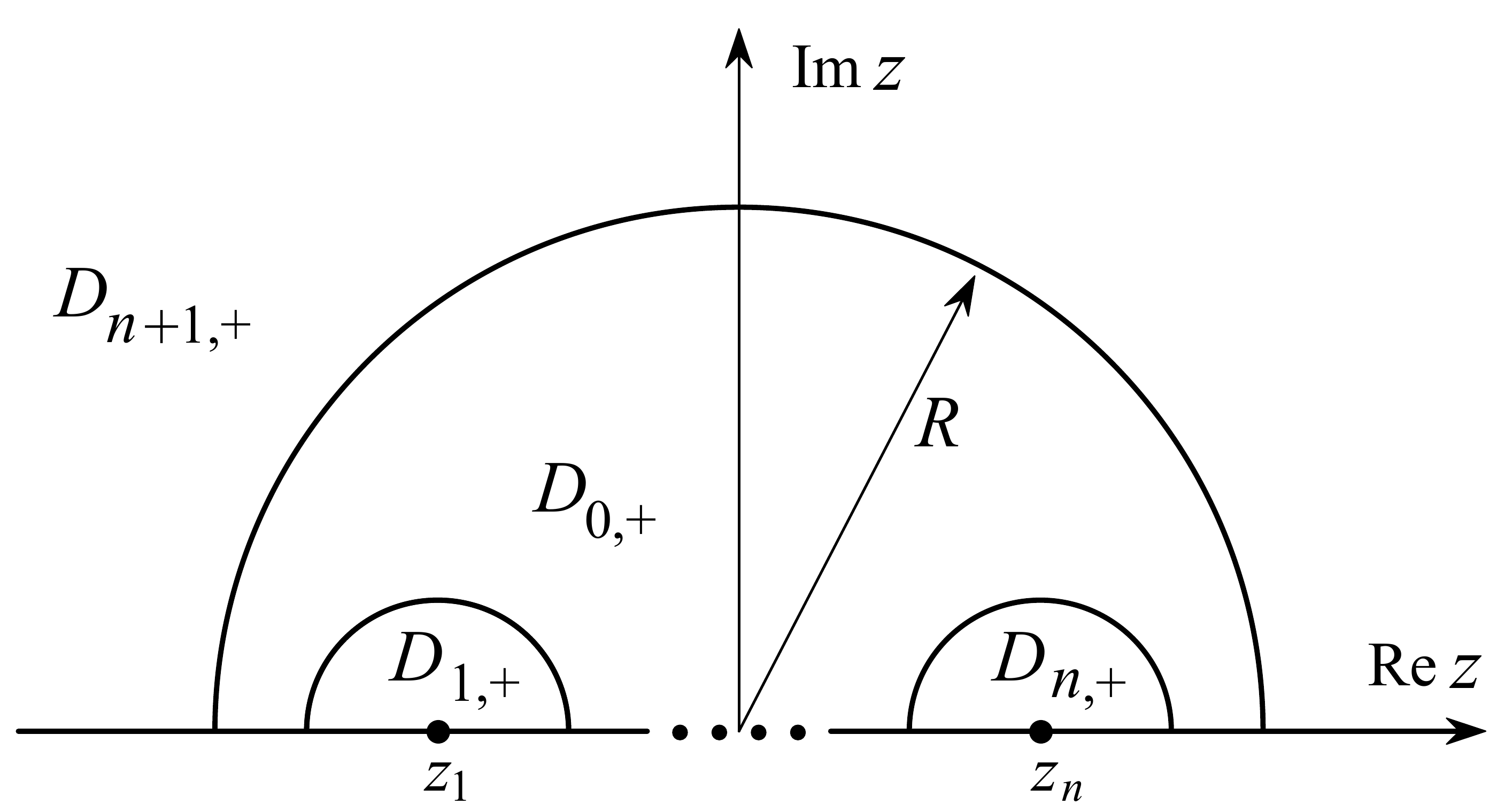}
\end{center}
\caption{Regions $D_{j,+}$}
\label{fig-2}
\end{figure}

Although the simple arguments of the previous sections lead to correct
convergence conditions, this naive order counting cannot be considered as a
rigorous proof of Statements 1, 2. In order to upgrade this simple order
counting argument to a careful proof, we use decomposition
(\ref{Pi-D-plus-splitting}) with regions $D_{j,+}$ defined in section
\ref{choce-D-b-section} and shown in Fig. \ref{fig-2}.

The integral controlling absolute convergence (\ref{abs-convergence}) can be
decomposed as
\begin{align}
&  \int_{\mathbb{C}_{+}}d^{2}z\,\left|  \left|  \operatorname{Im}\,z\right|
^{\alpha-1}\prod_{k=1}^{n}\left|  z-z_{k}\right|  ^{\gamma_{k}-\alpha
-1}P\left(  z,z^{\ast}\right)  \right| \nonumber\\
&  =\sum_{j=0}^{n}\int_{D_{n,+}}d^{2}z\,\left|  \left|  \operatorname{Im}
\,z\right|  ^{\alpha-1}\prod_{k=1}^{n}\left|  z-z_{k}\right|  ^{\gamma
_{k}-\alpha-1}P\left(  z,z^{\ast}\right)  \right|  \,.
\end{align}
so that the problem reduces to the analysis convergence conditions for each
separate integral
\begin{equation}
U_{n,+}\left(  \alpha,\left\{  \gamma_{k}\right\}  _{k=1}^{n}\right)
=\int_{D_{n,+}}d^{2}z\,\left|  \operatorname{Im}\,z\right|
^{\operatorname{Re}\left(  \alpha-1\right)  }\prod_{k=1}^{n}\left|
z-z_{k}\right|  ^{\operatorname{Re}\left(  \gamma_{k}-\alpha-1\right)
}\left|  P\left(  z,z^{\ast}\right)  \right|
\end{equation}
where we use compact notation (\ref{U-A-def}).

If one is interested in \emph{sufficient} convergence conditions for integrals
$U_{n,+}$ over separate regions $D_{n,+}$ then the results of our analysis
have a simple summary:

\begin{itemize}
\item convergence in $D_{0,+}$: condition (\ref{convergence-condition-1}),

\item convergence in $D_{k,+}$ ($1\leq k\leq n$): combination of conditions
(\ref{convergence-condition-1}) and (\ref{convergence-condition-2}),

\item convergence in $D_{n+1,+}$: combination of conditions
(\ref{convergence-condition-2}) and (\ref{convergence-condition-3}).
\end{itemize}

The combination of all these sufficient conditions leads to Statement 1.

In case of both sufficient and necessary condition formulated in Statement 2
the situation is more subtle (see sections devoted to the detailed analysis of
convergence conditions in each separate region).

\subsection{Convergence in region $D_{0,+}$}

\label{D-0-plus-convergence-section}

\subsubsection{Plan}

Region $D_{0,+}$ is defined by eq. (\ref{D-plus-def-via-D-b}) with $D_{0}$
defined by eq. (\ref{D-0-def}).
\begin{equation}
D_{0,+}=\left\{  z:\left(  |z|\leq R\right)  \,\&\,\left(  \forall
k|z-z_{k}|>r_{k}\right)  \&\left(  0<\operatorname{Im}\text{ }\,z\right)
\right\}  \,. \label{D-0-plus-def}
\end{equation}
We are interested in the convergence region of integral
\begin{equation}
U_{0,+}\left(  \alpha,\left\{  \gamma_{k}\right\}  _{k=1}^{n}\right)
=\int_{D_{0,+}}d^{2}z\,\left|  \operatorname{Im}\,z\right|
^{\operatorname{Re}\left(  \alpha-1\right)  }\left[  \prod_{k=1}^{n}\left|
z-z_{k}\right|  ^{\operatorname{Re}\left(  \gamma_{k}-\alpha-1\right)
}\right]  \left|  P\left(  z,z^{\ast}\right)  \right|  \,.
\label{U-0-plus-int}
\end{equation}
Here we use compact notation (\ref{U-A-def}).

Our aim is

1) to prove that condition
\begin{equation}
\operatorname{Re}\alpha>0 \label{convergence-condition-1-4}
\end{equation}
is sufficient for the convergence of this integral,

2) to prove that in the case of factorizable polynomials $P\left(  z,z^{\ast
}\right)  $ of Statement 2 obeying eqs. (\ref{P-T-factorizable}),
(\ref{T-z-k-nonzero}) condition (\ref{convergence-condition-1-4}) is not only
sufficient for the convergence of (\ref{U-0-plus-int}) but also necessary.

\subsubsection{Splitting $D_{0,+}$ in subregions}

\begin{figure}[ptb]
\begin{center}
\includegraphics[
width=9cm
]
{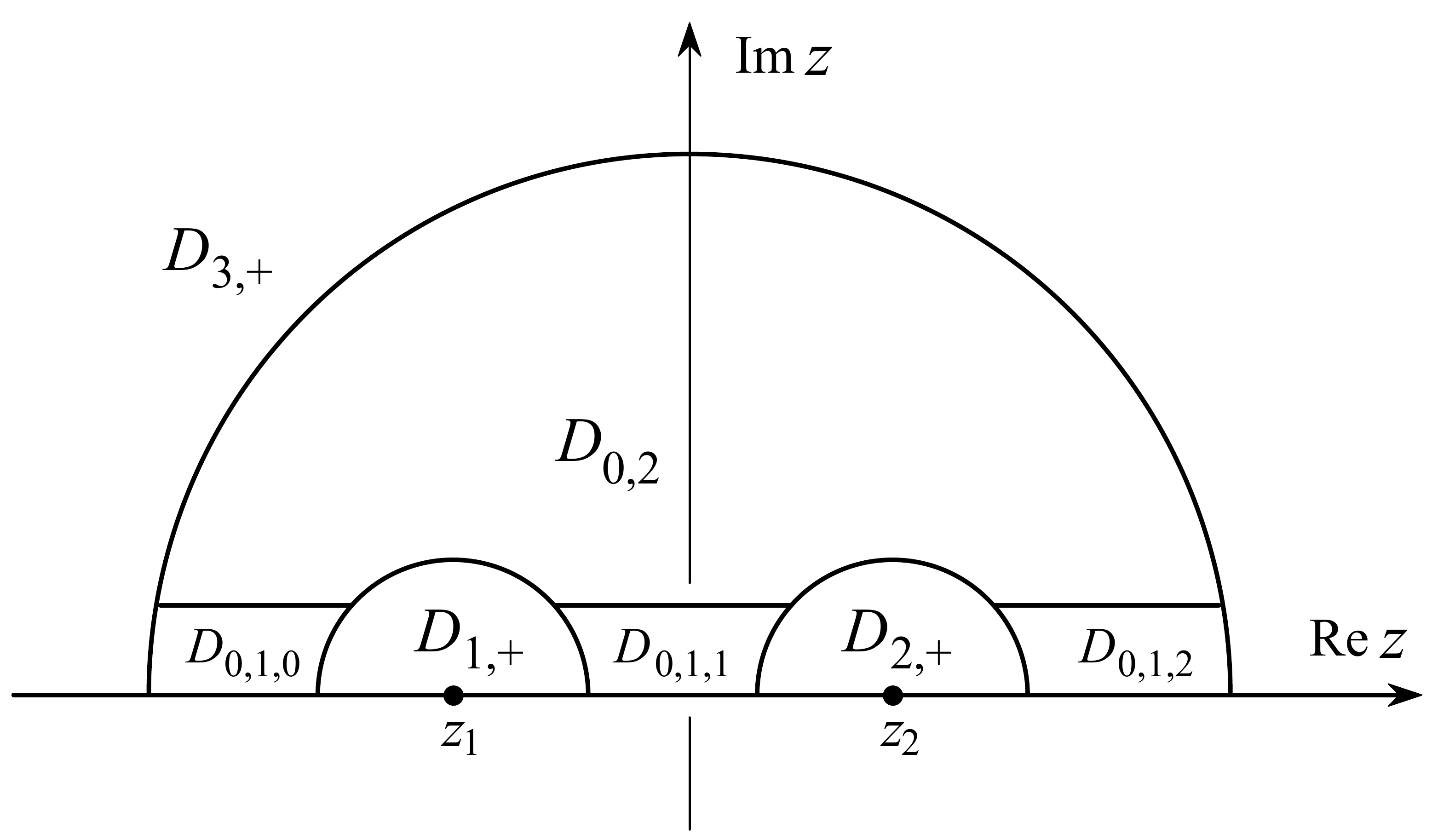}
\end{center}
\caption{Regions $D_{j,+}$, $D_{0,2}$, and $D_{0,1,j}$ for function $\Pi
_{P}^{(n)}$ with $n=2$.}
\label{fig-3}
\end{figure}

In order to proceed we must divide region $D_{0,+}$ (\ref{D-0-plus-def}) in
subregions. Let us split region $D_{0,+}$ by cutting $D_{0,+}$ with line
\[
\operatorname{Im}z=\eta\,.
\]
We choose $\eta$ so that
\begin{equation}
\eta>0
\end{equation}
and for all $k$ ($1\leq k\leq n$) we have
\begin{equation}
\eta<r_{k}\,. \label{eps-constraint-1}
\end{equation}
This cut splits $D_{0,+}$ in $n+2$ disconnected components. An example of the
new region structure is shown in Fig.~\ref{fig-3} for the case $n=2$.

Now we turn to the formal description of arising subregions. First we
introduce notation
\begin{align}
D_{0,1}  &  =D_{0,+}\cap\left\{  z:\operatorname{Im}\text{ }\,z<\eta\right\}
\nonumber\\
&  =\left\{  z:\left(  |z|\leq R\right)  \,\&\,\left(  \forall k|z-z_{k}
|>r_{k}\right)  \&\left(  0<\operatorname{Im}\text{ }\,z<\eta\right)
\right\}  \,, \label{D-0-1-def}
\end{align}
\begin{align}
D_{0,2}  &  =D_{0,+}\cap\left\{  z:\operatorname{Im}\text{ }\,z\geq
\eta\right\} \nonumber\\
&  =\left\{  z:\left(  |z|\leq R\right)  \,\&\,\left(  \forall k|z-z_{k}
|>r_{k}\right)  \&\left(  \operatorname{Im}\text{ }\,z\geq\eta\right)
\right\}
\end{align}
Obviously
\begin{equation}
D_{0,+}=D_{0,1}\cup D_{0,2}\,\,,
\end{equation}
\begin{equation}
D_{0,1}\cap D_{0,2}=\emptyset\,.
\end{equation}
As discussed below (and shown in Fig. ~\ref{fig-3}), region $D_{0,1}$ is
disconnected but we still have the decomposition of function $U_{0,+}$
(\ref{U-0-plus-int})
\begin{equation}
U_{0,+}\left(  \alpha,\left\{  \gamma_{k}\right\}  _{k=1}^{n}\right)
=U_{0,1}\left(  \alpha,\left\{  \gamma_{k}\right\}  _{k=1}^{n}\right)
+U_{0,2}\left(  \alpha,\left\{  \gamma_{k}\right\}  _{k=1}^{n}\right)  \,
\end{equation}
where $U_{0,1}$ and $U_{0,2}$ are defined by eq. (\ref{U-A-def}).

Note that

\begin{itemize}
\item the integrand of $U_{0,2}$ has no singularities in the integration
region $D_{0,2}$ (and in its small vicinity),

\item region $D_{0,2}$ has a finite size.
\end{itemize}

Hence the integral defining $U_{0,2}\left(  \alpha,\left\{  \gamma
_{k}\right\}  _{k=1}^{n}\right)  $ is convergent for any complex
$\alpha,\left\{  \gamma_{k}\right\}  _{k=1}^{n}$. Therefore integrals
(\ref{U-A-def}) defining functions $U_{0,+}$ and $U_{0,1}$ have a common
convergence region in the space of parameters $\alpha,\left\{  \gamma
_{k}\right\}  _{k=1}^{n}$.

\subsubsection{Region $D_{0,1}$}

Thus the problem is reduced to the analysis of the convergence region in the
$\alpha,\left\{  \gamma_{k}\right\}  _{k=1}^{n}$ space for integral

\begin{equation}
U_{0,1}\left(  \alpha,\left\{  \gamma_{k}\right\}  _{k=1}^{n}\right)
=\int_{D_{0,1}}d^{2}z\,\left|  \operatorname{Im}\,z\right|
^{\operatorname{Re}\left(  \alpha-1\right)  }\left[  \prod_{k=1}^{n}\left|
z-z_{k}\right|  ^{\operatorname{Re}\left(  \gamma_{k}-\alpha-1\right)
}\right]  \left|  P\left(  z,z^{\ast}\right)  \right|  \,. \label{U-0-1-int}
\end{equation}

Region $D_{0,1}$ (\ref{D-0-1-def}) consists of $n+1$ disconnected components
which will be denoted $D_{0,1,j}$ ($0\leq1\leq n$). Fig. \ref{fig-3}
illustrates the case of $n=2$.

The formal description of regions $D_{0,1,j}$ follows from the definition of
$D_{0,1}$ (\ref{D-0-1-def}) and from constraints (\ref{z-k-monotonic-order}),
(\ref{r-k-inside-R}), (\ref{D-circles-empty-intersection}),
(\ref{eps-constraint-1}):

\begin{align}
D_{0,1,j} &  =\left\{  z=x+iy:\left(  x_{j}^{-}\left(  y\right)  <x<x_{j}
^{+}\left(  y\right)  \right)  \,\&\,\left(  0<y<\eta\right)  \right\}  \,,\\
(0 &  \leq j\leq n)\,.
\end{align}
Here
\begin{equation}
x_{0}^{-}(y)=-\sqrt{R^{2}-y^{2}}\,,
\end{equation}
\begin{equation}
x_{0}^{+}(y)=z_{1}-\sqrt{r_{1}^{2}-y^{2}}\,,
\end{equation}
\begin{equation}
x_{n}^{-}(y)=z_{n}+\sqrt{r_{n}^{2}-y^{2}}\,,
\end{equation}
\begin{equation}
x_{n}^{+}(y)=\sqrt{R^{2}-y^{2}}\,\,,
\end{equation}

\begin{align}
\text{for }1  &  \leq j\leq n-1:\nonumber\\
x_{j}^{-}(y)  &  =z_{j}+\sqrt{r_{j}^{2}-y^{2}},\\
x_{j}^{+}(y)  &  =z_{j+1}-\sqrt{r_{j+1}^{2}-y^{2}}\,.
\end{align}
All $x_{j}^{\pm}(y)$ with $0\leq j\leq n$ have a universal form
\begin{align}
\text{for }0  &  \leq j\leq n:\nonumber\\
x_{j}^{\pm}\left(  y\right)   &  =X_{j}^{\pm}+\sigma_{j}^{\pm}\sqrt{\left(
\rho_{j}^{\pm}\right)  ^{2}-y^{2}} \label{x-pm-j-general-def}
\end{align}
with obvious expressions from real parameters $X_{j}^{\pm}$, $\rho_{j}^{\pm}$
and for sign factors $\sigma_{j}^{\pm}$ taking values $\pm1$.

Thus we have
\begin{equation}
D_{0,1}\mathbb{=}\bigcup_{j=0}^{n}D_{0,1,j}\,,
\end{equation}
\begin{equation}
\text{if}\,j\neq k\,\text{then}\,D_{0,1,j}\cap D_{0,1,k}=\emptyset\,.
\end{equation}
This leads to decomposition
\begin{equation}
U_{0,1}\left(  \alpha,\left\{  \gamma_{k}\right\}  _{k=1}^{n}\right)
=\sum_{j=0}^{n}U_{0,1,j}\left(  \alpha,\left\{  \gamma_{k}\right\}  _{k=1}
^{n}\right)  \,. \label{U-0-1-decomposition}
\end{equation}
As usual, functions $U_{0,1,j}$ are defined by general relation (\ref{U-A-def}).

We see that the problem of the convergence region in the $\alpha,\left\{
\gamma_{k}\right\}  _{k=1}^{n}$ space for integral $U_{0,1,j}$ reduces to the
problem of the convergence regions for all separate integrals $U_{0,1,j}$.

We have according to (\ref{U-A-def})
\begin{align}
U_{0,1,j}\left(  \alpha,\left\{  \gamma_{k}\right\}  _{k=1}^{n}\right)   &
=\int_{D_{0,1,j}}d^{2}z\,\left|  \operatorname{Im}\,z\right|
^{\operatorname{Re}\left(  \alpha-1\right)  }\left[  \prod_{k=1}^{n}\left|
z-z_{k}\right|  ^{\operatorname{Re}\left(  \gamma_{k}-\alpha-1\right)
}\right]  \left|  P\left(  z,z^{\ast}\right)  \right| \nonumber\\
&  =\int_{0}^{\eta}dy\,\left|  y\right|  ^{\alpha-1}\int_{x_{j}^{-}(y)}
^{x_{j}^{+}(y)}dx\prod_{k=1}^{n}\left|  \left(  x-z_{k}\right)  ^{2}
+y^{2}\right|  ^{(\gamma_{k}-\alpha-1)/2}\nonumber\\
&  \times\left|  P\left(  x+iy,x-iy\right)  \right|  \,. \label{U-0-1-j-int}
\end{align}
The integrand of
\begin{equation}
G_{j}(y;\alpha,\left\{  \gamma_{k}\right\}  _{k=1}^{n})=\int_{x_{j}^{-}
(y)}^{x_{j}^{+}(y)}dx\prod_{k=1}^{n}\left|  \left(  x-z_{k}\right)  ^{2}
+y^{2}\right|  ^{(\gamma_{k}-\alpha-1)/2}\left|  P\left(  x+iy,x-iy\right)
\right|  \label{G-j-int}
\end{equation}
has no singularities in the integration region. The integration limits are
also regular functions. The only object that can slightly violate regularity
is absolute value $\left|  P\left(  x+iy,x-iy\right)  \right|  $ which has
cusps at zeros of polynomial $P\left(  x+iy,x-iy\right)  $. Anyway
$G_{j}(y;\alpha,\left\{  \gamma_{k}\right\}  _{k=1}^{n})$ is a continuous
function of all its arguments. Therefore integral representation
\begin{equation}
U_{0,1,j}\left(  \alpha,\left\{  \gamma_{k}\right\}  _{k=1}^{n}\right)
=\int_{0}^{\eta}dy\,\left|  y\right|  ^{\alpha-1}G_{j}(y;\alpha,\left\{
\gamma_{k}\right\}  _{k=1}^{n}) \label{U-1-j-int-via-G}
\end{equation}
guarantees that this integral is convergent if condition
(\ref{convergence-condition-1-4}) holds.

\subsubsection{Sufficient conditions for absolute convergence}

Once we have proved that condition $\operatorname{Re}\alpha>0$ is sufficient
for the convergence of all integrals $U_{0,1,j}$, we can trace back our
arguments which have reduced the problem of convergence of the original
integral (\ref{U-0-plus-int}) to the problem of convergence of integrals
$U_{0,1,j}$. Thus we have proved that condition
(\ref{convergence-condition-1-4}) is sufficient for the convergence of
integral $U_{0,+}$ (\ref{U-0-plus-int}).

\subsubsection{Comments on necessary conditions for absolute convergence}

Now we turn to the question about a necessary condition for convergence the
convergence of $U_{0,1,j}$. If function $G_{j}(y;\alpha,\left\{  \gamma
_{k}\right\}  _{k=1}^{n})$ could vanish at $y=0$ then integral
(\ref{U-1-j-int-via-G}) would be convergent at some negative values of
$\operatorname{Re}\alpha$ so that condition $\operatorname{Re}\alpha>0$ would
not be necessary for convergence. But can $G_{j}(y;\alpha,\left\{  \gamma
_{k}\right\}  _{k=1}^{n})$ vanish at $y=0$? In this case integral
representation (\ref{G-j-int}) would lead to
\begin{equation}
P\left(  x,x\right)  =0
\end{equation}
in a finite interval of real $x$. This vanishing is possible if $P$ has the
form
\begin{equation}
P\left(  z,z^{\ast}\right)  =\left(  z-z^{\ast}\right)  R(z,z^{\ast})
\end{equation}
where $R(z,z^{\ast})$ is a polynomial of $z,z^{\ast}$. Symmetry
(\ref{P-symmetric-1}) leads to
\begin{equation}
R(z,z^{\ast})=-R(z^{\ast},z)
\end{equation}
which results in decomposition
\begin{equation}
R(z,z^{\ast})=\left(  z-z^{\ast}\right)  Q(z,z^{\ast})
\end{equation}
where $Q$ is a symmetric polynomial
\[
Q(z,z^{\ast})=Q(z^{\ast},z)\,.
\]

Thus
\begin{equation}
P\left(  z,z^{\ast}\right)  =\left(  z-z^{\ast}\right)  ^{2}Q(z,z^{\ast
})=-4\left(  \operatorname{Im}z\right)  ^{2}Q(z,z^{\ast})\,.
\end{equation}
Thus the problem with the interpretation of (\ref{convergence-condition-1-4})
as a necessary condition for convergence arises only if polynomial $P\left(
z,z^{\ast}\right)  $ has a factor of $\left(  \operatorname{Im}z\right)  ^{2}
$. This problem is obvious from the very beginning because extra factors
$\left(  \operatorname{Im}z\right)  ^{2}$ coming from $P\left(  z,z^{\ast
}\right)  $ effectively can be interpreted as a shift of parameter $\alpha$
and a change of polynomial $P$ in integral (\ref{Pi-P-n-def}).

Anyway this problem does not affect our final results: Statement 1 provides
only sufficient conditions for the absolute convergence and Statement 2 deals
with factorizable polynomials (\ref{P-T-factorizable}) which cannot contain
factors of $\left(  z-z^{\ast}\right)  ^{2}$.

\subsection{Convergence in regions $D_{j,+}$ ($1\leq j\leq n$)}

\label{convergence-in-D-j-semicricle-section}

Regions $D_{j}$ with $1\leq j\leq n$ are circles (\ref{D-k-def}). Region
$D_{j,+}$ is the corresponding semicircle in the upper complex semiplane.
Therefore
\begin{align}
&  \int_{D_{j,+}}d^{2}z\,\left|  \operatorname{Im}\,z\right|
^{\operatorname{Re}\left(  \alpha-1\right)  }\left[  \prod_{k=1}^{n}\left|
z-z_{k}\right|  ^{\operatorname{Re}\left(  \gamma_{k}-\alpha-1\right)
}\right]  \left|  P\left(  z,z^{\ast}\right)  \right|  \nonumber\\
&  =\int_{\left|  z-z_{j}\right|  <r_{j},\operatorname{Im}z>0}d^{2}z\,\left|
\operatorname{Im}\,z\right|  ^{\operatorname{Re}\left(  \alpha-1\right)
}\left[  \prod_{k=1}^{n}\left|  z-z_{k}\right|  ^{\operatorname{Re}\left(
\gamma_{k}-\alpha-1\right)  }\right]  \left|  P\left(  z,z^{\ast}\right)
\right|  \nonumber\\
&  =\int_{\left|  z\right|  <r_{j},\operatorname{Im}z>0}d^{2}z\,\left|
\operatorname{Im}\,z\right|  ^{\operatorname{Re}\left(  \alpha-1\right)
}\left[  \prod_{k=1}^{n}\left|  z+z_{j}-z_{k}\right|  ^{\operatorname{Re}
\left(  \gamma_{k}-\alpha-1\right)  }\right]  \nonumber\\
&  \times\left|  P\left(  z+z_{j},z^{\ast}+z_{j}\right)  \right|  \,.
\end{align}
Next we factor the integrand in singular and regular parts
\begin{equation}
F_{\text{sing}}=\,\left|  \operatorname{Im}\,z\right|  ^{\operatorname{Re}
\left(  \alpha-1\right)  }\left|  z\right|  ^{\operatorname{Re}\left(
\gamma_{j}-\alpha-1\right)  }\,,
\end{equation}
\begin{equation}
F_{\text{reg}}=\left|  P\left(  z+z_{j},z^{\ast}+z_{j}\right)  \right|
\prod_{\substack{1\leq j\leq n\\j\neq k}}\left|  z+z_{j}-z_{k}\right|
^{\operatorname{Re}\left(  \gamma_{k}-\alpha-1\right)  }\,
\end{equation}
so that
\begin{align}
&  \int_{D_{j,+}}d^{2}z\,\left|  \operatorname{Im}\,z\right|
^{\operatorname{Re}\left(  \alpha-1\right)  }\left[  \prod_{k=1}^{n}\left|
z-z_{k}\right|  ^{\operatorname{Re}\left(  \gamma_{k}-\alpha-1\right)
}\right]  \left|  P\left(  z,z^{\ast}\right)  \right|  \nonumber\\
&  =\int_{\left|  z\right|  <r_{j},\operatorname{Im}z>0}d^{2}zF_{\text{sing}
}F_{\text{reg}}\,.\label{int-F-F}
\end{align}

In order to proceed we need

\textbf{Statement 3.} Integral
\begin{equation}
\int_{\left|  z\right|  <r_{j},\operatorname{Im}z>0}d^{2}z\,\left|
\operatorname{Im}\,z\right|  ^{\operatorname{Re}\left(  \alpha-1\right)
}\left|  z\right|  ^{\operatorname{Re}\left(  \gamma_{j}-\alpha-1\right)  }
\end{equation}
is convergent if and only if two conditions hold
\begin{equation}
\operatorname{Re}\alpha>0\,, \label{Re-alpha-positive-7}
\end{equation}
\begin{equation}
\operatorname{Re}\gamma_{j}>0\,. \label{Re-gamma-positive-7}
\end{equation}

This Statement follows from the results of Appendix
\ref{integral-I-alpha-gamma-section}.

Combining Statement 3 and regularity of $F_{\text{reg}}$ in the integration
region, we conclude that conditions (\ref{Re-alpha-positive-7}),
(\ref{Re-gamma-positive-7}) are sufficient the convergence of integral
(\ref{int-F-F}). This is the contribution of region $D_{j,+}$ to the full set
of sufficient convergence conditions of Statement 1.

If
\begin{equation}
P\left(  z_{j},z_{j}^{\ast}\right)  \neq0 \label{P-z-j-nonzero}
\end{equation}
then we can choose $r_{j}$ so small that in the integration region $\left|
P\left(  z,z^{\ast}\right)  \right|  $ is bounded by two nonzero positive
constants
\begin{equation}
\left|  z\right|  <r_{j}\quad\Longrightarrow\quad0<C_{1}<P\left(  z,z^{\ast
}\right)  <C_{2}\,. \label{P-C1-C2-bound}
\end{equation}
Combining this double bound with Statement 3, we see that in case
(\ref{P-z-j-nonzero}) conditions (\ref{Re-alpha-positive-7}),
(\ref{Re-gamma-positive-7}) are both necessary and sufficient for the
convergence of integral (\ref{int-F-F}). This argument provides the
contribution of region $D_{j,+}$ to the proof of Statement 2 of section
\ref{results-convergence-conditions-section}.

\subsection{Convergence in region $D_{n+1,+}$}

\label{Convergence-D-n-plus-1-section}

Region $D_{n+1,+}$ is defined by eqs. (\ref{D-n-plus-1-def}),
(\ref{D-plus-def-via-D-b})
\begin{equation}
D_{n+1,+}=\left\{  z:\left(  |z|>R\right)  \,\&\,\left(  \operatorname{Im}
z>0\right)  \right\}  \,.
\end{equation}
Using inversion
\begin{equation}
z^{\prime}=-\frac{1}{z}\,, \label{z-inversion-minus}
\end{equation}
we can map region $D_{n+1,+}$ to semicircle
\begin{equation}
\left\{  z:\left(  |z|<1/R\right)  \,\&\,\left(  \operatorname{Im}z>0\right)
\right\}  \,.
\end{equation}
Now the problem of the convergence in region $D_{n+1,+}$ reduces to the same
type as the convergence in semicircle regions $D_{j,+}$ ($1\leq j\leq n$)
which was considered in section \ref{convergence-in-D-j-semicricle-section}.

\begin{align}
&  \int_{D_{n+1,+}}d^{2}z\,\left|  \operatorname{Im}\,z\right|
^{\operatorname{Re}\left(  \alpha-1\right)  }\left[  \prod_{k=1}^{n}\left|
z-z_{k}\right|  ^{\operatorname{Re}\left(  \gamma_{k}-\alpha-1\right)
}\right]  \left|  P\left(  z,z^{\ast}\right)  \right| \nonumber\\
&  =\int_{\left|  z\right|  <1/R,\operatorname{Im}z>0}\frac{d^{2}z\,}{|z|^{4}
}\left|  \operatorname{Im}\,\frac{1}{z}\right|  ^{\operatorname{Re}\left(
\alpha-1\right)  }\left[  \prod_{k=1}^{n}\left|  -\frac{1}{z}-z_{k}\right|
^{\operatorname{Re}\left(  \gamma_{k}-\alpha-1\right)  }\right]  \left|
P\left(  -\frac{1}{z},-\frac{1}{z^{\ast}}\right)  \right| \nonumber\\
&  =\int_{\left|  z\right|  <1/R,\operatorname{Im}z>0}\frac{d^{2}z\,}{|z|^{4}
}\left[  |z|^{-2}\left|  \operatorname{Im}\,z\right|  \right]
^{\operatorname{Re}\left(  \alpha-1\right)  }\nonumber\\
&  \times\left[  \prod_{k=1}^{n}\left(  \left|  \frac{z_{k}}{z}\right|
\left|  z+z_{k}^{-1}\right|  \right)  ^{\operatorname{Re}\left(  \gamma
_{k}-\alpha-1\right)  }\right]  P\left(  -\frac{1}{z},-\frac{1}{z^{\ast}
}\right)  \,. \label{int-D-n-plus-1-1}
\end{align}
Then
\begin{align}
&  \int_{D_{n+1,+}}d^{2}z\,\left|  \operatorname{Im}\,z\right|
^{\operatorname{Re}\left(  \alpha-1\right)  }\left[  \prod_{k=1}^{n}\left|
z-z_{k}\right|  ^{\operatorname{Re}\left(  \gamma_{k}-\alpha-1\right)
}\right]  \left|  P\left(  z,z^{\ast}\right)  \right| \nonumber\\
&  =\left[  \prod_{k=1}^{n}\left|  z_{k}\right|  ^{\operatorname{Re}\left(
\gamma_{k}-\alpha-1\right)  }\right]  \int_{\left|  z\right|
<1/R,\operatorname{Im}z>0}d^{2}z\,|z|^{-4-2\operatorname{Re}\left(
\alpha-1\right)  -N_{P}-\sum_{k=1}^{n}\operatorname{Re}\left(  \gamma
_{k}-\alpha-1\right)  }\nonumber\\
&  \times\left|  \operatorname{Im}\,z\right|  ^{\operatorname{Re}\left(
\alpha-1\right)  }\left[  \prod_{k=1}^{n}\left|  z+z_{k}^{-1}\right|
^{\operatorname{Re}\left(  \gamma_{k}-\alpha-1\right)  }\right]  |z|^{N_{P}
}\left|  P\left(  -1/z,-1/z^{\ast}\right)  \right| \nonumber\\
&  =\int_{\left|  z\right|  <1/R,\operatorname{Im}z>0}d^{2}z\tilde
{F}_{\text{sing}}\tilde{F}_{\text{reg}} \label{int-D-n-plus-1-2}
\end{align}
where
\begin{equation}
\tilde{F}_{\text{sing}}=|z|^{\left(  n-2\right)  \operatorname{Re}\left(
\alpha+1\right)  -N_{P}-\sum_{k=1}^{n}\operatorname{Re}\gamma_{k}}\left|
\operatorname{Im}\,z\right|  ^{\operatorname{Re}\left(  \alpha-1\right)  }\,,
\end{equation}
\begin{equation}
\tilde{F}_{\text{reg}}=\left[  \prod_{k=1}^{n}\left|  z_{k}\right|
^{\operatorname{Re}\left(  \gamma_{k}-\alpha-1\right)  }\right]  \left[
\prod_{k=1}^{n}\left|  z+z_{k}^{-1}\right|  ^{\operatorname{Re}\left(
\gamma_{k}-\alpha-1\right)  }\right]  |z|^{N_{P}}\left|  P\left(
-1/z,-1/z^{\ast}\right)  \right|  \,.
\end{equation}

Here we have the same structure of the integral as in eq. (\ref{int-F-F}).
Note that factor
\[
|z|^{N_{P}}\left|  P\left(  1/z,1/z^{\ast}\right)  \right|
\]
may have a soft singularity (divergent derivatives) at $z\rightarrow0$ but
still is bounded in the integration limit

\begin{equation}
\left|  z\right|  <R^{-1}\Longrightarrow\quad|z|^{N_{P}}\left|  P\left(
1/z,1/z^{\ast}\right)  \right|  <\text{const\thinspace.}
\end{equation}
Therefore for the derivation of \emph{sufficient} convergence conditions we
still can use methods applied earlier to integral (\ref{int-F-F}). Note that
parameter $\gamma_{j}$ appearing in eq. (\ref{int-F-F}) is now replaced by
\begin{equation}
\gamma_{j}\rightarrow\left(  n-1\right)  \operatorname{Re}\left(
\alpha+1\right)  -N_{P}-\sum_{k=1}^{n}\operatorname{Re}\gamma_{k}\,.
\end{equation}
Making this replacement in convergence old sufficient conditions
(\ref{Re-alpha-positive-7}), (\ref{Re-gamma-positive-7}) for integral
(\ref{int-F-F}), we arrive at \emph{sufficient} conditions for the new
integral (\ref{int-D-n-plus-1-1})
\begin{equation}
\operatorname{Re}\alpha>0\,, \label{condition-1-3}
\end{equation}
\begin{equation}
\left(  n-1\right)  \operatorname{Re}\left(  \alpha+1\right)  -N_{P}
-\sum_{k=1}^{n}\operatorname{Re}\gamma_{k}>0\,. \label{condition-3-3}
\end{equation}
The last condition can be rearranged to the form
(\ref{convergence-condition-3}).

Thus \emph{sufficient} convergence conditions in region $D_{n+1,+}$ lead to
conditions (\ref{convergence-condition-1}) and (\ref{convergence-condition-3})
in the full set of \emph{sufficient} convergence conditions of Statement 1.

Now we turn to the \emph{necessary} convergence conditions. In case of
semicircle regions $D_{j,+}$ ($1\leq j\leq n$) the derivation of necessary
convergence conditions was based on assumption (\ref{P-z-j-nonzero}) and on
bound (\ref{P-C1-C2-bound}). In case of region $D_{n+1,+}$ the role of
polynomial $P\left(  z,z^{\ast}\right)  $ is played by function
\begin{equation}
Q\left(  z,z^{\ast}\right)  =|z|^{N_{P}}\left|  P\left(  -1/z,-1/z^{\ast
}\right)  \right|  \,.
\end{equation}
Generally speaking, this function is not a polynomial and it is not regular at
$z\rightarrow0$ (although it is bounded at $z\rightarrow0$) so that we cannot
derive the analog of bound (\ref{P-C1-C2-bound}). But in the special case of
factorizable polynomials $P$ (\ref{P-T-factorizable}) $Q\left(  z,z^{\ast
}\right)  $ is a polynomial with property
\begin{equation}
\left[  Q\left(  z,z^{\ast}\right)  \right]  _{z=0}\neq0
\end{equation}
so that the methods used in regions $D_{j,+}$ ($1\leq j\leq n$) for the
derivation of necessary convergence conditions work also in our case. Thus in
the case of factorizable polynomials $P$ (\ref{P-T-factorizable}) conditions
(\ref{condition-1-3}) (\ref{condition-3-3}) are both necessary and sufficient
for the convergence of integral (\ref{int-D-n-plus-1-1}).

This completes the analysis of the contribution of region $D_{n+1,+}$ to the
full set of necessary and sufficient convergence conditions of Statement 2.

\section{Analytical continuation}

\label{Analytical-continuation-section}

\setcounter{equation}{0} 

\subsection{Starting analytical continuation from the convergence region}

Thus we have proved that integral (\ref{Pi-P-n-def}) defining function
$\Pi_{P}^{\left(  n\right)  }$ is convergent in the region $U$ of parameters
$\alpha,\left\{  \gamma_{k}\right\}  _{k=1}^{n}$ specified by conditions
(\ref{convergence-condition-1}) -- (\ref{convergence-condition-3}) of
Statement 1. This convergence region is non-empty. Obviously function $\Pi
_{P}^{\left(  n\right)  }$ is holomorphic in this $\alpha,\left\{  \gamma
_{k}\right\}  _{k=1}^{n}$ region of $\mathbb{C}^{n+1}$. The next step of the
work is to study the analytical continuation of $\Pi_{P}^{\left(  n\right)  }$
from this region $U\subset\mathbb{C}^{n+1}$ to the full space $\mathbb{C}
^{n+1}$ and to prove announced meromorphic structure
(\ref{K-rough-pole-structure}).

Note that region $U$ is connected. This excludes a possible unpleasant
situation when analytical continuations starting from different disconnected
regions could lead to different analytical continuations.

The above analysis of conditions (\ref{convergence-condition-1}) --
(\ref{convergence-condition-3}) for the absolute convergence of the integral
defining function $\Pi_{P}^{\left(  n\right)  }$ proceeded in terms of
integral (\ref{Pi-P-n-def}) over complex semiplane $\mathbb{C}_{+}$. When it
comes to the problem of analytical continuation of $\Pi_{P}^{\left(  n\right)
}$ in $\alpha,\left\{  \gamma_{k}\right\}  _{k=1}^{n}$ from the convergence
region $U$ to the full complex space $\mathbb{C}^{n+1}$ of parameters
$\alpha,\left\{  \gamma_{k}\right\}  _{k=1}^{n}$, it is more convenient to
work with the equivalent representation for $\Pi_{P}^{\left(  n\right)  }$
with $z$ integral (\ref{Pi-P-n-plane}) running over full complex plane
$\mathbb{C}$. The advantage of this representation becomes clear at later
stages of the work (see section \ref{Meromorphic-structure-D-z-j-r-j-section})
but it makes sense to pass from the original semiplane integral representation
for $\Pi_{P}^{\left(  n\right)  }$ to the plane representation
(\ref{Pi-P-n-plane}) right now.

Our first step is to split integral (\ref{Pi-P-n-plane}) in the sum of
integrals over regions $D_{j}$ (\ref{Pi-D-b}) and to study the problem of
analytical continuation in $\alpha,\left\{  \gamma_{k}\right\}  _{k=1}^{n}$
for each separate integral
\begin{equation}
S_{j}\left(  \alpha,\left\{  \gamma_{k}\right\}  _{k=1}^{n}\right)
=\int_{D_{j}}d^{2}z\,\left|  \operatorname{Im}\,z\right|  ^{\alpha-1}
\prod_{k=1}^{n}\left|  z-z_{k}\right|  ^{\gamma_{k}-\alpha-1}P\left(
z,z^{\ast}\right)  \,. \label{Pi-b-def}
\end{equation}
In order to simplify notation, on the LHS we omit the dependence on quantities
$z_{k},P$ which are kept fixed in our analytical continuation.

Alternatively we can work with regions $D_{j,+}$ defined by
(\ref{D-plus-def-via-D-b}) and shown in Fig. \ref{fig-2}:
\begin{equation}
S_{j,+}\left(  \alpha,\left\{  \gamma_{k}\right\}  _{k=1}^{n}\right)
=\int_{D_{j,+}}d^{2}z\,\left|  \operatorname{Im}\,z\right|  ^{\alpha-1}
\prod_{k=1}^{n}\left|  z-z_{k}\right|  ^{\gamma_{k}-\alpha-1}P\left(
z,z^{\ast}\right)  \,. \label{S-j-plus-def}
\end{equation}
Obviously
\begin{equation}
S_{j}\left(  \alpha,\left\{  \gamma_{k}\right\}  _{k=1}^{n}\right)
=2S_{j,+}\left(  \alpha,\left\{  \gamma_{k}\right\}  _{k=1}^{n}\right)  \,.
\label{S-j-j-plus}
\end{equation}

\begin{itemize}
\item in the convergence region of $\alpha,\left\{  \gamma_{k}\right\}
_{k=1}^{n}$ space,

\item after the analytical continuation in $\alpha,\left\{  \gamma
_{k}\right\}  _{k=1}^{n}$.
\end{itemize}

The choice of $D_{j}$-decomposition or $D_{j,+}$-decomposition is a matter of convenience.

For each separate region $D_{j}$ we will prove that function $S_{j}\left(
\alpha,\left\{  \gamma_{k}\right\}  _{k=1}^{n}\right)  $ originally defined in
the convergence region can be analytically continued to a meromorphic function
with the pole structure
\begin{equation}
S_{j}\left(  \alpha,\left\{  \gamma_{k}\right\}  _{k=1}^{n}\right)
=\Gamma_{j}\left(  \alpha,\left\{  \gamma_{k}\right\}  _{k=1}^{n}\right)
H_{j}\left(  \alpha,\left\{  \gamma_{k}\right\}  _{k=1}^{n}\right)
\label{Pi-b-Gamma-H-decomposition}
\end{equation}
where

\begin{itemize}
\item $H_{j}\left(  \alpha,\left\{  \gamma_{k}\right\}  _{k=1}^{n}\right)  $
is an entire function of $\alpha,\left\{  \gamma_{k}\right\}  _{k=1}^{n}$
(i.e. holomorphic in $\mathbb{C}^{n+2}$),

\item $\Gamma_{j}\left(  \alpha,\left\{  \gamma_{k}\right\}  _{k=1}
^{n}\right)  $ is a certain product of Euler $\Gamma$ functions depending on
linear combinations of variables $\alpha,\left\{  \gamma_{k}\right\}
_{k=1}^{n}$.
\end{itemize}

After we complete the proof of the pole structure
(\ref{Pi-b-Gamma-H-decomposition}) of functions $S_{j}\left(  \alpha,\left\{
\gamma_{k}\right\}  _{k=1}^{n}\right)  $, it is straightforward to derive pole
structure (\ref{K-rough-pole-structure}) from
\begin{equation}
\Pi_{P}^{(n)}\left(  \alpha,\left\{  \gamma_{k}\right\}  _{k=1}^{n}\right)
=\sum_{j}S_{j}\left(  \alpha,\left\{  \gamma_{k}\right\}  _{k=1}^{n}\right)
=\sum_{j}\Gamma_{i}\left(  \alpha,\left\{  \gamma_{k}\right\}  _{k=1}
^{n}\right)  H_{b}\left(  \alpha,\left\{  \gamma_{k}\right\}  _{k=1}
^{n}\right)  \,.
\end{equation}

\subsection{Results for $\Gamma_{j}$}

For functions $\Gamma_{j}\left(  \alpha,\left\{  \gamma_{k}\right\}
_{k=1}^{n}\right)  $ appearing in decomposition
(\ref{Pi-b-Gamma-H-decomposition}) one can derive the following expressions
\begin{equation}
\Gamma_{0}\left(  \alpha,\left\{  \gamma_{k}\right\}  _{k=1}^{n}\right)
=\Gamma\left(  \frac{\alpha}{2}\right)  \,, \label{Gamma-0}
\end{equation}

\begin{equation}
\Gamma_{j}\left(  \alpha,\left\{  \gamma_{k}\right\}  _{k=1}^{n}\right)
=\Gamma\left(  \frac{\alpha}{2}\right)  \Gamma\left(  \frac{\gamma_{j}}
{2}\right)  \quad(0\leq j\leq n)\,, \label{Gamma-j-intermediate}
\end{equation}
\begin{equation}
\Gamma_{n+1}\left(  \alpha,\left\{  \gamma_{k}\right\}  _{k=1}^{n}\right)
=\Gamma\left(  \frac{\alpha}{2}\right)  \Gamma\left(  \frac{1}{2}\left(
-2M_{P}-1-\alpha-\sum_{k=1}^{n}\left(  \gamma_{k}-\alpha-1\right)  \right)
\right)  \,. \label{Gamma-n-plus-1}
\end{equation}
Parameter $M_{P}$ appearing on the RHS eq. (\ref{Gamma-n-plus-1}) related to
polynomial $P$ of functions $\Pi_{P}^{(n)}$ is defined by eq.
(\ref{M-P-min-def}).

\subsection{Rough pole structure}

It should be stressed that all factorized decompositions of meromorphic
function into regular and pole factors

\begin{itemize}
\item intermediate result (\ref{Pi-b-Gamma-H-decomposition}),

\item final result (\ref{K-rough-pole-structure}).
\end{itemize}

are somewhat rough in the sense that poles of Gamma functions may be sometimes
compensated by zeros of regular factors. In particular, the derivation of
(\ref{K-rough-pole-structure}) from (\ref{Pi-b-Gamma-H-decomposition}) is
straightforward if one combines all singular $\Gamma$ factors appearing on the
RHS of eqs. (\ref{Gamma-0}) -- (\ref{Gamma-n-plus-1}) in one common product
\begin{equation}
\Gamma\left(  \frac{\alpha}{2}\right)  \Gamma\left(  \frac{1}{2}\left(
-2M_{P}-1-\alpha-\sum_{k=1}^{n}\left(  \gamma_{k}-\alpha-1\right)  \right)
\right)  \prod_{j=1}^{n}\Gamma\left(  \frac{\gamma_{j}}{2}\right)
\label{Gamma-rough-factor}
\end{equation}
appearing on the RHS of (\ref{K-rough-pole-structure}). Therefore
representation (\ref{K-rough-pole-structure}) creates an illusion that the
structure of singularities is more severe than it really is. Although final
representation (\ref{K-rough-pole-structure}) exaggerates the pole
singularities, this exaggeration does not interfere with our final aim, the
proof that our analytical regularization provides a finite result for
$f_{2}(C)$, because `exaggerated pole factor' (\ref{Gamma-rough-factor}) is
regular at the final point of the analytical continuation used in eqs.
(\ref{I-1-ren-via-Pi}), (\ref{I-2-ren-via-Pi}).

\section{Calculation of singular factors $\Gamma_{j}$}

\setcounter{equation}{0} 

\subsection{Preliminary remarks}

Now we turn to the calculation of expressions for $\Gamma_{j}$ announced in
eqs. (\ref{Gamma-0}) -- (\ref{Gamma-n-plus-1}). These expressions hint that
Gamma functions appearing in $\Gamma_{j}$ are determined by $z$ singularities
of the integrand of (\ref{Pi-b-def}) in region $D_{j}$:
\begin{equation}
\operatorname{Im}\,z\rightarrow0\quad\Longrightarrow\quad\Gamma\left(
\frac{\alpha}{2}\right)  \,,
\end{equation}
\begin{equation}
z\rightarrow z_{k}\quad\Longrightarrow\quad\Gamma\left(  \frac{\gamma_{j}}
{2}\right)  \,,
\end{equation}
\begin{equation}
|z|\rightarrow\infty\quad\Longrightarrow\quad\Gamma\left(  \frac{1}{2}\left(
-2M_{P}-1-\alpha-\sum_{k=1}^{n}\left(  \gamma_{k}-\alpha-1\right)  \right)
\right)  \,.
\end{equation}

Note that region $D_{0}$ contains only the singularity at $\operatorname{Im}
\,z\rightarrow0$ so that $\Gamma_{0}$ has only one Gamma function
$\Gamma\left(  \alpha/2\right)  $.

Region $D_{j}$ with $1\leq j\leq n$ has two singularities ($\operatorname{Im}
\,z\rightarrow0$ and $z\rightarrow z_{k}$) which lead to two associated Gamma
functions $\Gamma\left(  \alpha/2\right)  $ and $\Gamma\left(  \gamma
_{j}/2\right)  $ in eq. (\ref{Gamma-j-intermediate}).

Region $D_{n+1}$ has also two singularities ($\operatorname{Im}\,z\rightarrow
0$ and $|z|\rightarrow\infty$) which lead to the two Gamma function in
$\Gamma_{n+1}$ appearing on the RHS of eq. (\ref{Gamma-n-plus-1}).

\subsection{Calculation of $\Gamma_{0}$}

\label{Gamma-0-section}

The case of $\Gamma_{0}$ is the simplest because in $D_{0}$ we have only one
singularity $\operatorname{Im}\,z\rightarrow0$ and expect only one associated
Euler Gamma function $\Gamma\left(  \alpha/2\right)  $ in $\Gamma_{0}\left(
\alpha,\left\{  \gamma_{k}\right\}  _{k=1}^{n}\right)  $. Function $S_{0}$ is
defined by eq. (\ref{Pi-b-def}) with $j=0$
\begin{equation}
S_{0}\left(  \alpha,\left\{  \gamma_{k}\right\}  _{k=1}^{n}\right)
=\int_{D_{0}}d^{2}z\,\left|  \operatorname{Im}\,z\right|  ^{\alpha-1}
\prod_{k=1}^{n}\left|  z-z_{k}\right|  ^{\gamma_{k}-\alpha-1}P\left(
z,z^{\ast}\right)  \,. \label{Pi-0-def}
\end{equation}
but we prefer to work with its $D_{0,+}$ analog (\ref{S-j-plus-def})
\begin{equation}
S_{0,+}\left(  \alpha,\left\{  \gamma_{k}\right\}  _{k=1}^{n}\right)
=\int_{D_{0,+}}d^{2}z\,\left|  \operatorname{Im}\,z\right|  ^{\alpha-1}
\prod_{k=1}^{n}\left|  z-z_{k}\right|  ^{\gamma_{k}-\alpha-1}P\left(
z,z^{\ast}\right)  \,. \label{S-0-plus-def}
\end{equation}
obeying relation (\ref{S-j-j-plus})
\begin{equation}
S_{0}\left(  \alpha,\left\{  \gamma_{k}\right\}  _{k=1}^{n}\right)
=2S_{0,+}\left(  \alpha,\left\{  \gamma_{k}\right\}  _{k=1}^{n}\right)  \,,
\end{equation}

As discussed in the proof of Statement 1, integral (\ref{S-0-plus-def}) is
absolutely convergent in the region constrained by condition
(\ref{convergence-condition-1})
\begin{equation}
\operatorname{Re}\alpha>0\,.\label{Re-alpha-positive-for-D-cent-2}
\end{equation}
The same obviously holds for integral (\ref{Pi-0-def}).

Our aim of is to prove that function $S_{0}\left(  \alpha,\left\{  \gamma
_{k}\right\}  _{k=1}^{n}\right)  $ can be analytically continued to a
meromorphic function in $\alpha,\left\{  \gamma_{k}\right\}  $ with pole
structure (\ref{Pi-b-Gamma-H-decomposition}), (\ref{Gamma-0})
\begin{equation}
S_{0}\left(  \alpha,\left\{  \gamma_{k}\right\}  _{k=1}^{n}\right)
=2S_{0,+}\left(  \alpha,\left\{  \gamma_{k}\right\}  _{k=1}^{n}\right)
=\Gamma\left(  \frac{\alpha}{2}\right)  H_{0}\left(  \alpha,\left\{
\gamma_{k}\right\}  _{k=1}^{n}\right)  \label{Pi-0-H-0}
\end{equation}
where $H_{0}\left(  \alpha,\left\{  \gamma_{k}\right\}  _{k=1}^{n}\right)  $
is an entire function.

In order to derive meromorphic factorization (\ref{Pi-0-H-0}), we will split
$D_{0}$ in subregions. This splitting is essentially the same as in our
analysis of convergence conditions in section
\ref{D-0-plus-convergence-section}.

Now we can reuse the work of section \ref{D-0-plus-convergence-section}
inverting replacements (\ref{U-to-S-dic-1}) -- (\ref{U-to-S-dic-4}). We have
\begin{equation}
S_{0,+}\left(  \alpha,\left\{  \gamma_{k}\right\}  _{k=1}^{n}\right)
=S_{0,1}\left(  \alpha,\left\{  \gamma_{k}\right\}  _{k=1}^{n}\right)
+S_{0,2}\left(  \alpha,\left\{  \gamma_{k}\right\}  _{k=1}^{n}\right)  \,.
\label{Pi-0-decomposition-1}
\end{equation}
This equation is first derived in absolute convergence region
(\ref{Re-alpha-positive-for-D-cent-2}) of integral (\ref{S-0-plus-def}). Note
that the integrand (\ref{S-A-def}) for $S_{0,2}$ is regular in associated
region $D_{0,2}$ (and in its small vicinity) so that
\begin{equation}
S_{0,2}\left(  \alpha,\left\{  \gamma_{k}\right\}  _{k=1}^{n}\right)
=\text{entire function in }\mathbb{C}^{n+1}\,.
\end{equation}
Combining this fact with decomposition (\ref{Pi-0-decomposition-1}), we
conclude that the problem of the derivation of representation (\ref{Pi-0-H-0})
for the analytical continuation of $S_{0}$ reduces to the derivation its
analog for $S_{0,1}$:
\begin{equation}
S_{0,1}\left(  \alpha,\left\{  \gamma_{k}\right\}  _{k=1}^{n}\right)
=\Gamma\left(  \frac{\alpha}{2}\right)  H_{0,1}\left(  \alpha,\left\{
\gamma_{k}\right\}  _{k=1}^{n}\right)  \label{Pi-0-1-H-0-1}
\end{equation}
where $H_{0,1}\left(  \alpha,\left\{  \gamma_{k}\right\}  _{k=1}^{n}\right)  $
is an entire function.

Due to constraint (\ref{eps-constraint-1}) region $D_{0,1}$ consists of $n+1$
disconnected components $D_{0,1,j}$.

Using (\ref{S-A-def}) we define functions $S_{0,1,j}$ associated with regions
$D_{0,1,j}$. Then
\begin{equation}
S_{0,1}\left(  \alpha,\left\{  \gamma_{k}\right\}  _{k=1}^{n}\right)
=\sum_{j=0}^{n}S_{0,1,j}\left(  \alpha,\left\{  \gamma_{k}\right\}  _{k=1}
^{n}\right)  \,.
\end{equation}
We have associated functions $S_{0,1,j}$ (\ref{S-A-def})
\begin{align*}
&  S_{0,1,j}\left(  \alpha,\left\{  \gamma_{k}\right\}  _{k=1}^{n}\right)  \\
&  =\int_{D_{0,1,j}}d^{2}z\,\left|  \operatorname{Im}\,z\right|  ^{\alpha
-1}\prod_{k=1}^{n}\left|  z-z_{k}\right|  ^{\gamma_{k}-\alpha-1}P\left(
z,z^{\ast}\right)  \,
\end{align*}
\begin{equation}
=\int_{0}^{\eta}dy\,\left|  y\right|  ^{\alpha-1}\int_{x_{j}^{-}(y)}
^{x_{j}^{+}(y)}dx\prod_{k=1}^{n}\left|  \left(  x-z_{k}\right)  ^{2}
+y^{2}\right|  ^{(\gamma_{k}-\alpha-1)/2}P\left(  x+iy,x-iy\right)
\label{Pi-0-1-j-calc-1}
\end{equation}
This integral representation is absolute convergent in the region
(\ref{Re-alpha-positive-for-D-cent-2})
\begin{equation}
\operatorname{Re}\alpha>0\label{Re-alpha-positive-for-D-cent-4}
\end{equation}
inherited from the absolute convergence region of integral (\ref{Pi-0-def}).

\bigskip Next we want to continue functions $S_{0,1,j}$ analytically to
arbitrary complex $\alpha,\left\{  \gamma_{k}\right\}  _{k=1}^{n}$.

The problem of the derivation of meromorphic decomposition (\ref{Pi-0-1-H-0-1}
) reduces to the problem of derivation of meromorphic structure
\begin{equation}
S_{0,1,j}\left(  \alpha,\left\{  \gamma_{k}\right\}  _{k=1}^{n}\right)
=\Gamma\left(  \frac{\alpha}{2}\right)  H_{0,1,j}\left(  \alpha,\left\{
\gamma_{k}\right\}  _{k=1}^{n}\right)  \quad(0\leq j\leq n)
\label{Pi-0-1-1-H-0-1-1}
\end{equation}
where $H_{0,1,j}\left(  \alpha,\left\{  \gamma_{k}\right\}  _{k=1}^{n}\right)
$ are entire functions.

Using symmetry (\ref{P-symmetric-1}) of the polynomial $P\left(  z,z^{\ast
}\right)  $ we can write
\begin{equation}
\left[  P\left(  z,z^{\ast}\right)  \right]  _{z=x+iy}=Q\left(  x,y^{2}
\right)
\end{equation}
where $Q\left(  x,q\right)  $ is a polynomial of its arguments $x,q$.

Therefore in absolute convergence region (\ref{Re-alpha-positive-for-D-cent-4}
) we derive from (\ref{Pi-0-1-j-calc-1})
\begin{align}
&  S_{0,1,j}\left(  \alpha,\left\{  \gamma_{k}\right\}  _{k=1}^{n}\right)
\nonumber\\
&  =\int_{0}^{\eta}dy\,\,y^{\alpha-1}\int_{x_{j}^{-}(y)}^{x_{j}^{+}(y)}
dx\prod_{k=1}^{n}\left|  \left(  x-z_{k}\right)  ^{2}+y^{2}\right|
^{(\gamma_{k}-\alpha-1)/2}Q\left(  x,y^{2}\right)  \,.
\end{align}

Changing integration variable
\begin{equation}
y=\sqrt{q}
\end{equation}
we find
\begin{align}
&  S_{0,1,j}\left(  \alpha,\left\{  \gamma_{k}\right\}  _{k=1}^{n}\right)
\nonumber\\
&  =\frac{1}{2}\int_{0}^{\sqrt{\eta}}dq\,\,q^{(\alpha/2)-1}\int_{x_{j}
^{-}\left(  q^{1/2}\right)  }^{x_{j}^{+}\left(  q^{1/2}\right)  }dx\prod
_{k=1}^{n}\left|  \left(  x-z_{k}\right)  ^{2}+q\right|  ^{(\gamma_{k}
-\alpha-1)/2}Q\left(  x,q\right)  \,.
\end{align}
Let us define
\begin{equation}
F_{0,1,j}(q)=\frac{1}{2}\int_{x_{j}^{-}\left(  q^{1/2}\right)  }^{x_{j}
^{+}\left(  q^{1/2}\right)  }dx\prod_{k=1}^{n}\left|  \left(  x-z_{k}\right)
^{2}+q\right|  ^{(\gamma_{k}-\alpha-1)/2}Q\left(  x,q\right)  \,.
\label{F-0-1-def}
\end{equation}
Then
\begin{equation}
S_{0,1,j}\left(  \alpha,\left\{  \gamma_{k}\right\}  _{k=1}^{n}\right)
=\int_{0}^{\sqrt{\eta}}dq\,\,q^{(\alpha/2)-1}F_{0,1,j}\left(  q\right)  \,.
\label{S-0-1-j-via-F}
\end{equation}
Note that integration limits $x_{j}^{+}\left(  q^{1/2}\right)  $ on the RHS of
(\ref{F-0-1-def}) given by (\ref{x-pm-j-general-def})
\begin{equation}
x_{j}^{\pm}\left(  q^{1/2}\right)  =X_{j}^{\pm}+\sigma_{j}^{\pm}\sqrt{\left(
\rho_{j}^{\pm}\right)  ^{2}-q}
\end{equation}
are infinitely differentiable functions of $q$ in the integration region of
(\ref{S-0-1-j-via-F})
\begin{equation}
0\leq q\leq\sqrt{\eta}
\end{equation}
because we have
\begin{equation}
\eta<\rho_{j}^{\pm}
\end{equation}
since $\eta$ obeys constraint (\ref{eps-constraint-1}).

Next, the integrand of (\ref{F-0-1-def}) is also an infinitely differentiable
function of $x,q$ in the integration region
\begin{equation}
x_{j}^{-}\left(  q^{1/2}\right)  <x<x_{j}^{+}\left(  q^{1/2}\right)  \,,
\end{equation}
\begin{equation}
0\leq q\leq\sqrt{\eta}
\end{equation}
because

1) points $z_{k}$ are outside of the integration region so that all factors
$\left|  \left(  x-z_{k}\right)  ^{2}+q\right|  ^{(\gamma_{k}-\alpha-1)/2}$
are regular in this integration region,

2) $Q\left(  x,q\right)  $ is a polynomial of $x$ and $q$.

Thus both integrand and integration limits on the RHS of (\ref{F-0-1-def}) are
infinitely differentiable functions. Therefore $F_{0,1,j}(q)$ is also an
infinitely differentiable function of $q$ in the range $0\leq q\leq\sqrt{\eta
}$.

Now the problem reduces to the study of analytical continuation of integral
(\ref{F-0-1-def}) in $\alpha$ starting from the convergence region
(\ref{Re-alpha-positive-for-D-cent-4}). This analytical continuation can be
done iteratively integrating by parts
\[
S_{0,1,j}\left(  \alpha,\left\{  \gamma_{k}\right\}  _{k=1}^{n}\right)
=\int_{0}^{\sqrt{\eta}}dq\,\,q^{(\alpha/2)-1}F_{0,1,j}\left(  q\right)
\]
\begin{gather}
=\left(  \frac{\alpha}{2}\right)  ^{-1}\left.  q^{\alpha/2}F_{0,1,j}\left(
q\right)  \right|  _{0}^{\sqrt{\eta}}-\left(  \frac{\alpha}{2}\right)
^{-1}\int_{0}^{\sqrt{\eta}}dq\,\,q^{\alpha/2}F_{0,1,j}^{\prime}\left(
q\right) \nonumber\\
=\left(  \frac{\alpha}{2}\right)  ^{-1}\left.  q^{\alpha/2}F_{0,1,j}\left(
q\right)  \right|  _{0}^{\sqrt{\eta}}-\left(  \frac{\alpha}{2}\right)
^{-1}\left(  \frac{\alpha}{2}+1\right)  ^{-1}\left.  q^{\left(  \alpha
/2\right)  +1}F_{0,1,j}^{\prime}\left(  q\right)  \right|  _{0}^{\sqrt{\eta}
}\nonumber\\
+\left(  \frac{\alpha}{2}\right)  ^{-1}\left(  \frac{\alpha}{2}+1\right)
^{-1}\int_{0}^{\sqrt{\eta}}dq\,\,q^{\left(  \alpha/2\right)  +1}
F_{0,1,j}^{\prime\prime}\left(  q\right)  =\ldots
\end{gather}
The poles appearing in this analytical continuation may lie only at points
\begin{equation}
\frac{\alpha}{2}=0,-1,-2,...
\end{equation}
This proves pole structure (\ref{Pi-0-1-1-H-0-1-1}) of $S_{0,1,j}$. This also
completes the derivation of the chain of related meromorphic representations
for $S_{0,1}$ (\ref{Pi-0-1-H-0-1}) and for $S_{0}$ (\ref{Pi-0-H-0}) as well as
the derivation of expression (\ref{Gamma-0}) for $\Gamma_{0}$.

\subsection{Calculation of $\Gamma_{j}$ ($1\leq j\leq n$)}

\label{Meromorphic-structure-D-z-j-r-j-section}

We want to derive meromorphic structure (\ref{Pi-b-Gamma-H-decomposition}),
(\ref{Gamma-j-intermediate}) of functions $S_{j}$ ($1\leq j\leq n$). We start
from eq. (\ref{Pi-b-def})
\begin{align}
S_{j}\left(  \alpha,\left\{  \gamma_{k}\right\}  _{k=1}^{n}\right)   &
=\int_{D_{j}}d^{2}z\,\left|  \operatorname{Im}\,z\right|  ^{\alpha-1}\left[
\prod_{k=1}^{n}\left|  z-z_{k}\right|  ^{\gamma_{k}-\alpha-1}\right]  P\left(
z,z^{\ast}\right)  \nonumber\\
&  =\int_{\left|  z-z_{j}\right|  <r_{j}}d^{2}z\,\left|  \operatorname{Im}
\,z\right|  ^{\alpha-1}\left[  \prod_{k=1}^{n}\left|  z-z_{k}\right|
^{\gamma_{k}-\alpha-1}\right]  P\left(  z,z^{\ast}\right)  \nonumber\\
&  =\int_{\left|  z\right|  <r_{j}}d^{2}z\,\left|  \operatorname{Im}
\,z\right|  ^{\alpha-1}\left[  \prod_{k=1}^{n}\left|  z+z_{j}-z_{k}\right|
^{\gamma_{k}-\alpha-1}\right]  \nonumber\\
&  \times P\left(  z+z_{j},z^{\ast}+z_{j}\right)  \,.
\end{align}
Let us introduce notation
\begin{equation}
M\left(  \alpha,\left\{  \gamma_{k}\right\}  ,\left\{  z_{k}\right\}
;z,z^{\ast}\right)  =\,P\left(  z+z_{j},z^{\ast}+z_{j}^{\ast}\right)
\prod_{\substack{k=1\\k\neq j}}^{n}\left|  z+z_{j}-z_{k}\right|  ^{\gamma
_{k}-\alpha-1}\,.\label{M-explicit-def}
\end{equation}
Then
\begin{equation}
S_{j}\left(  \alpha,\left\{  \gamma_{k}\right\}  _{k=1}^{n}\right)
=\int_{\left|  z\right|  <r_{j}}d^{2}z\,\left|  \operatorname{Im}\,z\right|
^{\alpha-1}\left|  z\right|  ^{\gamma_{j}-\alpha-1}M\left(  \alpha,\left\{
\gamma_{k}\right\}  ,\left\{  z_{k}\right\}  ;z,z^{\ast}\right)
\,.\label{int-D-z-k-r-k-M}
\end{equation}
Note that function $M\left(  \alpha,\left\{  \gamma_{k}\right\}  ,\left\{
z_{k}\right\}  ;z,z^{\ast}\right)  $

\begin{itemize}
\item is infinitely differentiable in $\operatorname{Re}z$ and in
$\operatorname{Im}z$ in a vicinity of the integration region $\left|
z\right|  <r_{j}$,

\item is an entire function of $\alpha,\left\{  \gamma_{k}\right\}  $ at any
fixed $z$ in the region $\left|  z\right|  <r_{k}$.
\end{itemize}

Remember that in our choice of regions we are free to choose $r_{k}$ as small
as we like. Therefore we can replace $M\left(  \alpha,\left\{  \gamma
_{k}\right\}  ,\left\{  z_{k}\right\}  ;z,z^{\ast}\right)  $ by its Taylor
series in small $z,z^{\ast}$ or in small $\operatorname{Re}z,\operatorname{Im}
z$ and concentrate on the analytical continuation in variables $\alpha
,\gamma_{j}$ keeping other parameters $\gamma_{k}$ with $k\neq j$ fixed.

Thus we can turn to a simplified problem of analytical continuation in
$\alpha,\gamma$ for function $\tilde{S}\left(  \alpha,\gamma\right)  $ defined
by integral
\begin{equation}
\tilde{S}\left(  \alpha,\gamma\right)  =\int_{\left|  z\right|  <\rho}
d^{2}z\,\left|  \operatorname{Im}\,z\right|  ^{\alpha-1}\left|  z\right|
^{\gamma-\alpha-1}\tilde{M}\left(  \operatorname{Re}z,\operatorname{Im}
z\right)  \label{Pi-tilde-integral}
\end{equation}
with $\tilde{M}$ is represented by a convergent power series
\begin{equation}
\tilde{M}\left(  x,y\right)  =\sum_{m=0}^{\infty}\sum_{n=0}^{\infty}
a_{mn}x^{m}y^{n}\,.
\end{equation}
In this analysis we are free to choose parameter $\rho$ as small as we like in
order to improve the convergence of series whenever this is needed.

Absolute-convergence (sufficient) condition for integral
(\ref{Pi-tilde-integral}) is determined by Statement 3 of section
\ref{convergence-in-D-j-semicricle-section}
\begin{align}
\text{Re\thinspace}\alpha &  >0\,,\label{Re-alpha-positive-5}\\
\text{Re\thinspace}\gamma &  >0\,. \label{Re-gamma-positive-5}
\end{align}

Working in this convergence region we can use the symmetry of integration
region under reflections
\begin{equation}
x\rightarrow-x\,,
\end{equation}
\begin{equation}
y\rightarrow-y\,.
\end{equation}
which leads to
\begin{equation}
\tilde{S}\left(  \alpha,\gamma\right)  =\int_{\left|  z\right|  <\rho}
d^{2}z\,\left|  \operatorname{Im}\,z\right|  ^{\alpha-1}\left|  z\right|
^{\gamma-\alpha-1}\frac{1}{4}\sum_{\varepsilon_{1}=\pm1}\sum_{\varepsilon
_{2}=\pm1}\tilde{M}\left(  \varepsilon_{1}\operatorname{Re}z,\varepsilon
_{2}\operatorname{Im}z\right)  \,.
\end{equation}
After this symmetrization only the even powers of the Taylor series
survive:\newline
\[
\frac{1}{4}\sum_{\varepsilon_{1}=\pm1}\sum_{\varepsilon_{2}=\pm1}\tilde
{M}\left(  \varepsilon_{1}x,\varepsilon_{2}y\right)  =\sum_{m=0}^{\infty}
\sum_{n=0}^{\infty}a_{2m,2n}x^{2m}y^{2n}\,.
\]

Thus\newline
\begin{equation}
\tilde{S}\left(  \alpha,\gamma\right)  =\int_{\left|  z\right|  <\rho}
d^{2}z\,\left|  \operatorname{Im}\,z\right|  ^{\alpha-1}\left|  z\right|
^{\gamma-\alpha-1}\sum_{m=0}^{\infty}\sum_{n=0}^{\infty}a_{2m,2n}\left(
\operatorname{Re}z\right)  ^{2m}\left(  \operatorname{Im}z\right)  ^{2n}\,.
\end{equation}
Next we can express
\begin{equation}
\left(  \operatorname{Re}z\right)  ^{2}=\left|  z^{2}\right|  -\left(
\operatorname{Im}z\right)  ^{2}
\end{equation}
and rearrange the power series in terms of new variables $\left|
z^{2}\right|  $ and $\left(  \operatorname{Im}z\right)  ^{2}$
\begin{equation}
\sum_{m=0}^{\infty}\sum_{n=0}^{\infty}a_{2m,2n}\left(  \operatorname{Re}
z\right)  ^{2m}\left(  \operatorname{Im}z\right)  ^{2n}=\sum_{k=0}^{\infty
}\sum_{l=0}^{\infty}d_{kl}\left(  \operatorname{Im}z\right)  ^{2k}\left|
z\right|  ^{2l}\,.
\end{equation}
Thus
\begin{equation}
\tilde{S}\left(  \alpha,\gamma\right)  =\sum_{k=0}\sum_{l=0}d_{kl}
\int_{\left|  z\right|  <\rho}d^{2}z\,\left|  \operatorname{Im}\,z\right|
^{\alpha-1+2k}\left|  z\right|  ^{\gamma-\alpha-1+2l}\,.
\end{equation}
This result was derived in convergence region (\ref{Re-alpha-positive-5}),
(\ref{Re-gamma-positive-5}).

The integrals can be computed using (\ref{I-alpha-gamma-def}),
(\ref{I-alpha-gamma-res})
\begin{equation}
\int_{\left|  z\right|  <\rho}d^{2}z\,\left|  \operatorname{Im}\,z\right|
^{\text{\thinspace}\alpha-1+2k}\left|  z\right|  ^{\gamma-\alpha
-1+2l}=\frac{2\sqrt{\pi}}{\gamma+2k+2l}\frac{\Gamma\left(  \frac{\alpha}
{2}+k\right)  }{\Gamma\left(  \frac{\alpha+1}{2}+k\right)  }\rho
^{\gamma+2\left(  k+l\right)  }\,.
\end{equation}
Hence
\begin{equation}
\tilde{S}\left(  \alpha,\gamma\right)  =\sum_{k=0}^{\infty}\sum_{l=0}^{\infty
}d_{kl}\frac{1}{\frac{\gamma}{2}+\left(  k+l\right)  }\frac{\Gamma\left(
\frac{\alpha}{2}+k\right)  }{\Gamma\left(  \frac{\alpha+1}{2}+k\right)  }
\rho^{\gamma+2\left(  k+l\right)  }\,. \label{int-M-series-res}
\end{equation}
Now we can start with analytical continuation in $\alpha$ and $\gamma$. There
are two explicit factors obstructing analytical continuation in $\alpha$ and
$\gamma$
\begin{equation}
\Gamma\left(  \frac{\alpha}{2}+k\right)  \frac{1}{\frac{\gamma}{2}+\left(
k+l\right)  }\,.
\end{equation}
These singularities can be absorbed in
\begin{equation}
\Gamma\left(  \frac{\alpha}{2}\right)  \Gamma\left(  \frac{\gamma}{2}\right)
\,.
\end{equation}
In other words, we can rewrite (\ref{int-M-series-res}) in the form
\begin{equation}
\tilde{S}\left(  \alpha,\gamma\right)  =\Gamma\left(  \frac{\alpha}{2}\right)
\Gamma\left(  \frac{\gamma}{2}\right)  \tilde{H}\left(  \alpha,\gamma\right)
\label{Pi-tilde-H-tilde}
\end{equation}
where
\begin{equation}
\tilde{H}\left(  \alpha,\gamma\right)  =\sum_{k=0}^{\infty}\sum_{l=0}^{\infty
}d_{kl}\frac{1}{\Gamma\left(  \frac{\gamma}{2}\right)  \left[  \frac{\gamma
}{2}+\left(  k+l\right)  \right]  }\frac{\Gamma\left(  \frac{\alpha}
{2}+k\right)  }{\Gamma\left(  \frac{\alpha}{2}\right)  }\frac{1}{\Gamma\left(
\frac{\alpha+1}{2}+k\right)  }\rho^{\gamma+2\left(  k+l\right)  }\,.
\label{H-tilde-series}
\end{equation}
Here all factors
\begin{equation}
\frac{1}{\Gamma\left(  \frac{\gamma}{2}\right)  \left[  \frac{\gamma}
{2}+\left(  k+l\right)  \right]  }\,,
\end{equation}
\begin{equation}
\frac{\Gamma\left(  \frac{\alpha}{2}+k\right)  }{\Gamma\left(  \frac{\alpha
}{2}\right)  }=\text{Polynomial(}\alpha\text{)\thinspace,}
\end{equation}
\begin{equation}
\frac{1}{\Gamma\left(  \frac{\alpha+1}{2}+k\right)  }
\end{equation}
are entire functions of $\alpha,\gamma_{j}$ so that the full product
\begin{equation}
\frac{1}{\Gamma\left(  \frac{\gamma}{2}\right)  \left[  \frac{\gamma}
{2}+\left(  k+l\right)  \right]  }\frac{\Gamma\left(  \frac{\alpha}
{2}+k\right)  }{\Gamma\left(  \frac{\alpha}{2}\right)  }\frac{1}{\Gamma\left(
\frac{\alpha+1}{2}+k\right)  }\rho^{\gamma+2\left(  k+l\right)  }
\end{equation}
is also an entire function of $\alpha,\left\{  \gamma_{k}\right\}  $.
Therefore in eq. (\ref{H-tilde-series}) we have a power series in variable
$\rho$ (which can be chosen arbitrarily small) of entire functions in
$\alpha,\gamma$ so that $\tilde{H}\left(  \alpha,\gamma\right)  $ is also an
entire function.

Thus analytical continuation of function $\tilde{S}\left(  \alpha
,\gamma\right)  $ defined by integral(\ref{Pi-tilde-integral}) in its
convergence region leads to a meromorphic function with the pole structure
described by representation (\ref{Pi-tilde-H-tilde}).

Returning to our original problem of the analytical continuation of function
$S_{j}\left(  \alpha,\left\{  \gamma_{k}\right\}  _{k=1}^{n}\right)  $
(\ref{int-D-z-k-r-k-M}) in $\alpha$ and in $\gamma_{j}$ we conclude that this
function has the pole structure
\begin{equation}
S_{j}\left(  \alpha,\left\{  \gamma_{k}\right\}  _{k=1}^{n}\right)
=\Gamma\left(  \frac{\alpha}{2}\right)  \Gamma\left(  \frac{\gamma_{j}}
{2}\right)  H_{j}\left(  \alpha,\left\{  \gamma_{k}\right\}  _{k=1}
^{n}\right)  \label{Pi-j-circle-meromorphic-structure}
\end{equation}
where $H_{j}\left(  \alpha,\left\{  \gamma_{k}\right\}  _{k=1}^{n}\right)  $
is an entire function. Thus we have derived representation
(\ref{Pi-b-Gamma-H-decomposition}) for $S_{j}$ ($1\leq j\leq n$) and
corresponding expression (\ref{Gamma-j-intermediate}) for $\Gamma_{j}$.

\subsection{Calculation of $\Gamma_{n+1}$}

We want to derive meromorphic structure (\ref{Pi-b-Gamma-H-decomposition}),
(\ref{Gamma-j-intermediate}) of $S_{n+1}$. We have
\begin{align}
S_{n+1}\left(  \alpha,\left\{  \gamma_{k}\right\}  _{k=1}^{n}\right)   &
=\int_{D_{n+1}}d^{2}z\,\left|  \operatorname{Im}\,z\right|  ^{\alpha-1}\left[
\prod_{k=1}^{n}\left|  z-z_{k}\right|  ^{\gamma_{k}-\alpha-1}\right]  P\left(
z,z^{\ast}\right) \nonumber\\
&  =\int_{\left|  z\right|  >R}d^{2}z\,\left|  \operatorname{Im}\,z\right|
^{\alpha-1}\left[  \prod_{k=1}^{n}\left|  z-z_{k}\right|  ^{\gamma_{k}
-\alpha-1}\right]  P\left(  z,z^{\ast}\right)  \,.
\end{align}
We proceed similarly to our work in section
\ref{Convergence-D-n-plus-1-section} where we studied convergence condition
for region $D_{n+1,+}$. First we perform inversion (\ref{z-inversion-minus})
and derive by analogy with (\ref{int-D-n-plus-1-1}), (\ref{int-D-n-plus-1-2})
\begin{align}
&  S_{n+1}\left(  \alpha,\left\{  \gamma_{k}\right\}  _{k=1}^{n}\right)
\nonumber\\
&  =\left[  \prod_{k=1}^{n}\left|  z_{k}\right|  ^{\gamma_{k}-\alpha
-1}\right]  \int_{\left|  z\right|  <1/R}d^{2}z\,|z|^{\left(  n-2\right)
\left(  \alpha+1\right)  -\sum_{k=1}^{n}\gamma_{k}}\nonumber\\
&  \times\left|  \operatorname{Im}\,z\right|  ^{\alpha-1}\left[  \prod
_{k=1}^{n}\left|  z+z_{k}^{-1}\right|  ^{\left(  \gamma_{k}-\alpha-1\right)
}\right]  P\left(  -1/z,-1/z^{\ast}\right)  \,. \label{S-n-plus-1-calc-1}
\end{align}
Next we use definition (\ref{M-P-min-def}) of integer parameter $M_{P}$
\begin{equation}
R(z,z^{\ast})\equiv(zz^{\ast})^{M_{P}}P\left(  -1/z,-1/z^{\ast}\right)
=\text{Polynomial}(z,z^{\ast})\,.
\end{equation}
Hence
\begin{align}
&  S_{n+1}\left(  \alpha,\left\{  \gamma_{k}\right\}  _{k=1}^{n}\right)
\nonumber\\
&  =\left[  \prod_{k=1}^{n}\left|  z_{k}\right|  ^{\gamma_{k}-\alpha
-1}\right]  \int_{\left|  z\right|  <1/R}d^{2}z\,|z|^{-2M_{P}+(n-1)\left(
\alpha+1\right)  -\sum_{k=1}^{n}\gamma_{k}}\nonumber\\
&  \times\left|  \operatorname{Im}\,z\right|  ^{\alpha-1}\left[  \prod
_{k=1}^{n}\left|  z+z_{k}^{-1}\right|  ^{\left(  \gamma_{k}-\alpha-1\right)
}\right]  R(z,z^{\ast})\,. \label{Pi-n-plus-1-calc-1}
\end{align}
This integral has the same structure as integral (\ref{int-D-z-k-r-k-M}) but
with replacement
\begin{equation}
\gamma_{j}\rightarrow-2M_{P}+(n-1)\left(  1+\alpha\right)  -\sum_{k=1}
^{n}\gamma_{k}\,. \label{gamma-j-replacement-fo-Pi-n-plus-1}
\end{equation}
Therefore analytical continuation of new integral (\ref{Pi-n-plus-1-calc-1})
is described by representation (\ref{Pi-j-circle-meromorphic-structure}) with
replacement (\ref{gamma-j-replacement-fo-Pi-n-plus-1})
\begin{align}
&  S_{n+1}\left(  \alpha,\left\{  \gamma_{k}\right\}  _{k=1}^{n}\right)
\nonumber\\
&  =\Gamma\left(  \frac{\alpha}{2}\right)  \Gamma\left(  \frac{1}{2}\left(
-2M_{P}+(n-1)\left(  1+\alpha\right)  -\sum_{k=1}^{n}\gamma_{k}\right)
\right)  H_{n+1}\left(  \alpha,\left\{  \gamma_{k}\right\}  _{k=1}^{n}\right)
\,.
\end{align}
Here $H_{n+1}\left(  \alpha,\left\{  \gamma_{k}\right\}  _{k=1}^{n}\right)  $
is an entire function. Thus we have derived representation
(\ref{Pi-b-Gamma-H-decomposition}) for $S_{n+1}$ with $\Gamma_{n+1}$ given by
(\ref{Gamma-n-plus-1}).

\section{Analytical renormalization produces finite results}

\label{Analytical-renormalization-finite-section}

\setcounter{equation}{0} 

\subsection{Is analytical continuation to the physical point finite?}

Now we want to check that the renormalization procedure suggested in ref.
\cite{Pobylitsa-2019} for the calculation of two-loop EST correction
$f_{2}(C)$ (\ref{f2-via-I-I2-ren}) really renormalizes
ultraviolet\ divergences and provides a finite expression for $f_{2}(C)$. Eq.
(\ref{f2-via-I-I2-ren}) expresses $f_{2}(C)$ via functions $I_{1}^{\text{ren}
}$ and $I_{2}^{\text{ren}}$ so that we must test that our final expressions
for $I_{1}^{\text{ren}}$ (\ref{I-1-ren-via-Pi}) and for $I_{2}^{\text{ren}}$
(\ref{I-2-ren-via-Pi}) are finite. In other words we must test that functions
$\Pi_{P}^{(n)}$ in eqs. (\ref{I-1-ren-via-Pi}) and (\ref{I-2-ren-via-Pi}) are
finite at points appearing in (\ref{I-1-ren-via-Pi}), (\ref{I-2-ren-via-Pi}).

In other words, we must check that arguments of functions $\Pi_{P}^{(n)}$ in
eqs. (\ref{I-1-ren-via-Pi}) and (\ref{I-2-ren-via-Pi}) do not overlap with
poles of $\Gamma$ functions that describe pole structure
(\ref{K-rough-pole-structure}) of $\Pi_{P}^{(n)}$.

\subsection{Case of $I_{1}^{\text{ren}}$}

Function $\Pi_{1}^{(n_{\text{v}}-1)}$ appearing in expression
(\ref{I-1-ren-via-Pi}) for $I_{1}^{\text{ren}}$ has the following pole
representation (\ref{K-rough-pole-structure})
\begin{align}
&  \Pi_{1}^{(n_{\text{v}}-1)}\left(  -3,\left\{  2\beta_{k}-2\right\}
_{k=1}^{n_{\text{v}}-1},\left\{  z_{k}\right\}  _{k=1}^{n_{\text{v}}-1}\right)
\nonumber\\
&  =\Gamma\left(  -\frac{3}{2}\right)  \Gamma\left(  \frac{1}{2}\left(
-2M_{1}-2\left(  n_{\text{v}}-2\right)  -\sum_{k=1}^{n_{\text{v}}-1}\left(
2\beta_{k}-2\right)  \right)  \right)  \prod_{j=1}^{n_{\text{v}}-1}
\Gamma\left(  \beta_{k}-1\right) \nonumber\\
&  \times H_{1}^{\left(  n_{\text{v}}-1\right)  }\left(  -3,\left\{
2\beta_{k}-2\right\}  _{k=1}^{n_{\text{v}}-1},\left\{  z_{k}\right\}
_{k=1}^{n_{\text{v}}-1}\right)  \,. \label{Pi-for-I1}
\end{align}
Here $M_{1}$ is $M_{P}$ parameter (\ref{M-P-min-def}) for the trivial
polynomial $P=1$ so that
\begin{equation}
M_{1}=0\,.
\end{equation}
Now we can simplify
\begin{equation}
\Gamma\left(  \frac{1}{2}\left(  -2M_{1}-2\left(  n_{\text{v}}-2\right)
-\sum_{k=1}^{n_{\text{v}}-1}\left(  2\beta_{k}-2\right)  \right)  \right)
=\Gamma\left(  1-\sum_{k=1}^{n_{\text{v}}-1}\beta_{k}\right)  \,.
\end{equation}
Thus on the RHS of (\ref{Pi-for-I1}) we have the product of potentially
dangerous $\Gamma$ functions
\begin{equation}
\Gamma\left(  1-\sum_{k=1}^{n_{\text{v}}-1}\beta_{k}\right)  \prod
_{j=1}^{n_{\text{v}}-1}\Gamma\left(  \beta_{k}-1\right)  \,.
\label{Gamma-product-for-I1}
\end{equation}
According to (\ref{beta-n-v})
\begin{equation}
1-\sum_{k=1}^{n_{\text{v}}-1}\beta_{k}=\beta_{n_{\text{v}}}-1
\end{equation}
so that
\begin{equation}
\Gamma\left(  1-\sum_{k=1}^{n_{\text{v}}-1}\beta_{k}\right)  \prod
_{j=1}^{n_{\text{v}}-1}\Gamma\left(  \beta_{j}-1\right)  =\prod_{j=1}
^{n_{\text{v}}}\Gamma\left(  \beta_{j}-1\right)  \,.
\end{equation}

Parameters $\beta_{k}$ ($1\leq k\leq n_{\text{v}}$) are related to inside
angles of the polygon $\theta_{k}$ (\ref{theta-k-beta-k}) and belong to the
range (\ref{beta-k-range}) so that
\begin{align}
-1  &  <\beta_{k}-1<1\,,\\
\beta_{k}-1  &  \neq0\,.
\end{align}
Therefore for all $k$ in the range $1\leq k\leq n_{\text{v}}$ we have
\begin{equation}
\Gamma\left(  \beta_{k}-1\right)  =\text{finite\thinspace.}
\end{equation}
Thus all $\Gamma$ functions appearing in eq. (\ref{Pi-for-I1}) are regular.

\subsection{Case of $I_{2}^{\text{ren}}$}

\label{I2-finite-section}

Function $\Pi_{P_{2}}^{(n_{\text{v}}-1)}$ appearing in expression
(\ref{I-2-ren-via-Pi}) for $I_{2}^{\text{ren}}$ has the following pole
representation (\ref{K-rough-pole-structure})
\begin{align}
&  \Pi_{P_{2}}^{(n_{\text{v}}-1)}\left(  1,\left\{  2\beta_{k}-2\right\}
_{k=1}^{n_{\text{v}}-1},\left\{  z_{k}\right\}  _{k=1}^{n_{\text{v}}-1}\right)
\nonumber\\
&  =\Gamma\left(  \frac{1}{2}\right)  \Gamma\left(  \frac{1}{2}\left(
-2M_{P_{2}}+2\left(  n_{\text{v}}-2\right)  -\sum_{k=1}^{n_{\text{v}}
-1}\left(  2\beta_{k}-2\right)  \right)  \right)  \prod_{j=1}^{n_{\text{v}}
-1}\Gamma\left(  \beta_{j}-1\right) \nonumber\\
&  \times H_{P_{2}}^{\left(  n_{\text{v}}-1\right)  }\left(  1,\left\{
2\beta_{k}-2\right\}  _{k=1}^{n_{\text{v}}-1},\left\{  z_{k}\right\}
_{k=1}^{n_{\text{v}}-1}\right)  \,\,.
\end{align}
Polynomial $P_{2}$ is given by eqs. (\ref{P2-via-T2}), (\ref{T2-def}).

Using expression (\ref{M-P2-res}) for $M_{P_{2}}$, we find
\begin{equation}
\Gamma\left(  \frac{1}{2}\left(  -2M_{P_{2}}+2\left(  n_{\text{v}}-2\right)
-\sum_{k=1}^{n_{\text{v}}-1}\left(  2\beta_{k}-2\right)  \right)  \right)
=\Gamma\left(  1-\sum_{k=1}^{n_{\text{v}}-1}\beta_{k}\right)  \,.
\end{equation}
Thus we arrive at the same product of $\Gamma$ functions as in the case of
(\ref{Gamma-product-for-I1}) which is regular.

\section{Analytical renormalization and SC reparametrization}

\label{renormalization-invariance-inversion-section}

\setcounter{equation}{0} 

\subsection{Problem}

In section \ref{different-SC-parametrizations-problem} we mentioned a problem:
it is \emph{not obvious} that starting from \emph{different}
SC\ parametrizations of a \emph{fixed} polygon, one arrives at the same values
for $I_{m}^{\text{ren}}$ after our analytical renormalization formulated in
terms of SC parameters. Indeed our renormalization procedure for
$I_{m}^{\text{ren}}$ is formulated in terms of SC parameters and not directly
in terms of the geometry of the polygon.

Remember that our analytical renormalization is formulated in terms of a
special class of SC$_{\infty}$\ mappings defined by eq. (\ref{SC-diff-eq}) and
assuming that one SC\ vertex is kept at infinity. In section \ref{SC-section}
it was suggested to refer to this type of SC\ transformation as SC$_{\infty}$ mapping.

The following statement shows that our procedure of analytical renormalization
is independent of the choice of the SC$_{\infty}$ mapping for a given polygon:

\textbf{Statement 4.} If two SC$_{\infty}$ mappings $\zeta^{(a)}\left(
z\right)  $ (\ref{SC-diff-eq}) with SC$_{\infty}$\ parameters
\begin{equation}
\left\{  \tilde{A}^{(a)},\left\{  \beta_{k}^{(a)}\right\}  _{k=1}
^{n_{\text{v}}-1},\left\{  z_{k}^{(a)}\right\}  _{k=1}^{n_{\text{v}}
-1}\right\}  \quad(a=1,2) \label{SC-parameters-5}
\end{equation}
map the upper complex semiplane to the same polygon then for quantities
$I_{m}^{\text{ren}}$ given by (\ref{I-1-ren-via-Pi}), (\ref{I-2-ren-via-Pi})
we have
\begin{equation}
I_{m}^{\text{ren}}\left(  \tilde{A}^{(1)},\left\{  \beta_{k}^{(1)}\right\}
_{k=1}^{n_{\text{v}}-1},\left\{  z_{k}^{(1)}\right\}  _{k=1}^{n_{\text{v}}
-1}\right)  =I_{m}^{\text{ren}}\left(  \tilde{A}^{(2)},\left\{  \beta
_{k}^{(2)}\right\}  _{k=1}^{n_{\text{v}}-1},\left\{  z_{k}^{(2)}\right\}
_{k=1}^{n_{\text{v}}-1}\right)  \,. \label{I-ren-invariance}
\end{equation}

\subsection{Permutation symmetry}

\label{permutation-symmetry-section}

In our previous work we assumed that real points $\left\{  z_{k}\right\}
_{k=1}^{n}$ are monotonically ordered (\ref{z-k-monotonic-order}). This
assumption simplifies intermediate technical calculations when one is
interested in such properties of functions $\Pi_{P}^{(n)}$ like convergence
region of their integral representations or in the analytical continuation of
$\Pi_{P}^{(n)}$ in $\alpha,\gamma_{k}$ at fixed $z_{k}$.

Functions $\Pi_{P}^{(n)}$ have another important property: symmetry under
permutations of their arguments including permutations of $z_{k}$. This
symmetry property will simplify our following work. Therefore in part of our
work devoted to the derivation of relation (\ref{I-ren-invariance}) we do not
impose constraint (\ref{z-k-monotonic-order}).

Let $R$ be an arbitrary permutation of indices $k=1,2,\ldots,n$. Functions
$\Pi_{P}^{(n)}$ obey relation
\begin{equation}
\Pi_{P}^{(n)}\left(  \alpha,\left\{  \gamma_{k}\right\}  _{k=1}^{n},\left\{
z_{k}\right\}  _{k=1}^{n}\right)  =\Pi_{P}^{(n)}\left(  \alpha,\left\{
\gamma_{R\left(  k\right)  }\right\}  _{k=1}^{n},\left\{  z_{R\left(
k\right)  }\right\}  _{k=1}^{n}\right)  \label{P-P-n-permutation-symmetry}
\end{equation}
The method of derivation is standard: one first proves this relation in the
region of parameters $\alpha,\left\{  \gamma_{k}\right\}  _{k=1}^{n}$ where
the integrals (\ref{Pi-P-n-def}) representing LHS\ and RHS\ are convergent.
After that one can continue identity (\ref{P-P-n-permutation-symmetry})
analytically in $\alpha,\left\{  \gamma_{k}\right\}  _{k=1}^{n}$ at fixed
$z_{k}$. This program meets no problems because

\begin{itemize}
\item the set of convergence constraints (\ref{convergence-condition-1}) --
(\ref{convergence-condition-3}) on $\alpha,\left\{  \gamma_{k}\right\}
_{k=1}^{n}$ is symmetric with respect to permutations of $\gamma_{k}$,

\item pole structure of $\Pi_{P}^{(n)}$ (\ref{K-rough-pole-structure}) is also
symmetric with respect to permutations of $\gamma_{k}$.
\end{itemize}

One should distinguish

\begin{itemize}
\item properties of functions $\Pi_{P}^{(n)}$ with respect to permutations,

\item application of these properties of $\Pi_{P}^{(n)}$ to the calculation of
$I_{m}^{\text{ren}}$,

\item resulting symmetry properties of $I_{m}^{\text{ren}}\left(
\tilde{A},\left\{  \beta_{k}\right\}  _{k=1}^{n_{\text{v}}-1},\left\{
z_{k}\right\}  _{k=1}^{n_{\text{v}}-1}\right)  $.
\end{itemize}

In other words, we should not mix

\begin{itemize}
\item intrinsic symmetry properties of functions $\Pi_{P}^{(n)}$,

\item intrinsic symmetry properties of problems that are solved using
functions $\Pi_{P}^{(n)}$.
\end{itemize}

Our work consists of several stages:
\begin{equation}
C\rightarrow\zeta(z)\rightarrow\tilde{A},\left\{  \beta_{k}\right\}
_{k=1}^{n_{\text{v}}-1},\left\{  z_{k}\right\}  _{k=1}^{n_{\text{v}}
-1}\rightarrow I_{m}^{\text{ren}}\left(  \tilde{A},\left\{  \beta_{k}\right\}
_{k=1}^{n_{\text{v}}-1},\left\{  z_{k}\right\}  _{k=1}^{n_{\text{v}}
-1}\right)
\end{equation}

Permutations play an important role at each stage. For example, SC$_{\infty}$
mapping $\zeta(z)$ satisfies differential equation (\ref{SC-diff-eq}). The
RHS\ of this differential equation is a function of $\left\{  \beta_{k}
,z_{k}\right\}  _{k=1}^{n_{\text{v}}-1}$ which is symmetric under permutations
of $k$.

Next, in relation (\ref{P-P-n-permutation-symmetry}) $P\left(  z,z^{\ast
}\right)  $ stands for an arbitrary polynomial. We have the same polynomial on
the LHS and on the RHS of (\ref{P-P-n-permutation-symmetry}). When it comes to
the calculation of $I_{2}^{\text{ren}}$ based on eq. (\ref{I-2-ren-via-Pi})
then we use polynomial $P_{2}$ (\ref{P2-T2-expanded}) which has property
\begin{equation}
P_{2}\left(  z,z^{\ast};\left\{  \beta_{R\left(  k\right)  }\right\}
_{k=1}^{n_{\text{v}}-1},\left\{  z_{R\left(  k\right)  }\right\}
_{k=1}^{n_{\text{v}}-1}\right)  =P_{2}\left(  z,z^{\ast};\left\{  \beta
_{k}\right\}  _{k=1}^{n_{\text{v}}-1},\left\{  z_{k}\right\}  _{k=1}
^{n_{\text{v}}-1}\right)  \label{P2-symmetric}
\end{equation}
that follows from eqs. (\ref{T2-def}), (\ref{P2-T2-expanded}).

If one combines

\begin{itemize}
\item permutation symmetry property (\ref{P-P-n-permutation-symmetry}) of
function $\Pi_{P}^{(n)}$,

\item \emph{other} permutation symmetry properties like symmetry of the RHS of
differential equation (\ref{SC-diff-eq}) and symmetry of polynomial $P_{2}$
(\ref{P2-symmetric})
\end{itemize}

then one easily derives from eqs. (\ref{I-1-ren-via-Pi}),
(\ref{I-2-ren-via-Pi})
\begin{equation}
I_{m}^{\text{ren}}\left(  \tilde{A},\left\{  \beta_{k}\right\}  _{k=1}
^{n_{\text{v}}-1},\left\{  z_{k}\right\}  _{k=1}^{n_{\text{v}}-1}\right)
=I_{m}^{\text{ren}}\left(  \tilde{A},\left\{  \beta_{R\left(  k\right)
}\right\}  _{k=1}^{n_{\text{v}}-1},\left\{  z_{R\left(  k\right)  }\right\}
_{k=1}^{n_{\text{v}}-1}\right)  \,.
\end{equation}
for any permutation $R$.

\subsection{SC$_{\infty}$\ reparametrization step by step}

\label{strategy-SC-invariance-section}

A careful proof of Statement 4 requires some work.

For a give polygon there is an infinite set of SC$_{\infty}$ mappings of the
upper complex semiplane to this polygon. Within this infinite set one define
the concept of SC$_{\infty}$ mappings differing by an \emph{elementary
change}. When can define these elementary changes so that any two SC$_{\infty
}$ mappings $\zeta^{(1)}$ and $\zeta^{(2)}$ representing the same polygon can
connected by a finite chain of sequential elementary changes:
\begin{equation}
\zeta^{(1)}\rightarrow\zeta^{(I)}\rightarrow\zeta^{(II)}\rightarrow
...\rightarrow\zeta^{(2)}\,. \label{zeta-SC-infnity-chain}
\end{equation}
Therefore it is sufficient to prove identity (\ref{I-ren-invariance}) only for
the case when mappings $\zeta^{(1)}$ and $\zeta^{(2)}$ differ by an elementary change.

A careful definition of \emph{elementary changes} involves several subtleties
but the general idea is rather simple: elementary changes include

1) translation in the $z^{\prime}=z+b$,

2) dilation $z^{\prime}=az$,

3) inversion $z^{\prime}=-1/z$.

\subsection{Linear fractional transformations}

\label{linear-fractional-section}

As already mentioned, linear fractional transformations
(\ref{z-prime-z-linear-fractional})
\begin{equation}
V\left(  z\right)  =\frac{az+b}{cz+d}\quad(a,b,c,d\,\in\,\mathbb{R})
\label{G-z-linfrac}
\end{equation}
map upper complex semiplane $\mathbb{C}_{+}$ to itself. Therefore two
\emph{general} SC (not necessarily SC$_{\infty}$) mappings $\zeta^{(1)}$ and
$\zeta^{(2)}$ map $z$ semiplane $\mathbb{C}_{+}$ to the same polygon if and
only if they are connected by a linear fractional transformation $V$
(\ref{z-prime-z-linear-fractional})
\begin{equation}
\zeta^{(2)}(z)=\zeta^{(1)}\left(  V(z)\right)
\end{equation}
or in short
\begin{equation}
\zeta^{(2)}=\zeta^{(1)}\circ V\,. \label{zeta-2-1-G}
\end{equation}

In principle, this statement also holds for SC$_{\infty}$ mappings
$\zeta^{(1)}$ and $\zeta^{(2)}$ but there is one subtlety: if

\begin{itemize}
\item $\zeta^{(1)}$ is SC$_{\infty}$ mapping,

\item $V$ is linear fractional mapping (\ref{z-prime-z-linear-fractional})
\end{itemize}

then $\zeta^{(1)}\circ V$ is SC mapping but not necessarily SC$_{\infty}$ mapping.

Therefore the precise statement is

\textbf{Statement 5.}

If two SC$_{\infty}$ transformations $\zeta^{(1)}$ and $\zeta^{(2)}$ map
$\mathbb{C}_{+}$ to the same polygon then there exists such linear fractional
transformation that relation (\ref{zeta-2-1-G}) holds.

One can be also interested in a different question: starting from a give
SC$_{\infty}$ mapping which linear fractional transformations $V$ can be used
for $\zeta^{(2)}$ to be also an SC$_{\infty}$ mapping. The answer is obvious:
SC$_{\infty}$\ mappings have one SC\ vertex at infinity. Linear fractional
transformation $V$ must respect this property.

Hence we have proved

\textbf{Statement 6.}

Let $\zeta^{(1)}$ be SC$_{\infty}$ mapping with $n_{\text{v}}-1$ finite
vertices $\left\{  z_{k}^{(1)}\right\}  _{k=1}^{n_{\text{v}}-1}$ and one
hidden vertex at infinity. Then $\zeta^{(1)}\circ V$ will be SC$_{\infty}$
mapping for those and only for those linear fractional mappings $V$ which have
the property
\begin{equation}
V(\infty)=\infty
\end{equation}
or
\begin{equation}
V(\infty)=z_{k}\,\,\text{for some }k\,\text{in the range }1\leq k\leq
n_{\text{v}}-1\,.
\end{equation}

According to (\ref{G-z-linfrac})
\begin{equation}
V(\infty)=\frac{a}{c}\,.
\end{equation}
Thus we have proved

\textbf{Statement 7.}

Let $\zeta^{(1)}$ be SC$_{\infty}$ mapping with $n_{\text{v}}-1$ finite
vertices $\left\{  z_{k}^{(1)}\right\}  _{k=1}^{n_{\text{v}}-1}$. Then
$\zeta^{(1)}\circ V$ will be SC$_{\infty}$ mapping only for the following set
of linear fractional transformations $V(z)$

\begin{itemize}
\item case 1
\end{itemize}

\begin{equation}
V(z)=az+b\,\quad(a>0,\,b\,\in\,\mathbb{R}) \label{statement-8-case-1}
\end{equation}

\begin{itemize}
\item case 2
\[
\text{for some }k\,\text{in the range }1\leq k\leq n_{\text{v}}-1:
\]
\begin{equation}
V\left(  z\right)  =z_{k}-\frac{a}{z+b}\quad(a>0,\,b\,\in\,\mathbb{R})\,.
\label{statement-8-case-2}
\end{equation}
\end{itemize}

\subsection{Elementary changes of SC$_{\infty}$ mappings}

In section \ref{strategy-SC-invariance-section} it was suggested to define
elementary changes of SC$_{\infty}$\ mappings in such a way that any two
SC$_{\infty}$ mappings $\zeta^{(1)}$ and $\zeta^{(2)}$ of $\mathbb{C}_{+}$ to
the same polygon can be connected by a chain of elementary changes
(\ref{zeta-SC-infnity-chain}) so that at each step of this chain we have
SC$_{\infty}$ mappings $\zeta^{(I)},\zeta^{(II)},\ldots$ to the same polygon.

Statement 7 suggest the following definition of \emph{elementary changes} from
SC$_{\infty}$ mapping $\zeta^{(1)}(z)$ to SC$_{\infty}$ mapping $\zeta^{(2)}(z)$

1) translation
\begin{equation}
\zeta^{(2)}(z)=\zeta^{(1)}\left(  z+b\right)  \quad\left(  b\in\mathbb{R}
\right)  \,, \label{dilatation-elementary}
\end{equation}

2) dilation
\begin{equation}
\zeta^{(2)}(z)=\zeta^{(1)}\left(  az\right)  \quad\left(  a>0\right)  \,,
\label{translation-elementary}
\end{equation}

3) inversion
\begin{equation}
\zeta^{(2)}(z)=\zeta^{(1)}(-1/z)\,. \label{inversion-elementary}
\end{equation}

Making chain (\ref{zeta-SC-infnity-chain}) of translation
(\ref{translation-elementary}) of dilation (\ref{dilatation-elementary}), one
can generate case (\ref{statement-8-case-1}) of Statement 7.

Inversion (\ref{inversion-elementary}) corresponds to the special case
\begin{align}
a  &  =1\,,\\
b  &  =0\,,\\
z_{k}  &  =0\quad\text{in }\zeta^{(1)}(z)\, \label{z-k-0-for-F1}
\end{align}
of case (\ref{statement-8-case-2}) in Statement 7. Making a chain containing
inversion, translations and dilations, one can generate the general
\emph{non-elementary} modification of SC$_{\infty}$ mapping.

Note that in cases of translations (\ref{translation-elementary}) and dilation
(\ref{dilatation-elementary}) it is sufficient to assume that only
$\zeta^{(1)}(z)$ is SC$_{\infty}$ mapping. Mappings $\zeta^{(2)}$ defined by
relations (\ref{translation-elementary}), (\ref{dilatation-elementary}) will
be automatically SC$_{\infty}$ mappings.

In case of inversion (\ref{inversion-elementary}) the situation is different.
Staring from SC$_{\infty}$ mappings $\zeta^{(1)}(z)$ and applying
(\ref{inversion-elementary}), we may arrive at non-SC$_{\infty}$ mapping
$\zeta^{(2)}(z)$. Only if condition (\ref{z-k-0-for-F1}) holds for mapping
$\zeta^{(1)}$ (i.e. $\zeta^{(1)}$ is defined by eq. (\ref{SC-diff-eq}) with
$z_{k}=0$ for some $k$) then $\zeta^{(2)}(z)$ will be SC$_{\infty}$ mapping.

Anyway the set of eqs. (\ref{translation-elementary}),
(\ref{dilatation-elementary}) and (\ref{inversion-elementary}) provides a
complete set of elementary changes and the problem of the proof of
(\ref{I-ren-invariance}) reduces to the case of SC$_{\infty}$ mappings
connected by elementary changes.

Let us formulate precisely what we must prove.

\textbf{Statement 8.}

If

1) two SC$_{\infty}$ mappings $\zeta^{(a)}(z)$ ($a=1,2$) are generated by eq.
(\ref{SC-diff-eq}) with parameters
\begin{equation}
\left\{  \tilde{A}^{(a)},\left\{  \beta_{k}^{(a)}\right\}  _{k=1}
^{n_{\text{v}}-1},\left\{  z_{k}^{(a)}\right\}  _{k=1}^{n_{\text{v}}
-1}\right\}  \quad(a=1,2)\,,
\end{equation}

2) these two SC$_{\infty}$\ mappings $\zeta^{(a)}(z)$ obey one of relations
(\ref{translation-elementary}), (\ref{dilatation-elementary}) and
(\ref{inversion-elementary})

then
\begin{equation}
I_{m}^{\text{ren}}\left(  \tilde{A}^{(1)},\left\{  \beta_{k}^{(1)}\right\}
_{k=1}^{n_{\text{v}}-1},\left\{  z_{k}^{(1)}\right\}  _{k=1}^{n_{\text{v}}
-1}\right)  =I_{m}^{\text{ren}}\left(  \tilde{A}^{(2)},\left\{  \beta
_{k}^{(2)}\right\}  _{k=1}^{n_{\text{v}}-1},\left\{  z_{k}^{(2)}\right\}
_{k=1}^{n_{\text{v}}-1}\right)  \,.
\end{equation}

If we prove Statement 8 then it will be the end of the work.

In order to prove Statement 8, we must consider three different cases:
(\ref{translation-elementary}), (\ref{dilatation-elementary}) and
(\ref{inversion-elementary}). In the case of translations Statement 8 is
trivial. The case of dilation requires some simple work. Only in the case of
inversion (\ref{inversion-elementary}) the proof of Statement 8 requires a
certain effort. Below we concentrate on this problematic case.

\subsection{Invariance of $I_{m}^{\text{ren}}$ with respect to inversion}

\subsubsection{Main result}

\label{invariance-main-result-section}

Thus we must prove Statement 8 for the case of inversion
(\ref{inversion-elementary}), e.g. we must prove

\textbf{Statement 9.}

If

1) two SC$_{\infty}$ mappings $\zeta^{(a)}(z)$ ($a=1,2$) are generated by eq.
(\ref{SC-diff-eq}) with parameters
\begin{equation}
\left\{  \tilde{A}^{(a)},\left\{  \beta_{k}^{(a)}\right\}  _{k=1}
^{n_{\text{v}}-1},\left\{  z_{k}^{(a)}\right\}  _{k=1}^{n_{\text{v}}
-1}\right\}  \quad(a=1,2)\,,
\end{equation}

2) these two SC$_{\infty}$\ mappings $\zeta^{(a)}(z)$ obey relation
\begin{equation}
\zeta^{(2)}(z)=\zeta^{(1)}(-1/z) \label{inversion-elementary-2}
\end{equation}
then
\begin{equation}
I_{m}^{\text{ren}}\left(  \tilde{A}^{(1)},\left\{  \beta_{k}^{(1)}\right\}
_{k=1}^{n_{\text{v}}-1},\left\{  z_{k}^{(1)}\right\}  _{k=1}^{n_{\text{v}}
-1}\right)  =I_{m}^{\text{ren}}\left(  \tilde{A}^{(2)},\left\{  \beta
_{k}^{(2)}\right\}  _{k=1}^{n_{\text{v}}-1},\left\{  z_{k}^{(2)}\right\}
_{k=1}^{n_{\text{v}}-1}\right)  \,. \label{I-m-ren-invariant-inversion-2}
\end{equation}

Statement 9 assumes that both $\zeta^{(1)}(z)$ and $\zeta^{(2)}(z)$ are
SC$_{\infty}$ mappings. As already discussed, combining this with
(\ref{inversion-elementary-2}), one arrives at (\ref{z-k-0-for-F1}), i.e.
\begin{equation}
z_{k_{1}}^{(1)}=0\quad\text{for some }k_{1}\text{ in the range }1\leq
k_{1}\leq n_{\text{v}}-1\,. \label{z-k-1-zero}
\end{equation}
Since relation (\ref{inversion-elementary-2}) is symmetric under exchange
$\zeta^{(1)}\leftrightarrow\zeta^{(2)}$, we also have
\begin{equation}
z_{k_{2}}^{(2)}=0\quad\text{for some }k_{2}\text{ in the range }1\leq
k_{2}\leq n_{\text{v}}-1\,. \label{z-k-2-zero}
\end{equation}

According to results of section \ref{permutation-symmetry-section} we are free
to change the numeration of SC$_{\infty}$\ parameters $\left\{  z_{k}
^{(1)},\beta_{k}^{(1)}\right\}  _{k=1}^{n_{\text{v}}-1}$ as we like so that we
can simplify property (\ref{z-k-1-zero}) $z_{k_{1}}^{(1)}=0$ to $z_{1}
^{(1)}=0$. We can apply the same argument to the set of parameters and reduce
property (\ref{z-k-2-zero}) $z_{k_{2}}^{(2)}=0$ to $z_{1}^{(2)}=0$.

Thus we can proceed assuming that
\begin{equation}
z_{1}^{(1)}=z_{1}^{(2)}=0\,. \label{z-1-2-1-zero}
\end{equation}
Remember that $\zeta^{(1)}$ and $\zeta^{(2)}$ map $\mathbb{C}_{+}$ to the same
polygon. We can again use use the freedom of numeration of parameters
$\left\{  z_{k}^{(a)},\beta_{k}^{(a)}\right\}  _{k=1}^{n_{\text{v}}-1}$ so
that
\begin{equation}
\zeta^{(1)}\left(  z_{k}^{(1)}\right)  =\zeta^{(2)}\left(  z_{k}^{(2)}\right)
\quad\text{if }\,2\leq k\leq n_{\text{v}}-1\,. \label{vertex-correspondence-1}
\end{equation}

Combining (\ref{vertex-correspondence-1}) and (\ref{inversion-elementary-2}),
we find
\begin{equation}
z_{k}^{(2)}=-1/z_{k}^{(1)}\,. \label{z-k-1-2-inv}
\end{equation}

\textbf{Conclusion.} It is sufficient to prove Statement 9 for the case when
parameters $z_{k}^{(a)}$ obey relations (\ref{z-1-2-1-zero}),
(\ref{z-k-1-2-inv}).

\subsubsection{Straightforward way is not efficient}

Our aim is to derive relation (\ref{I-m-ren-invariant-inversion-2}) assuming
assumptions of Statement 9. At this step one may think about expressing
$I_{m}^{\text{ren}}$ via functions $\Pi_{P}^{(n_{\text{v}}-1)}$ according to
eqs. (\ref{I-1-ren-via-Pi}), (\ref{I-2-ren-via-Pi}) and to reduce the
derivation of (\ref{I-m-ren-invariant-inversion-2}) to the derivation of
certain relations for functions $\Pi_{P}^{(n_{\text{v}}-1)}$. In principle,
one can prove Statement 9 in this way. However,

\begin{itemize}
\item this straightforward method requires a rather boring calculation,

\item one can learn too little from this calculation because the final result
comes as a sort of miracle.
\end{itemize}

It is much more instructive to use another way.

\subsubsection{Alternative representation for analytical regularization}

Let us derive another representation for $I_{m}^{\text{ren}}$

\begin{itemize}
\item which is equivalent to (\ref{I-1-ren-via-Pi}), (\ref{I-2-ren-via-Pi}) representation

\item but allows for a much better tracing of the covariance with respect to
inversion (and with respect to other linear fractional transformations).
\end{itemize}

Note that we want to use the \emph{same analytical regularization} as before.
But we do not want to work in terms of functions $\Pi_{P}^{(n_{\text{v}}-1)}$.
We prefer to work in terms of functions that carry the same information as
$\Pi_{P}^{(n_{\text{v}}-1)}$ (including analytical properties) but in a better representation

\begin{itemize}
\item explicitly showing \emph{invariance} of the original nonregularized
expressions for $I_{m}$ with respect to inversion (and other fractional transformations),

\item demonstrating how analytical regularization \emph{modifies} the behavior
of $I_{m}$ under inversion,

\item showing how invariance with respect to inversion \emph{is restored}
after the analytical renormalization.
\end{itemize}

In principle, one can derive this new covariant representation for
$I_{m}^{\text{ren}}$ directly from our expressions $I_{m}^{\text{ren}}$ from
(\ref{I-1-ren-via-Pi}), (\ref{I-2-ren-via-Pi}) via functions $\Pi
_{P}^{(n_{\text{v}}-1)}$.

But it is much more instructive an much easier to read this new covariant
representation from the results of ref. \cite{Pobylitsa-2019}. Eqs. (4.30) and
(4.31) in ref. \cite{Pobylitsa-2019} provide non-regularized and
ultraviolet\ divergent expressions for $I_{1}$ and $I_{2}$:

\begin{equation}
I_{1}=\int_{\mathbb{C}_{+}}d^{2}z\left|  \frac{d\zeta}{dz}\right|
^{-2}\left|  \operatorname{Im}\,z\right|  ^{-4}, \label{I-mixed-1}
\end{equation}

\begin{equation}
I_{2}=\int_{\mathbb{C}_{+}}d^{2}z\left|  \frac{d\zeta}{dz}\right|
^{-2}\left|  \left\{  \zeta,z\right\}  \right|  ^{2}. \label{I-holo-1}
\end{equation}

Here $\zeta(z)$ is a conformal mapping from $z$ semiplane $\mathbb{C}_{+}$ to
the polygon placed on the plane of complex variable $\zeta$. On the RHS of
(\ref{I-holo-1}) there appears Schwarz derivative $\left\{  \zeta,z\right\}  $
defined by

\begin{equation}
\left\{  \zeta,z\right\}  =\frac{\zeta^{\prime\prime\prime}}{\zeta^{\prime}
}-\frac{3}{2}\left(  \frac{\zeta^{\prime\prime}}{\zeta^{\prime}}\right)  ^{2}
\label{Schwarz-derivative-def-3}
\end{equation}
where the prime stands for derivative $d/dz$:
\begin{equation}
\zeta^{\prime}=\frac{d\zeta}{dz},\quad\zeta^{\prime\prime}=\frac{d^{2}\zeta
}{dz^{2}},\quad z^{\prime\prime\prime}=\frac{d^{3}\zeta}{dz^{3}}.
\end{equation}

Representations (\ref{I-mixed-1}) and (\ref{I-holo-1}) have one bad and one
good feature:

\begin{itemize}
\item \emph{Integrals} (\ref{I-mixed-1}) and (\ref{I-holo-1}) are
\emph{divergent} on the boundary of $\mathbb{C}_{+}$ (i.e. on the real axis
and at and at infinity).

\item The \emph{integrands} of integrals (\ref{I-mixed-1}) and (\ref{I-holo-1}
) are \emph{invariant} with respect to linear fractional transformations
(\ref{G-z-linfrac}) if one combines the integrand and the Jacobian
corresponding to the change of the integration variable:
\end{itemize}

\begin{equation}
\left|  \frac{d\zeta}{dz^{\prime}}\right|  ^{-2}\left|  \operatorname{Im}
\,z^{\prime}\right|  ^{-4}d^{2}z^{\prime}=\left|  \frac{d\zeta}{dz}\right|
^{-2}\left|  \operatorname{Im}\,z\right|  ^{-4}d^{2}z\,,
\label{I1-reg-z-covariance}
\end{equation}
\begin{equation}
\left|  \frac{d\zeta^{\prime}}{dz}\right|  ^{-2}\left|  \left\{
\zeta,z^{\prime}\right\}  \right|  ^{2}d^{2}z^{\prime}=\left|  \frac{d\zeta
}{dz}\right|  ^{-2}\left|  \left\{  \zeta,z\right\}  \right|  ^{2}d^{2}z\,.
\label{I2-reg-z-covariance}
\end{equation}
This invariance appears naturally in the context of ref. \cite{Pobylitsa-2019}
. Relation (\ref{I1-reg-z-covariance}) is trivial. Relation
(\ref{I2-reg-z-covariance}) follows from standard properties of Schwarz
derivative (\ref{Schwarz-derivative-def-3}) with respect to linear fractional transformations.

Let us introduce compact notation
\begin{equation}
J_{1}\left[  \zeta,z\right]  =\left|  \frac{d\zeta}{dz}\right|  ^{-2}\left|
\operatorname{Im}\,z\right|  ^{-4}\,,
\end{equation}
\begin{equation}
J_{2}\left[  \zeta,z\right]  =\left|  \frac{d\zeta}{dz}\right|  ^{-2}\left|
\left\{  \zeta,z\right\}  \right|  ^{2}\,.
\end{equation}
Then
\begin{equation}
I_{m}=\int_{\mathbb{C}_{+}}d^{2}zJ_{m}\left[  \zeta,z\right]
\,.\label{I-m-non-reg}
\end{equation}
Under linear fractional transformations (\ref{G-z-linfrac})
\begin{equation}
J_{m}\left[  \zeta^{\prime},z^{\prime}\right]  d^{2}z^{\prime}=J_{m}\left[
\zeta,z\right]  d^{2}z\,.\label{J-m-covariance}
\end{equation}

Analytical regularization corresponds to the introduction of the temporary
factor
\begin{equation}
\,\left|  \operatorname{Im}\,z\right|  ^{\lambda}\prod_{k=1}^{n_{\text{v}}
-1}\left|  z-z_{k}\right|  ^{\mu_{k}}
\end{equation}
in the integrands of the non-regularized version (\ref{I-m-non-reg})
\begin{equation}
I_{m}^{\text{reg}}\left(  \lambda,\left\{  \mu_{k}\right\}  _{k=1}
^{n_{\text{v}}-1};\left\{  z_{k}\right\}  _{k=1}^{n_{\text{v}}-1}
|\zeta\right)  =\int_{\mathbb{C}_{+}}d^{2}zJ_{m}\left[  \zeta,z\right]
\,\left|  \operatorname{Im}\,z\right|  ^{\lambda}\prod_{k=1}^{n_{\text{v}}
-1}\left|  z-z_{k}\right|  ^{\mu_{k}}\,. \label{I1-reg-lambda-mu}
\end{equation}
Here it is assumed that $\zeta(z)$ is SC$_{\infty}$ mapping of the form
(\ref{SC-diff-eq}) and parameters $z_{k}$ are taken from (\ref{SC-diff-eq}).
This regularized version $I_{m}^{\text{reg}}$ depends on conformal mapping
$\zeta(z)$ representing the polygon so that we include $\zeta$ in the list of
arguments of $I_{m}^{\text{reg}}$.

Some comments must be made about the list arguments for $I_{m}^{\text{reg}}$
which contains real parameters $\left\{  z_{k}\right\}  _{k=1}^{n_{\text{v}
}-1}$ and SC$_{\infty}$ mapping $\zeta$ (understood as a function and not as a
complex variable). Obviously objects $\left\{  z_{k}\right\}  _{k=1}
^{n_{\text{v}}-1}$ and $\zeta$ are not independent: they are connected by
equation (\ref{SC-diff-eq}). Nevertheless in our work with relations
containing different SC$_{\infty}$ mappings $\zeta^{(a)}$ it is convenient to
include both $\left\{  z_{k}^{\left(  a\right)  }\right\}  _{k=1}
^{n_{\text{v}}-1}$ and $\zeta^{\left(  a\right)  }$ in the argument list of
$I_{m}^{\text{reg}}$ for a careful tracing of $a$-dependences.

If

\begin{itemize}
\item one computes $J_{m}\left[  \zeta,z\right]  $ in terms of elementary functions,

\item properly represents parameters $\lambda,\left\{  \mu_{k}\right\}
_{k=1}^{n_{\text{v}}-1}$ as a linear combinations of native parameters
$\alpha,\left\{  \nu_{k}\right\}  _{k=1}^{n_{\text{v}}-1}$
\end{itemize}

then one will see that (\ref{I1-reg-lambda-mu}) provides the same analytical
regularization as eqs. (\ref{I-1-ren-via-Pi}), (\ref{I-2-ren-via-Pi}).

However, there is no need to do this work because

\begin{itemize}
\item the above statements are a sort of reverse engineering of the work that
has already been done in ref. \cite{Pobylitsa-2019};

\item in order to prove Statement 9, we do not need explicit expressions for
$\lambda,\left\{  \mu_{k}\right\}  _{k=1}^{n_{\text{v}}-1}$ via $\alpha
,\left\{  \nu_{k}\right\}  _{k=1}^{n_{\text{v}}-1}$; it is sufficient to know
that functions $I_{m}^{\text{reg}}$ contain essentially the same pole
structure as functions $\Pi_{P}^{(n)}$ so that both $I_{m}^{\text{reg}}$ and
$\Pi_{P}^{(n)}$ are regular at the `final physical point' corresponding to the
transition from regularization to renormalization.
\end{itemize}

The rest of the work is straightforward. One must compute integrals
(\ref{I1-reg-lambda-mu}) in the region of $\lambda,\left\{  \mu_{k}\right\}
_{k=1}^{n_{\text{v}}-1}$ where these integrals are convergent and then one
must perform analytical continuation to `physical point' $\lambda=0$, $\mu
_{k}=0$:
\begin{equation}
I_{m}^{\text{ren}}=\left[  I_{m}^{\text{reg}}\left(  \lambda,\left\{  \mu
_{k}\right\}  _{k=1}^{n_{\text{v}}-1}\right)  ;\left\{  z_{k}\right\}
_{k=1}^{n_{\text{v}}-1}|\zeta\right]  _{\text{analyt. cont. }\lambda
\rightarrow0,\mu_{k}\rightarrow0}\,. \label{I-ren-lambda-mu}
\end{equation}

\subsubsection{Inversion and alternative representation for analytical regularization}

Now we return to the proof of Statement 9. We want to prove that SC$_{\infty}$
mappings $\zeta^{(1)}(z)$ and $\zeta^{(2)}(z)$ connected by relation
(\ref{inversion-elementary}) lead to the same $I_{m}^{\text{ren}}$. We can
apply eq. (\ref{I-ren-lambda-mu}) to both SC$_{\infty}$ mappings $\zeta^{(1)}$
and $\zeta^{(2)}$:

\begin{align}
&  I_{m}^{\text{reg}}\left(  \lambda,\left\{  \mu_{k}\right\}  _{k=1}
^{n_{\text{v}}-1};\left\{  z_{k}^{(a)}\right\}  _{k=1}^{n_{\text{v}}-1}
|\zeta^{(a)}\right)  \nonumber\\
&  =\int_{\mathbb{C}_{+}}d^{2}zJ_{m}\left[  \zeta^{(a)},z\right]  \left|
\operatorname{Im}\,z\right|  ^{\lambda}\prod_{k=1}^{n_{\text{v}}-1}\left|
z-z_{k}^{(i)}\right|  ^{\mu_{k}}.\label{I1-via-F-i}
\end{align}
Our aim is to show that
\begin{equation}
I_{m}^{\text{ren}}\left(  \zeta^{(1)}\right)  =I_{m}^{\text{ren}}\left(
\zeta^{(2)}\right)  \,.\label{I-m-ren-F1-F2-equal}
\end{equation}
Combining (\ref{I-ren-lambda-mu}), (\ref{I-m-ren-F1-F2-equal}) and
\emph{Conclusion} in the end of section \ref{invariance-main-result-section},
we see that the proof of Statement 9 reduces to

\textbf{Statement 10.}

If

1) $\zeta^{(1)}$ and $\zeta^{(2)}$ are SC$_{\infty}$ mappings related by
inversion:
\begin{equation}
\zeta^{(2)}(z)=\zeta^{(1)}(-1/z)\,,
\end{equation}

2) $\left\{  z_{k}^{(a)}\right\}  $ are parameters of SC$_{\infty}$ mapping
$\zeta^{(a)}$ as they appear in eq. (\ref{SC-diff-eq}),

3) $z_{k}^{(a)}$ obey relations

\begin{equation}
z_{1}^{(2)}=z_{1}^{(1)}=0\,, \label{z2-z1-1-3}
\end{equation}
\begin{equation}
z_{k}^{(2)}=-1/z_{k}^{(1)}\quad(2\leq k\leq n_{\text{v}}-1)\,.
\label{z2-z1-2-3}
\end{equation}
then
\begin{align}
&  \left[  I_{m}^{\text{reg}}\left(  \lambda,\left\{  \mu_{k}\right\}
_{k=1}^{n_{\text{v}}-1};\left\{  z_{k}^{(1)}\right\}  _{k=1}^{n_{\text{v}}
-1}|\zeta^{(1)}\right)  \right]  _{\text{analyt. cont. }\lambda\rightarrow
0,\,\mu_{k}\rightarrow0}\nonumber\\
&  =\left[  I_{m}^{\text{reg}}\left(  \lambda,\left\{  \mu_{k}\right\}
_{k=1}^{n_{\text{v}}-1};\left\{  z_{k}^{(2)}\right\}  _{k=1}^{n_{\text{v}}
-1}|\zeta^{(2)}\right)  \right]  _{\text{analyt. cont. }\lambda\rightarrow
0,\,\mu_{k}\rightarrow0}\,, \label{reg-F1-F2-equal}
\end{align}

Statement 10 will be derived from Statement 11 in section
\ref{st-10-from-st-11-section}.

\subsubsection{Transformation of $I_{m}^{\text{reg}}$ under inversion}

In order to derive Statement 10 we need

\textbf{Statement 11.}

If one

a) makes assumptions 1, 2, 3 of Statement 10,

b) considers $\lambda^{(1)},\left\{  \mu_{k}^{(1)}\right\}  _{k=1}
^{n_{\text{v}}-1}$ as independent complex variables,

c) considers $\lambda^{(2)},\left\{  \mu_{k}^{(2)}\right\}  _{k=1}
^{n_{\text{v}}-1}$ as functions of variables $\lambda^{(1)},\left\{  \mu
_{k}^{(1)}\right\}  _{k=1}^{n_{\text{v}}-1}$ defined by relations

\begin{equation}
\lambda^{(2)}=\lambda^{(1)}\,, \label{lampbda-2-1}
\end{equation}
\begin{equation}
\mu_{1}^{(2)}=-2\lambda^{(1)}-\sum_{k=1}^{n_{\text{v}}-1}\mu_{k}^{(1)}\,,
\label{mu-2-via-mu1-1}
\end{equation}
\begin{equation}
\mu_{k}^{(2)}=\mu_{k}^{(1)}\quad(2\leq k\leq n_{\text{v}}-1)
\label{mu-2-via-mu1-2}
\end{equation}

then the following identity holds
\begin{align}
&  I_{m}^{\text{reg}}\left(  \lambda^{(1)},\left\{  \mu_{k}^{(1)}\right\}
_{k=1}^{n_{\text{v}}-1};\left\{  z_{k}^{(1)}\right\}  _{k=1}^{n_{\text{v}}
-1}|\zeta^{(1)}\right) \nonumber\\
&  =\left(  \prod_{k=2}^{n_{\text{v}}-1}\left|  z_{k}^{(1)}\right|  ^{\mu
_{k}^{(1)}}\right)  I_{m}^{\text{reg}}\left(  \lambda^{(2)},\left\{  \mu
_{k}^{(2)}\right\}  _{k=1}^{n_{\text{v}}-1};\left\{  z_{k}^{(2)}\right\}
_{k=1}^{n_{\text{v}}-1}|\zeta^{(2)}\right)  \,.
\label{I-m-F1-F2-invesrion-res}
\end{align}

\textbf{Remark.} LHS and RHS of eq. (\ref{I-m-F1-F2-invesrion-res})\ are
meromorphic functions of independent complex variables $\lambda^{(1)},\left\{
\mu_{k}^{(1)}\right\}  _{k=1}^{n_{\text{v}}-1}$ at fixed mappings $\zeta
^{(a)}$ and fixed $\left\{  z_{k}^{(a)}\right\}  _{k=1}^{n_{\text{v}}-1}$.

\subsubsection{From regularization to renormalization}

\label{st-10-from-st-11-section}

Statement 10 is a trivial consequence of Statement 11. Indeed, point
\begin{equation}
\lambda^{(1)}=0,\,\,\mu_{k}^{(1)}=0\quad(1\leq k\leq n_{\text{v}}-1)
\end{equation}
is a regular point of the meromorphic function represented by identity
(\ref{I-m-F1-F2-invesrion-res}). According to eqs. (\ref{lampbda-2-1}) --
(\ref{mu-2-via-mu1-2}) at this point we have
\begin{equation}
\lambda^{(2)}=0,\,\,\mu_{k}^{(2)}=0\quad(1\leq k\leq n_{\text{v}}-1)\,.
\end{equation}

This completes the derivation of Statement 10 from Statement 11.

Now only one problem remains: we must prove Statement 11.

\subsubsection{Final step: proof of Statement 11}

Let us set in eq. \bigskip(\ref{I1-via-F-i})
\begin{equation}
a=1
\end{equation}
and let us add superscript (1) to variables $\lambda,\left\{  \mu_{k}\right\}
_{k=1}^{n_{\text{v}}-1}$:
\begin{align}
&  I_{m}^{\text{reg}}\left(  \lambda^{(1)},\left\{  \mu_{k}^{(1)}\right\}
_{k=1}^{n_{\text{v}}-1};\left\{  z_{k}^{(1)}\right\}  _{k=1}^{n_{\text{v}}
-1}|\zeta^{(1)}\right) \nonumber\\
&  =\int_{\mathbb{C}_{+}}d^{2}zJ_{m}\left[  \zeta^{(1)},z\right]  \,\left|
\operatorname{Im}\,z\right|  ^{\lambda^{(1)}}\prod_{k=1}^{n_{\text{v}}
-1}\left|  z-z_{k}^{(1)}\right|  ^{\mu_{k}^{(1)}}\,.
\end{align}
We know that the convergence region of this integral in the $\lambda
^{(1)},\left\{  \mu_{k}^{(1)}\right\}  _{k=1}^{n_{\text{v}}-1}$ space is
non-empty and we work in this convergence region. We change the integration
variable $z=-1/z^{\prime}$ and we use eqs. (\ref{J-m-covariance}),
(\ref{inversion-elementary}). We obtain
\begin{align}
&  I_{m}^{\text{reg}}\left(  \lambda^{(1)},\left\{  \mu_{k}^{(1)}\right\}
_{k=1}^{n_{\text{v}}-1};\left\{  z_{k}^{(1)}\right\}  _{k=1}^{n_{\text{v}}
-1}|\zeta^{(1)}\right) \nonumber\\
&  =\int_{\mathbb{C}_{+}}d^{2}z^{\prime}J_{m}\left[  \zeta^{(2)},z^{\prime
}\right]  \left|  \operatorname{Im}\,\left[  -\left(  z^{\prime}\right)
^{-1}\right]  \right|  ^{\lambda^{(1)}}\prod_{k=1}^{n_{\text{v}}-1}\left|
-\left(  z^{\prime}\right)  ^{-1}-z_{k}^{(1)}\right|  ^{\mu_{k}^{(1)}}.
\end{align}
Using (\ref{z2-z1-1-3}), (\ref{z2-z1-2-3}), we obtain
\begin{align}
&  \left|  \operatorname{Im}\,\left[  -\left(  z^{\prime}\right)
^{-1}\right]  \right|  ^{\lambda^{(1)}}\prod_{k=1}^{n_{\text{v}}-1}\left|
-\left(  z^{\prime}\right)  ^{-1}-z_{k}^{(1)}\right|  ^{\mu_{k}^{(1)}
}\nonumber\\
&  =\left|  \left|  z^{\prime}\right|  ^{-2}\operatorname{Im}\,z^{\prime
}\right|  ^{\lambda^{(1)}}\left|  z^{\prime}\right|  ^{-\mu_{1}^{(1)}}
\prod_{k=2}^{n_{\text{v}}-1}\left|  -\left(  z^{\prime}\right)  ^{-1}
-z_{k}^{(1)}\right|  ^{\mu_{k}^{(1)}}\nonumber\\
&  =\left|  \operatorname{Im}\,z^{\prime}\right|  ^{\lambda^{(1)}}\left|
z^{\prime}\right|  ^{-\mu_{1}^{(1)}-2\lambda^{(1)}}\prod_{k=2}^{n_{\text{v}
}-1}\left|  \left(  z^{\prime}\right)  ^{-1}z_{k}^{(1)}\right|  ^{\mu
_{k}^{(1)}}\prod_{k=2}^{n_{\text{v}}-1}\left|  z^{\prime}+\left(  z_{k}
^{(1)}\right)  ^{-1}\right|  ^{\mu_{k}^{(1)}}\nonumber\\
&  =\left(  \prod_{k=2}^{n_{\text{v}}-1}\left|  z_{k}^{(1)}\right|  ^{\mu
_{k}^{(1)}}\right)  \left|  \operatorname{Im}\,z^{\prime}\right|
^{\lambda^{(1)}}\left|  z^{\prime}\right|  ^{-2\lambda^{(1)}-\sum
_{k=1}^{n_{\text{v}}-1}\mu_{k}^{(1)}}\prod_{k=2}^{n_{\text{v}}-1}\left|
z^{\prime}-z_{k}^{(2)}\right|  ^{\mu_{k}^{(1)}}\nonumber\\
&  =\left(  \prod_{k=2}^{n_{\text{v}}-1}\left|  z_{k}^{(1)}\right|  ^{\mu
_{k}^{(1)}}\right)  \left|  \operatorname{Im}\,z^{\prime}\right|
^{\lambda^{(1)}}\left|  z^{\prime}-z_{1}^{(2)}\right|  ^{-2\lambda^{(1)}
-\sum_{k=1}^{n_{\text{v}}-1}\mu_{k}^{(1)}}\prod_{k=2}^{n_{\text{v}}-1}\left|
z^{\prime}-z_{k}^{(2)}\right|  ^{\mu_{k}^{(1)}}\,.
\end{align}
Hence
\begin{align}
&  I_{1}^{\text{reg}}\left(  \lambda^{(1)},\left\{  \mu_{k}^{(1)}\right\}
_{k=1}^{n_{\text{v}}-1};\left\{  z_{k}^{(1)}\right\}  _{k=1}^{n_{\text{v}}
-1}|\zeta^{(1)}\right) \nonumber\\
&  =\int_{\mathbb{C}_{+}}d^{2}z^{\prime}J_{m}\left[  \zeta^{(2)},z^{\prime
}\right] \nonumber\\
&  \times\left(  \prod_{k=2}^{n_{\text{v}}-1}\left|  z_{k}^{(1)}\right|
^{\mu_{k}^{(1)}}\right)  \left|  \operatorname{Im}\,z^{\prime}\right|
^{\lambda^{(1)}}\left|  z^{\prime}-z_{1}^{(2)}\right|  ^{-2\lambda^{(1)}
-\sum_{k=1}^{n_{\text{v}}-1}\mu_{k}^{(1)}}\prod_{k=2}^{n_{\text{v}}-1}\left|
z^{\prime}-z_{k}^{(2)}\right|  ^{\mu_{k}^{(1)}}\,.
\end{align}
Let us change the notation of the integration variable from $z^{\prime}$ to
$z$ and use (\ref{lampbda-2-1}) -- (\ref{mu-2-via-mu1-2})
\begin{align}
&  I_{m}^{\text{reg}}\left(  \lambda^{(1)},\left\{  \mu_{k}^{(1)}\right\}
_{k=1}^{n_{\text{v}}-1};\left\{  z_{k}^{(1)}\right\}  _{k=1}^{n_{\text{v}}
-1}|\zeta^{(1)}\right) \nonumber\\
&  =\left(  \prod_{k=2}^{n_{\text{v}}-1}\left|  z_{k}^{(1)}\right|  ^{\mu
_{k}^{(1)}}\right)  \int_{\mathbb{C}_{+}}d^{2}zJ_{m}\left[  \zeta
^{(2)},z\right] \nonumber\\
&  \times\left|  \operatorname{Im}\,z\right|  ^{\lambda^{(2)}}\prod
_{k=1}^{n_{\text{v}}-1}\left|  z-z_{k}^{(2)}\right|  ^{\mu_{k}^{(2)}}\,.
\label{I-m-sub-1}
\end{align}
Now let us set $a=2$ in eq. (\ref{I1-via-F-i})
\begin{align}
&  I_{m}^{\text{reg}}\left(  \lambda^{(2)},\left\{  \mu_{k}^{(2)}\right\}
_{k=1}^{n_{\text{v}}-1};\left\{  z_{k}^{(2)}\right\}  _{k=1}^{n_{\text{v}}
-1}|\zeta^{(2)}\right) \nonumber\\
&  =\int_{\mathbb{C}_{+}}d^{2}zJ_{m}\left[  \zeta^{(2)},z\right]  \,\left|
\operatorname{Im}\,z\right|  ^{\lambda^{(2)}}\prod_{k=1}^{n_{\text{v}}
-1}\left|  z-z_{k}^{(2)}\right|  ^{\mu_{k}^{(2)}}\,. \label{I-m-sub-2}
\end{align}
Comparing eqs. (\ref{I-m-sub-1}) and (\ref{I-m-sub-1}), we obtain identity
(\ref{I-m-F1-F2-invesrion-res}).

Thus Statement 11 is proved. This also completes the proof the main Statement 4.

\section{Conclusions}

The results of this work generalize observations made in ref.
\cite{Pobylitsa-2019} for triangular diagrams to the case of arbitrary
polygons. We have proved that analytical regularization formulated in terms of
SC$_{\infty}$\ mapping is internally consistent:

\begin{itemize}
\item Integrals representing functions $\Pi_{P}^{(n)}$ are convergent in a
non-empty region of complex parameters $\alpha,\left\{  \gamma_{k}\right\}
_{k=1}^{n}$.

\item After defining $\Pi_{P}^{(n)}$ in this convergence region one can
perform analytical continuation in $\mathbb{C}^{n+1}$ space of parameters
$\alpha,\left\{  \gamma_{k}\right\}  _{k=1}^{n}$.

\item Resulting function $\Pi_{P}^{(n)}$ is meromorphic in $\mathbb{C}^{n+1}$,
i.e. it has only poles but no branching singularities.

\item Analytical continuation of functions $\Pi_{P}^{(n)}$ to the physical
point (i.e. to the values of $\alpha,\left\{  \gamma_{k}\right\}  _{k=1}^{n}$
needed for the calculation of $I_{m}^{\text{ren}}$) is regular so that our
analytical renormalization provides finite results for renormalized quantities
$I_{m}^{\text{ren}}$.

\item The absence of branching singularities in $\Pi_{P}^{(n)}$ means that
analytical continuation of $\Pi_{P}^{(n)}$ from the convergence region to the
physical point does not depend on the path of continuation, i.e. our
analytical regularization has no ambiguities.

\item The final result for $I_{m}^{\text{ren}}$ is independent of the
SC$_{\infty}$\ parametrization of the polygon.
\end{itemize}

These results put analytical regularization in terms of SC$_{\infty}$ mapping
on solid ground.

\section{Acknowledgements}

I am grateful to many people with whom I discussed various aspects of this
work. I\ am indebted to my teacher and friend Victor Petrov who passed away on
September 22, 2021. The problem of confinement was his passion. Vitya's deep
understanding of the difference between the problem of confinement in the real
world and in abstract mathematical models made discussions with him
stimulating in both directions: his physical intuition suggested heuristic
solutions of nontrivial mathematical problems and helped filter physically
significant signals from the rattle of mathematical toys.

\appendix                                  \renewcommand{\theequation}
{\Alph{section}.\arabic{equation}} 

\section{Pole structure of $\Pi_{1}^{(2)}$}

\setcounter{equation}{0} 

\label{pole-structure-Pi-2-1-section}

In this appendix we check that function $\Pi_{1}^{(2)}$ has a pole structure
compatible with the general pole representation (\ref{K-rough-pole-structure}
). In this special case:
\begin{equation}
n=2,
\end{equation}
\begin{equation}
P(z,z^{\ast})=1
\end{equation}
function $\Pi_{P}^{(n)}$ was computed in ref. \cite{Pobylitsa-2019}:

\begin{align}
&  \Pi_{1}^{(2)}\left(  \alpha,\left\{  \gamma_{1},\gamma_{2}\right\}
,\left\{  z_{1},z_{2}\right\}  \right) \nonumber\\
&  =\left[  \frac{\sqrt{\pi}}{2}\left|  z_{1}-z_{2}\right|  ^{-\gamma_{3}
}\Gamma\left(  \frac{\alpha}{2}\right)  \prod_{k=1}^{3}\frac{\Gamma\left(
\frac{\gamma_{k}}{2}\right)  }{\Gamma\left(  \frac{\alpha+1}{2}-\frac{\gamma
_{k}}{2}\right)  }\right]  _{\gamma_{3}=\left(  \alpha+1\right)  -\left(
\gamma_{1}+\gamma_{2}\right)  }\,. \label{K-1-2-res}
\end{align}
The product on the RHS can be rearranged:
\begin{align}
&  \Pi_{1}^{(2)}\left(  \alpha,\left\{  \gamma_{1},\gamma_{2}\right\}
,\left\{  z_{1},z_{2}\right\}  \right) \nonumber\\
&  =\Gamma\left(  \frac{\alpha}{2}\right)  \Gamma\left(  \frac{\gamma_{1}}
{2}\right)  \Gamma\left(  \frac{\gamma_{2}}{2}\right)  \Gamma\left(
\frac{\alpha+1-\gamma_{1}+\gamma_{2}}{2}\right)  H_{1}^{(2)}\left(
\alpha,\left\{  \gamma_{1},\gamma_{2}\right\}  ,\left\{  z_{1},z_{2}\right\}
\right)  \,. \label{Pi-2-1-explicit}
\end{align}
Here
\begin{align}
&  H_{1}^{(2)}\left(  \alpha,\left\{  \gamma_{1},\gamma_{2}\right\}  ,\left\{
z_{1},z_{2}\right\}  \right) \nonumber\\
&  =\left\{  \frac{\sqrt{\pi}}{2}\left|  z_{1}-z_{2}\right|  ^{-\gamma_{3}
}\prod_{k=1}^{3}\left[  \Gamma\left(  \frac{\alpha+1}{2}-\frac{\gamma_{k}}
{2}\right)  \right]  ^{-1}\right\}  _{\gamma_{3}=\left(  \alpha+1\right)
-\left(  \gamma_{1}+\gamma_{2}\right)  }
\end{align}
is an entire function of $\alpha,\left\{  \gamma_{1},\gamma_{2}\right\}  $ (at
any fixed real $z_{1}\neq z_{2}$).

Expression (\ref{Pi-2-1-explicit}) agrees with the general representation
(\ref{K-rough-pole-structure}). Indeed, in the case of trivial polynomial
$P=1$ we have according to (\ref{M-P-min-def}):
\begin{equation}
M_{P}=0
\end{equation}
so that in eq. (\ref{K-rough-pole-structure}) the $M_{P}$ dependent $\Gamma$
function simplifies to
\begin{equation}
\Gamma\left(  \frac{1}{2}\left(  -2M_{P}+\left(  n-1\right)  \left(
\alpha+1\right)  -\sum_{k=1}^{n}\gamma_{k}\right)  \right)  _{M_{P}
=1,\,n=2}=\Gamma\left(  \frac{\alpha+1-\gamma_{1}+\gamma_{2}}{2}\right)  \,.
\end{equation}

\section{Integral $I\left(  \alpha,\gamma\right)  $}

\setcounter{equation}{0} 

\label{integral-I-alpha-gamma-section}

In this appendix we determine the convergence region of integral
\begin{equation}
I\left(  \alpha,\gamma\right)  =\int_{\left|  z\right|  <1}d^{2}z\,\left|
\operatorname{Im}\,z\right|  ^{\text{\thinspace}\alpha-1}\left|  z\right|
^{\gamma-\alpha-1} \label{I-alpha-gamma-def}
\end{equation}
and compute it in this convergence region.

In terms of real integration variables
\begin{equation}
x=\text{Re\thinspace}z\,,
\end{equation}
\begin{equation}
y=\operatorname{Im}\text{\thinspace}z\,
\end{equation}
we have
\begin{align}
I\left(  \alpha,\gamma\right)   &  =\int_{x^{2}+y^{2}<1}dxdy\,y^{\alpha
-1}\left(  x^{2}+y^{2}\right)  ^{\frac{\gamma-\alpha-1}{2}}\nonumber\\
&  =4\int_{x>0,y>0,x^{2}+y^{2}<1}dxdy\,y^{\alpha-1}\left(  x^{2}+y^{2}\right)
^{\frac{\gamma-\alpha-1}{2}}\,.
\end{align}
Next we change integration variables:
\begin{equation}
p=x^{2}\,,
\end{equation}
\begin{equation}
q=y^{2}\,.
\end{equation}
Then
\begin{equation}
I\left(  \alpha,\gamma\right)  =\int_{\substack{0<p,q\\p+q<1}
}dpdq\,q^{\frac{\alpha}{2}-1}p^{-1/2}\left(  p+q\right)  ^{\frac{\gamma
-\alpha-1}{2}}\,.
\end{equation}
Our next change of integration variables is
\begin{equation}
q=st\,, \label{change-q-s-t}
\end{equation}
\begin{equation}
p=\left(  1-s\right)  t\,. \label{change-p-s-t}
\end{equation}
with Jacobian
\begin{equation}
\frac{D\left(  p,q\right)  }{D\left(  t,s\right)  }=\det\left(
\begin{array}
[c]{cc}
1-s & -t\\
s & t
\end{array}
\right)  =t\,. \label{Jacobian-p-q-s-t}
\end{equation}
Then
\begin{align}
I\left(  \alpha,\gamma\right)   &  =\int_{0}^{1}ds\int_{0}^{1}dt\,t\left(
st\right)  ^{\frac{\alpha}{2}-1}\left[  \left(  1-s\right)  t\right]
^{-1/2}t^{\frac{\gamma-\alpha-1}{2}}\nonumber\\
&  =\int_{0}^{1}ds\int_{0}^{1}dt\,\frac{s^{\frac{\alpha}{2}-1}}{\left(
1-s\right)  ^{1/2}}t^{\frac{\gamma}{2}-1}=\left[  \int_{0}^{1}
dss^{\frac{\alpha}{2}-1}\left(  1-s\right)  ^{-1/2}\right]  \left[  \int
_{0}^{1}dt\,t^{\frac{\gamma}{2}-1}\right]  \,. \label{I-alpha-gamma-res-1}
\end{align}

The two integrals on the RHS\ are (absolutely) convergent if two conditions
hold:
\begin{equation}
\operatorname{Re}\alpha>0\,, \label{I-conv-alpha}
\end{equation}
\begin{equation}
\operatorname{Re}\gamma>0\,. \label{I-conv-gamma}
\end{equation}
The integrals on the RHS of (\ref{I-alpha-gamma-res-1}) can be easily computed
(in their convergence regions):

\begin{equation}
\int_{0}^{1}dss^{\frac{\alpha}{2}-1}\left(  1-s\right)  ^{-1/2}=B\left(
\frac{\alpha}{2},\frac{1}{2}\right)  =\sqrt{\pi}\frac{\Gamma\left(
\frac{\alpha}{2}\right)  }{\Gamma\left(  \frac{\alpha+1}{2}\right)  }\,,
\end{equation}
\begin{equation}
\int_{0}^{1}dt\,t^{\frac{\gamma}{2}-1}=\frac{2}{\gamma}\,.
\end{equation}
and we arrive at

\textbf{Statement.}

Integral
\begin{equation}
I\left(  \alpha,\gamma\right)  =\int_{\left|  z\right|  <1}d^{2}z\,\left|
\operatorname{Im}\,z\right|  ^{\text{\thinspace}\alpha-1}\left|  z\right|
^{\gamma-\alpha-1}
\end{equation}
is absolutely convergent if and only if
\begin{equation}
\text{Re\thinspace}\alpha>0\,,
\end{equation}
\begin{equation}
\text{Re\thinspace}\gamma>0
\end{equation}
and in this region of $\alpha,\gamma$
\begin{equation}
I\left(  \alpha,\gamma\right)  =\frac{2\sqrt{\pi}}{\gamma}\frac{\Gamma\left(
\frac{\alpha}{2}\right)  }{\Gamma\left(  \frac{\alpha+1}{2}\right)  }\,.
\label{I-alpha-gamma-res}
\end{equation}


\begin{thebibliography}{9}                                                                                                

\bibitem {Wilson-74}K.~Wilson, Phys. Rev. D 10 (1974) 2445.

\bibitem {Polyakov-1980}A.~M.~Polyakov, Nucl. Phys. B 164 (1980) 171.

\bibitem {Hooft-1974}G.~'t~Hooft, Nucl. Phys. B72 (1974) 461.

\bibitem {Veneziano-1976}G.~Veneziano, Nucl. Phys. B 117 (1976) 519.

\bibitem {MM-1981}Yu.~M.~Makeenko and A.~A.~Migdal, Nucl. Phys. B 188 (1981) 269.

\bibitem {Migdal-1984}A.~A.~Migdal, Phys. Rept. 102 (1984) 199.

\bibitem {Alvarez}O.~Alvarez, Phys. Rev. D24 (1981) 440.

\bibitem {Arvis-1983}J.~F.~Arvis, Phys. Lett. B 127 (1983) 106.

\bibitem {Ambjorn-Makeenko-2016}J.~Ambj\o rn and Y.~Makeenko, Phys. Lett. B756
(2016) 142, arXiv:1601.00540 [hep-th].

\bibitem {Makeenko-2012-a}Yu.~Makeenko, arXiv:1206.0922 [hep-th].

\bibitem {Makeenko-2009}Yu.~M.~Makeenko and P.~Olesen, Phys. Rev. D 80 (2009)
026002, arXiv:0903.4114 [hep-th].

\bibitem {Makeenko-2012-b}Yu.~M.~Makeenko, Phys. Part. Nucl. 45 (2014) 771,
arXiv:1208.1209 [hep-th].

\bibitem {Luscher:1980fr}M.~L\"{u}scher, K.~Symanzik and P.~Weisz, Nucl. Phys.
B 173 (1980) 365.

\bibitem {Luscher:1980ac}M.~L\"{u}scher, Nucl. Phys. B 180 (1981) 317.

\bibitem {Luscher:1980iy}M.~L\"{u}scher, G.~M\"{u}nster and P.~Weisz, Nucl.
Phys. B 180 (1981) 1.

\bibitem {Luscher:2004ib}M.~L\"{u}scher and P.~Weisz, JHEP 0407 (2004) 014, arXiv:hep-th/0406205.

\bibitem {Meyer:2006qx}H.~B.~Meyer, JHEP 05 (2006) 066, arXiv:hep-th/0602281.

\bibitem {Aharony:2009gg}O.~Aharony and E.~Karzbrun, JHEP 0906 (2009) 012,
arXiv:0903.1927 [hep-th].

\bibitem {Aharony:2010cx}O.~Aharony and M.~Field, JHEP 1101 (2011) 065,
arXiv:1008.2636 [hep-th].

\bibitem {Aharony:2010db}O.~Aharony and N.~Klinghoffer, JHEP 1012 (2010) 058,
arXiv:1008.2648 [hep-th].

\bibitem {Aharony:2011gb}O.~Aharony and M.~Dodelson, JHEP 1202 (2012) 008,
arXiv:1111.5758 [hep-th].

\bibitem {AFK-2012}O.~Aharony, M.~Field and N.~Klinghoffer, JHEP 1204 (2012)
048, arXiv:1111.5757 [hep-th].

\bibitem {AK-2013}O.~Aharony and Z.~Komargodski, JHEP 1305 (2013) 118,
arXiv:1302.6257 [hep-th].

\bibitem {BCGMP-12}M.~Bill\'{o}, M.~Caselle, F.~Gliozzi, M.~Meineri and
R.~Pellegrini, JHEP 05 (2012) 130, arXiv:1202.1984 [hep-th].

\bibitem {GM-12}F.~Gliozzi and M.~Meineri, JHEP 1208 (2012) 056,
arXiv:1207.2912 [hep-th].

\bibitem {Filk-preprint}T.~Filk, Regularization procedure for string
functionals, preprint BONN-HE-81-16 (Bonn U.), Sep 1981.

\bibitem {Dietz-83}K.~Dietz and T.~Filk, Phys. Rev. D 27 (1983) 2944.

\bibitem {BCVG-2010}M.~Bill\'{o}, M.~Caselle, V.~Verduci and M.~Zago, PoS
LATTICE2010 (2010) 273, arXiv:1012.3935 [hep-lat].

\bibitem {Billo:2011fd}M.~Bill\'{o}, M.~Caselle and R.~Pellegrini, JHEP 1201
(2012) 104, arXiv:1107.4356 [hep-th].

\bibitem {Gliozzi:2010zt}F.~Gliozzi, M.~Pepe and U.-J.~Wiese, JHEP 1011 (2010)
053, arXiv:1006.2252 [hep-lat].

\bibitem {Meyer-2010}H.~B.~Meyer, Phys. Rev. D 82 (2010) 106001, arXiv:hep-th/1008.1178.

\bibitem {BM-16}B.~B.~Brandt, M.~Meineri, Int. J. Mod. Phys. A31 (2016)
1643001, arXiv:1603.06969 [hep-th].

\bibitem {Athenodorou:2010cs}A.~Athenodorou, B.~Bringoltz and M.~Teper, JHEP
1102 (2011) 030, arXiv:1007.4720 [hep-lat].

\bibitem {Athenodorou:2011rx}A.~Athenodorou, B.~Bringoltz and M.~Teper, JHEP
1105 (2011) 042, arXiv:1103.5854 [hep-lat].

\bibitem {AT-2013}A.~Athenodorou and M.~Teper, JHEP 1306 (2013) 053,
arXiv:1303.5946 [hep-lat].

\bibitem {AT-2016-a}A.~Athenodorou and M.~Teper, JHEP 1610 (2016) 093,
arXiv:1602.07634 [hep-lat].

\bibitem {AT-2016-b}A.~Athenodorou and M.~Teper, JHEP 1702 (2017) 015,
arXiv:1609.03873 [hep-lat].

\bibitem {Caselle-2021}M. Caselle, Universe 7 (2021) 170, arXiv:2104.10486 [hep-lat].

\bibitem {Pobylitsa-2016}P.V. Pobylitsa, arXiv:1609.05869 [hep-lat].

\bibitem {Pobylitsa-2019}P.V. Pobylitsa, arXiv:1908.01724 [hep-th].

\bibitem {AS-93-journal}E.~Aurell and P.~Salomonson, Commun. Math. Phys. 165
(1994) 233.
\end{thebibliography}
\end{document}